\documentclass[12pt]{article}

\usepackage[dvips]{graphicx}
\usepackage{psfrag}
\usepackage{amssymb}
\usepackage{amsmath}
\usepackage[table]{xcolor}

\usepackage{booktabs}

\definecolor{lgris}{rgb}{0.95,0.95,0.95}
\definecolor{gr}{rgb}{0,0.67,0}
\definecolor{rd}{rgb}{0.75,0,0}

\renewcommand{\arraystretch}{1.2}

\newcommand{\ra}[1]{\renewcommand{\arraystretch}{#1}}
\newcommand{\rb}[1]{\renewcommand{\tabcolsep}{#1}}
\newcommand{\cen}[1]{\multicolumn{1}{c}{#1}}

\newcommand{\eq}[1]{\begin{equation} #1 \end{equation}}

\newcommand{\eqa}[1]{\begin{eqnarray} #1 \end{eqnarray}}

\newcommand{\mpa}[2]{\begin{minipage}{#1} \centering  #2 \end{minipage}}

\newcommand{\av}[1]{\langle #1 \rangle}
\newcommand{\sss}{\scriptscriptstyle}

\newcommand{\op}{\mathcal{O}}
\newcommand{\nn}{\nonumber}

\newcommand{\afb}{A_{\rm FB}}

\newcommand{\bin}{{\rm bin}}

\newcommand{\gr}[1]{{\color{gr} #1}}
\newcommand{\rd}[1]{{\color{rd} #1}}
\newcommand{\bl}[1]{{\color{blue} #1}}
\usepackage{color}

\newcommand{\Ceff}[1]{{\cal C}^{\rm eff}_{#1}}
\newcommand{\Cpeff}[1]{{\cal C}^{\rm eff\prime}_{#1}}
\newcommand{\C}[1]{{\cal C}_{#1}}
\newcommand{\Cp}[1]{{\cal C}^{\prime}_{#1}}

\newcommand{\intbin}[1]{\av{#1}_{\rm bin}}
\newcommand{\cp}{{\sss \rm CP}}

\begin{document}

{\footnotesize
\begin{flushright}
UAB-FT/732\\
LPT-ORSAY/13-24\\
MITP/13-020
\end{flushright}}

\vspace{1.5cm}
\begin{center}
{\Large\bf
Optimizing the basis of $B\to K^* \ell^+\ell^-$ observables\\[3mm]
in the full kinematic range
}
\end{center}

\vspace{3mm}

\begin{center}
{\sc  S\'ebastien Descotes-Genon$^{a}$, Tobias Hurth$^{b}$, Joaquim Matias$^{c}$\\ {\sf and} Javier Virto$^{c}$}\\[5mm]
{\em
$^{a}$ Laboratoire de Physique Th\'eorique, CNRS/Univ. Paris-Sud 11 (UMR 8627)\\ 91405 Orsay Cedex, France\\[2mm]
$^{b}$ PRISMA Cluster of Excellence \&	Institute for Physics (THEP)\\  Johannes Gutenberg University, D-55099 Mainz, Germany\\[2mm]
$^{c}$ Universitat Aut\`onoma de Barcelona, 08193 Bellaterra, Barcelona, Catalonia\\
}
\end{center}

\vspace{1mm}
\begin{abstract}\noindent
We discuss the observables for the $B\to K^*(\to K\pi) \ell^+\ell^-$ decay, focusing on both CP-averaged and CP-violating observables  at large and low hadronic recoil 
with special emphasis on their low sensitivity to form-factor uncertainties. 
We identify an optimal basis of observables that balances theoretical and experimental advantages, which will guide the New Physics searches in the short term.
We discuss some advantages of the observables in the basis, and in particular their
improved sensitivity to New Physics compared to other observables.
We present predictions within the Standard Model for the observables
of interest, integrated over the appropriate bins including lepton mass corrections.  Finally,~we present bounds on the S-wave contribution to the distribution coming from the $B\to K^*_0 \ell^+\ell^-$ decay, which will help to establish the systematic error associated to this pollution.
\end{abstract}

\thispagestyle{empty}

\newpage

\section{Introduction}

The recent results gathered by the B-factories and the LHCb experiment have greatly improved our knowledge concerning
the flavour structure of the fundamental theory that lies beyond the Standard Model (SM), leading to
a strongly constrained picture with only limited deviations from the SM.
Some examples of recent results are:  the  decreasing tension between $B \to \tau \nu$ and $\sin 2 \beta$ after the last Belle results~\cite{1208.4678}, the recent agreement with the SM of the semileptonic $a_s^l$ found in the last LHCb measurement~\cite{lhcbasls}, the consistency 
of the isospin asymmetry $A_I(B \to K^* \mu^+\mu^-)$~\cite{1205.3422} with its SM prediction~\cite{0212158}
 or the absence of large  deviations in $B_s \to \mu^+\mu^-$~\cite{1107.2304,1203.3976,1204.0735,1211.2674,1208.0934}, all of which have quietened down the hopes of seeing unambiguous
 signals of New Physics (NP). However, other observables are now exhibiting new discrepancies
 with SM, such as  the isospin asymmetry  $A_I(B \to K \mu^+\mu^-)$ \cite{1205.3422}, the longitudinal polarization fraction in $B_s\to K^*K^*$ \cite{1111.4183,1111.4882} or the pattern of $B\to D^{(*)}\tau\nu$ branching fractions~\cite{1205.5442,1203.2654}.

A recent experimental effort has brought a new player into the game,
the angular distribution of the flavour-changing neutral current decay $B \to K^*(\to K\pi)\mu^+\mu^-$,
  providing new and precise information on a set of important operators of the weak effective Hamiltonian: the electromagnetic ($O_7$) and semileptonic operators ($O_{9,10}$) together with their chirally-flipped counterparts ($O_{7',9',10'}$) and scalar/pseudoscalar operators ($O_{S,P,S',P'}$) and tensors. The main goal  of this paper is to describe 
  the 4-body angular distribution of the  decay $B \to K^*(\to K\pi)\mu^+\mu^-$ in an optimal way through 
   CP-conserving and CP violating observables covering the whole physical range for the dilepton invariant mass $q^2$ with limited sensitivity to long-distance (strong and SM) physics when possible, and thus enhanced sensitivity to short-distance (mainly weak and potentially NP) dynamics, but also excellent experimental accessibility. 
 Our main goal here is to extend our predictions for this optimal basis to the two available regions (low and high $q^2$, or equivalent large and low $K^*$ recoil), including the corresponding CP-violating observables. 
 
Such observables with little sensitivity to long-distance physics and enhanced potential in searches for NP can be seen as ``clean'' from the theoretical point of view, and they have been studied in depth during the last decade. 
For instance, a lot of effort has been put into the study of 
the zero of the forward-backward asymmetry  ($\afb$), because at the leading order, the position of this zero that depends in the SM on a combination of the Wilson coefficients $\Ceff9$ and $\Ceff7$, is independent of poorly known hadronic parameters (soft form factors) \cite{0106067}.
This idea was incorporated in the construction of the transverse asymmetry called $A_T^{(2)}$\cite{kruger} that exhibits 
the same cancellation of hadronic inputs
not only at one kinematic point but for {\it all} dilepton invariant mass  in the large $K^*$-recoil region. 
Soon after other observables, called by extension $A_T^{(3,4,5)}$, were proposed with a similar good behaviour \cite{matias1,matias2}. Even though conceptually important,
the zero of $\afb$ has been somehow superseded 
on one side by observables that provide similar SM tests over an extended $q^2$-range  and, on the other, by a clean version of $\afb$  (called $P_2=A_T^{\rm (re)}/2$ \cite{primary,becirevic}) that exhibitis the same zero as $\afb$.

A first guide for the construction of these observables is provided by effective theories available at low and high $q^2$, both based on an expansion in powers of $\Lambda/m_b$ to simplify 
the expression of the form factors and the amplitudes, either QCD factorisation/Soft Collinear Effective Theory at low $q^2$~\cite{0106067,LEET} or HQET at large $q^2$~\cite{grinstein+pirjol}. A second
 important guideline for the construction of observables was found when the symmetries of the angular distribution were identified
\cite{matias2,primary}, corresponding to transformations of the transversity amplitudes that leave the distribution invariant. The number of symmetries $n_S$ 
depends on the scenario considered (massive or massless leptons, presence or absence of scalar contributions) and it
is related to the number of  independent observables ($n_{obs}$) through $n_{obs}=2 n_A - n_S$, where $n_A$ is the number of amplitudes.  In the massless case, $n_{obs}=8$, or $n_{obs}=9$ if we include scalar operators. Including the mass terms leads to  $n_{obs}=10$ and $n_{obs}=12$ respectively.  Taking into account the CP-conjugated mode doubles the number of independent observables.  This number $n_{obs}$ defines the minimal  number of  observables required to extract {\it all} the information contained in the distribution.  Moreover, any angular observable can be reexpressed  in terms of this set of $n_{obs}$ observables, which has the properties of a \emph{basis} -- see Ref.~\cite{primary} for a detailed discussion of the different scenarios and associated symmetries.

A very accessible basis is given by the CP-\emph{S}ymmetric and CP-\emph{A}symmetric coefficients $S_i$ and $A_i$ defined in Ref.~\cite{buras}, but their strong sensitivity to the choice of soft form factors 
makes this basis less competitive for NP searches. 
The basis on which we will focus here represents a very good compromise between theoretical cleanliness and simplicity in their experimental accessibility  \cite{kruger,primary,becirevic,1207.2753}:
\eq{
\left\{\frac{d\Gamma}{dq^2}, A_{FB} \, {\rm or} \, F_L, P_1=A_T^2,  P_2=\frac{1}{2}A_T^{\rm re},  P_3=-\frac{1}{2}A_{T}^{\rm im},  P_4^\prime, P_5^\prime,  P_6^\prime \right\}\ ,
}
together with the corresponding CP-violating basis: 
\eq{
\left\{A_{CP}, A_{FB}^{CP} \, {\rm or} \, F_L^{CP}, P_1^{CP}, \, P_2^{CP}, \, P_3^{CP}, \, P_4^{\prime CP},\,  P_5^{\prime CP}, \, P_6^{\prime CP} \right\}\ .
}
At leading order (LO) in the low-$q^2$ effective theory (approximately from 0.1 to 8 GeV$^2$), this
basis of observables  is independent of 
soft form factors,
but in general it is not protected from form-factor uncertainties in the high-q$^2$ region. The SM predictions for the CP-average  basis of observables was computed in the massless limit and in the large-recoil region in Ref.~\cite{1207.2753}. Here we present our SM predictions for both bases including lepton mass corrections  and in both large- and low-recoil regions.  

One could consider other interesting bases, for example the
unprimed basis where $P_{4,5,6}$ are substituted for $P_{4,5,6}'$ (see for instance Ref.~\cite{primary}).
We do not consider this unprimed basis as optimal as the previous one, due to the difficulty to obtain these observables from experimental measurements: indeed, $P_{4,5}$ can be determined from the measured angular distribution only once one has determined $F_T$
(the transverse polarisation, also needed to extract $P_{4,5}^\prime$) but also $P_1$, 
reducing its discriminating power. 
Even though it is not optimal experimentally, this unprimed basis is interesting, as
some of  those unprimed observables are clean in both regions contrary to the primed ones. Therefore, they should be considered in the long run, as well as other
observables like $A_T^{(3,4,5)}$. 
In the current experimental situation, where the experimental statistics is likely to be higher in the large-recoil region that in the low-recoil case, it seems however more interesting to consider observables as accurately measured and as sensitive to NP as possible at low-$q^2$. In this sense, we believe that the basis presented above is currently the optimal one.
These unprimed observables  at large recoil are directly linked to 
a set of observables --called $H_T^{(i)}$-- proposed for the low recoil in a series of interesting papers \cite{bobeth,tensors}. These observables can be easily integrated  inside the following basis: $\{d\Gamma/dq^2, A_{FB}, P_1 , H_T^{(1,2,3,4,5)} \}$. Most of them can be identified  with the unprimed basis $P_i$ in the large recoil, for instance, $H_T^{(1,2)}$~\cite{bobeth} correspond in our notation to $P_{4,5}$~\cite{1207.2753}. We chart the correspondance in Table~\ref{TableObs}, providing an indication of their experimental accesibility as well as their sensitivity to form factors at low and large recoils.

\begin{table}
\ra{1.5}
\small
\begin{center}
\begin{tabular}{@{}ccccc@{}}
\toprule[1.1pt]
Observable  & \mpa{2cm}{Angular\\ coefficient} & \mpa{3cm}{Experimental\\ accessibility} & \mpa{3cm}{Clean at\\ Large Recoil} & \mpa{3cm}{Clean at\\ Low Recoil}  \\ [2mm]
\midrule
$P_1=A_T^{(2)}$ & $J_3$ & \bl{Measured} & \gr{Yes} & \rd{No} \\[1mm]
$P_2=\frac12 A_T^{(\rm re)}$ & $J_{6s}$ & \gr{Excellent} & \gr{Yes} &  \mpa{3cm}{Yes if $P_1\simeq 0$\\ not otherwise} \\[4mm]
$P_3=-\frac12A_T^{(\rm im)}$ & $J_{9}$ & \gr{Excellent} & \gr{Yes} & \mpa{3cm}{Yes if $P_1\simeq 0$\\ not otherwise}\\[4mm]
$P'_4$ & $J_{4}$ & \gr{Excellent} & \gr{Yes}  & \mpa{3cm}{Yes if $P_1\simeq 0$\\ not otherwise}\\[4mm]
$P'_5$ & $J_{5}$& \gr{Excellent} & \gr{Yes} &  \mpa{3cm}{Yes if $P_1\simeq 0$\\ not otherwise}\\[4mm]
$P'_6$ & $J_{7}$ & \gr{Excellent}& \gr{Yes} &  \mpa{3cm}{Yes if $P_1\simeq 0$\\ not otherwise}\\[4mm]
$P'_8$ & $J_{8}$ & \gr{Excellent}& \gr{Yes} & \mpa{3cm}{Yes if $P_1\simeq 0$\\ not otherwise}\\[4mm]
\midrule
$P_4=H_T^{(1)}$ & $J_{4}$ & \rd{Good} & \gr{Yes}  &  \gr{Yes} \\[1mm]
$P_5=H_T^{(2)}$ & $J_{5}$ & \rd{Good} & \gr{Yes}  & \gr{Yes} \\[1mm]
$P_6$ & $J_{7}$ & \rd{Good} & \gr{Yes}  & \gr{Yes}  \\[1mm]
$P_8=H_T^{(4)}$ & $J_{8}$ & \rd{Good} & \gr{Yes}  & \gr{Yes}   \\[1mm]
$H_T^{(3)}$ & $J_{6s}$ & \rd{Good} & \gr{Yes} & \gr{Yes}  \\[1mm]
$H_T^{(5)}$ & $J_{9}$ & \rd{Good} & \gr{Yes} & \gr{Yes}  \\[1mm]
\midrule
$F_L$ & $J_{2c}$ & \bl{Measured} & \rd{No} & \rd{No} \\[1mm]
$A_{FB}$ & $J_{6s},J_{6c}$ & \bl{Measured} & \rd{No} & \rd{No} \\[1mm]
\midrule
$S_i/A_i$ & $J_{i}$ & \gr{Excellent} & \rd{No} & \rd{No} \\[1mm]
\midrule
$A_T^{(3,4,5)}$ & All & \rd{Difficult} & \gr{Yes} & \rd{No} \\
\bottomrule[1.1pt]
\end{tabular}
\caption{
Experimental accessibility and theoretical cleanliness (at large and low recoils) of different observables. The statements apply both to CP-averaged observables $P_i$ and their CP-violating counterparts $P_i^{CP}$. We also indicate the angular coefficient used in the numerator to build the observable. These observables have been defined in Refs.~\cite{matias1,matias2,primary,becirevic,buras,1207.2753,bobeth,tensors}.
}
\label{TableObs}
\end{center}
\end{table}

The optimal basis should be complemented with two extra mass-dependent observables. There are two posibilities: (a) introducing the observables $M_1$ and $M_2$~\cite{primary} and the basis is then $\{\frac{d\Gamma}{dq^2},F_L,P_{1,2,3},P_{4,5,6}^\prime,M_1,M_2\}$   or (b) introducing two different definitions (see Ref.~\cite{1209.1525})  for the longitudinal (${\hat F}_L$ and ${\tilde F}_L$) and  the transverse polarization fractions (${\hat F}_T$ and ${\tilde F}_T$) such that the basis becomes
$\{{\hat F_T} \frac{d\Gamma}{dq^2},{\hat F_L} \frac{d\Gamma}{dq^2}, {\tilde F_T} \frac{d\Gamma}{dq^2}, {\tilde F_L} \frac{d\Gamma}{dq^2}, P_{1,2,3},P_{4,5,6}^\prime\}$. The ratios ${\hat F}_T/{\tilde F}_T$ and ${\hat F}_L/{\tilde F}_L$ can be mapped into the clean $M_1$ and $M_2$, respectively (see \cite{1209.1525}). In the presence of scalar operators,
a couple of scalar dependent observables $S_{1,2}$ can be introduced~\cite{primary}. However, given the current strong constraints on scalar Wilson coefficients from
radiative decays we will not consider them here.

As argumented above, there is an optimal basis to extract as much information on NP as possible from the $B\to K^*(\to K\pi) \ell^+\ell^-$ angular distribution considering the current experimental limitations of this analysis. Our goal in the present paper is to pave the way for further experimental analyses of these observables, by providing SM predictions and assessing their sensitivity to NP scenarios by checking their dependence on hadronic uncertainties, mainly the still poorly known  form factors and  the possibility of $S$-wave pollution.
In Section~2 we  discuss 
the construction of clean observables independently of the region (large or low recoil) and we provide all the details on our approach to form factors for both regions in Section~3. Considering the various determination of $B\to K^*$ form factors available in the literature, we discuss
the extension of form factor parametrizations to the low-recoil region that are validated (when possible) with lattice data.
The explicit definition of the observables in the optimal basis including binning effects are given in Section~4 and their SM prediction 
is provided in Section~5. For completeness we also provide predictions for other observables of interest in the appendix. In section 6 we discuss the impact of different choices for form factors on our basis, focusing on the large-recoil region to show their discriminating power considering some NP scenarios. In Section 7 we discuss the impact of the S-wave on the determination of observables and we present explicit bounds on the size of the polluting S-wave terms coming from the companion decay $B \to K_0^* \mu^+\mu^-$. 
This pollution can be eliminated, as pointed out in Ref.~\cite{1209.1525}, once there will be enough statistics to measure 
the folded distribution, including terms coming from the S-wave component. In Section~8 we present a comparison of our results with other results in the literature  
and we conclude in Section~9. The 
appendices contain a compendium of 
definitions for other observables of interest and
a set of tables and plots summarising
our SM predictions for all measured bins.

\section{{Clean observables: General arguments}}\label{sec:clean}

The differential decay rate of the process $\bar B_d \to \bar K^*(\to K\pi) \ell^+ \ell^-$ can be written as:
\eqa{\label{dist}
\frac{d^4\Gamma(\bar{B}_d)}{dq^2\,d\!\cos\theta_K\,d\!\cos\theta_l\,d\phi}&=&\frac9{32\pi} \bigg[
J_{1s} \sin^2\theta_K + J_{1c} \cos^2\theta_K + (J_{2s} \sin^2\theta_K + J_{2c} \cos^2\theta_K) \cos 2\theta_l\nn\\[1.5mm]
&&\hspace{-2.7cm}+ J_3 \sin^2\theta_K \sin^2\theta_l \cos 2\phi + J_4 \sin 2\theta_K \sin 2\theta_l \cos\phi  + J_5 \sin 2\theta_K \sin\theta_l \cos\phi \nn\\[1.5mm]
&&\hspace{-2.7cm}+ (J_{6s} \sin^2\theta_K +  {J_{6c} \cos^2\theta_K})  \cos\theta_l    
+ J_7 \sin 2\theta_K \sin\theta_l \sin\phi  + J_8 \sin 2\theta_K \sin 2\theta_l \sin\phi \nn\\[1.5mm]
&&\hspace{-2.7cm}+ J_9 \sin^2\theta_K \sin^2\theta_l \sin 2\phi \bigg]\,,
}
where the kinematical variables $\phi$, $\theta_\ell$, $\theta_K$, $q^2$ are defined as in Refs.~\cite{primary,buras,bobeth} : $\theta_\ell$ and $\theta_K$ describe the angles of emission
between $\bar{K}^{*0}$ and $\ell^-$ (in the di-meson rest frame) and between $\bar{K}^{*0}$ and $K^-$ (in the di-hadron rest frame)
respectively, whereas $\phi$ corresponds to the angle between the di-lepton and di-meson planes and $q^2$ to the di-lepton invariant mass. The decay rate $\bar \Gamma$ of the CP-conjugated process $B_d \to K^*(\to K\pi) \ell^+ \ell^-$ is obtained from Eq.~(\ref{dist}) by replacing $J_{1,2,3,4,7}\to \bar J_{1,2,3,4,7}$ and $J_{5,6,8,9}\to -\bar J_{5,6,8,9}$, where $\bar J$ is equal to $J$ with all weak phases conjugated. This convention corresponds to 
taking the same lepton $\ell^-$ for the definition of $\theta_\ell$ for both $B$ and $\bar B$  decays (see for example Ref.~\cite{0805.2525}). The usual convention among experimental collaborations is a different one, where $\theta_\ell$ in the $B$ decay is defined as the angle between $K^*$ and $\ell^+$. The translation between both conventions corresponds to the change $\theta_\ell \to \pi -\theta_\ell$, which means that in the experimental convention 
all $\bar{J}$ go with a positive sign in the distribution.  
The fact that the decay $B\to K^*\ell^+\ell^-$ is self-tagging ensures that the coefficients $J_i$ and $\bar J_i$ can be extracted independently, both for CP-averaged and CP-violating observables.

Currently, the LHCb experimental analysis of these angular observables deals with ``folded'' distributions, in order to exploit data as efficiently as possible before there is enough statistics for
a full angular analysis of this decay. In Ref.~\cite{LHCbinned} it has been shown that the identification of events with $\phi \leftrightarrow \phi + \pi$ leads to an angular distribution depending on 
a ``folded'' angle $\hat \phi \in [0,\pi]$ which {pins down} the coefficients $J_{1,2,3,6,9}$. Similar folded distributions can be constructed that depend on $J_{4,5}$ \cite{1209.1525}. The use of folded distributions is also optimal to isolate the S-wave pollution from scalar $K^*$ resonances, as has been discussed in Ref.~\cite{1209.1525}, as opposed to the use of uniangular distributions \cite{1207.4004} (see also Refs.~\cite{1111.1513,1210.5279}). We will come back to the issue of the S-wave interference in Section~\ref{sec:Swave}.

Once extracted, the coefficients $J_i$ must be interpreted. Assuming that the decay proceeds only via a (P-wave) $K^*$ resonance, these coefficients can be reexpressed in terms of transversity amplitudes $A_{0,\perp,||}^{L,R}$ describing both the chirality of the operator considered in the effective Hamiltonian and the polarisations of the $K^*$ meson and the intermediate virtual gauge boson decaying into $\ell^+\ell^-$. In addition we have two extra amplitudes $A_s$ and $A_t$ associated to the presence of scalars, pseudoscalars and lepton masses. All these amplitudes can be reexpressed in terms of short-distance Wilson coefficients of the effective Hamiltonian and long-distance quantities.
Long-distance quantities can be expressed in turn through form factors which are one of the main sources of uncertainties for the prediction of the coefficients ${J}_i$. The main operators entering the discussion are then the chromagnetic operator $O_7$ and the two semileptonic operators $O_9$ and $O_{10}$. At both ends of the dilepton mass range (low and high $q^2$, or equivalently large and low recoil of the emitted $K^*$ meson) one can perform a further expansion in inverse powers of quantities of order $m_b$ (following either QCD factorisation/Soft-Collinear Effective Theory or Heavy-Quark Effective Theory): the use of effective theories allows one to relate vector and tensor form factors and reduce the amount of hadronic inputs from external sources. Moreover, at low $q^2$, the formalism allows one to include the hard-gluon corrections from four-quark operators (not included in the analysis otherwise) \cite{0106067}.

These additional relations between form factors are particularly interesting to eliminate 
as much as possible hadronic uncertainties, in order to enhance the potential of this decay in
the search for New Physics. This leads us to define clean observables in both regions 
where one can use relations derived from effective theories.
The construction of clean observables is based on a cancellation of form factors at leading order in the relevant effective theory. The mechanisms are basically the same at high and low recoil, although the factorization of a single form factor multiplying each amplitude is achieved via different expansions -- the large energy limit of QCD factorisation (QCDF) at large recoil and the  heavy quark expansion at low recoil. At leading order the relevant transversity amplitudes are equivalent to the naive result in terms of $O_{7,9,10}$ form factors and are given by (see for example Ref.~\cite{kruger})
\eqa{
A_{\bot}^{L,R} &=& {\cal N}_\bot \Big[\C{9\mp10}^{+} V(q^2) + \C{7}^{+} T_1(q^2)\Big] + \op (\alpha_s,\Lambda/m_b\cdots)\label{asd1} \\[1mm]
A_{\|}^{L,R} &=& {\cal N}_\| \Big[\C{9\mp10}^{-} A_1(q^2) + \C{7}^{-} T_2(q^2)\Big] + \op (\alpha_s,\Lambda/m_b\cdots)\label{asd2} \\[1mm]
A_{0}^{L,R} &=& {\cal N}_0 \Big[\C{9\mp10}^{-} A_{12}(q^2) + \C{7}^{-} T_{23}(q^2)\Big]+ \op (\alpha_s,\Lambda/m_b\cdots)\label{asd3}
}
where $\C{9\mp10}^{\pm} \equiv [(\Ceff9\pm\Cpeff9)\mp (\Ceff{10}\pm\Cpeff{10})]/(m_B\pm m_{K^*})$ and $\C{7}^{\pm}\equiv2m_b/q^2 (\Ceff7\pm \Cpeff7)$. The ${\cal N}_i$ are different normalization factors, and $A_{12}$, $T_{23}$ are appropriate combinations of form factors: 
$$A_{12}\equiv (m_B^2-m_{K^*}^2)(m_B^2-m_{K^*}^2-q^2)A_1 - \lambda (m_B-m_{K^*})/(m_B+m_{K^*}) A_2$$ and 
$$T_{23}\equiv q^2(m_B^2 + 3m_{K^*}^2-q^2) T_2 - \lambda q^2/(m_B^2-m_{K^*}^2) T_3\ ,$$
with $\lambda=m_B^4+m_{K^*}^4+q^4-2 (m_B^2 m_{K^*}^2+m_{K^*}^2 q^2 + m_B^2 q^2)$.
These combinations appear naturally when the problem is expressed in terms of helicity amplitudes as shown in Ref.~\cite{camalich}.

The key observation is that the ratios $R_1=T_1/V$, $R_2=T_2/A_1$ and $\tilde R_3=T_{23}/A_{12}$ (a more extensive discussion on the form factors and their ratios will be given in Section~\ref{sec:ffs}) have well-defined limiting values in both regimes \cite{LEET,grinstein+pirjol}:
\eq{\label{effectiverelationships}
R_{1,2} = 1 + {\rm corrections}\ ,\quad \tilde R_3=\frac{q^2}{m_B^2}+ {\rm corrections}\ .
}
Using these ratios to eliminate $T_1,T_2,T_{23}$ in Eqs.~(\ref{asd1})-(\ref{asd3}), the transversity amplitudes can be written as (see for example Ref.~\cite{bobeth}):
\eqa{
A_{\bot}^{L,R} &=& X_\bot^{L,R}\, V(q^2)  + \op (\alpha_s,\Lambda/m_b\cdots)\label{abot}\\[1mm]
A_{\|}^{L,R} &=& X_\|^{L,R}\, A_1(q^2) + \op (\alpha_s,\Lambda/m_b\cdots)\\[1mm]
A_{0}^{L,R} &=& X_0^{L,R}\, A_{12}(q^2)+ \op (\alpha_s,\Lambda/m_b\cdots)\label{a0}
}
where $X_i$ are short-distance functions. The ellipses denote perturbative and power corrections that contain the corrections to the ratios (\ref{effectiverelationships}) as well as those in (\ref{asd1})-(\ref{asd3}). The fact that $L$ and $R$ transversity amplitudes are proportional to the same form factor allows one to build a number of clean observables by taking suitable ratios of angular coefficients. The expressions (\ref{abot})-(\ref{a0}) are true at low and large recoils. 
At low recoil we have no further relationships between form factors, contrary to the case of large recoil. Therefore, \emph{all observables that are clean at low recoil, are also clean at large recoil}. This is true in particular for the observables defined in Refs.~\cite{bobeth,tensors,bobeth2}.

At large recoil another relationship holds: $V$ and $A_1$ are related by (see for example Ref.~\cite{LEET}):
\eq{\label{effectiverelationshipslargerecoil}
2 E_{K^*} m_B V(q^2) = (m_B+m_{K^*})^2 A_1(q^2)\,+\, \op (\alpha_s,\Lambda/m_b\cdots)
} 
up to subleading corrections in the effective theory. This makes possible to build additional clean observables at large recoil that are not clean at low recoil, for example $P_1=A_T^{(2)}$ \cite{kruger,primary}, $A_T^{(\rm re)}$, $A_T^{(\rm im)}$ \cite{becirevic} or $P'_{4,5,6}$ \cite{1207.2753} (we will come back to these observables later in this article). According to the counting of Ref.~\cite{primary}, an optimal basis in the massless case will contain five observables clean in the full kinematic region, one observable clean only at large recoil, and two observables that depend on form factors. Similar countings can be performed in more general cases (mass terms, scalar operators, etc).

Clean observables are only independent of form factors at leading order in the corresponding effective-theory expansions. A residual sensitivity is introduced when subleading corrections are considered, which are
of two kinds: perturbative corrections (typically from hard-gluon exchanges) and non-perturbative corrections (higher orders in $1/m_b$ expansions). Even though these corrections are expected to be suppressed in the kinematical regions of interest, a reduction of such residual uncertainties should be attempted, in particular if New Physics contributions turn out to be rather small. In such a case, lattice determinations of $B\to K^*$ form factors with small uncertainties will be crucial, but the determination of $T_3,A_2,A_0$ seems particularly challenging~\cite{becirevic}. 

Alternatively, if sufficient statistics is collected at experimental facilities, the extraction of form factors from data with reasonable uncertainties becomes a possibility \cite{gudrun}. In this case the same argument concerning clean observables applies. From the chosen optimal basis, the observables having a significant sensitivity to form factors are used to extract the relevant form factors, whereas the clean observables are used to constrain the short distance physics. Furthermore, the ratios $R_i$ can be extracted which provide a test of the relationships derived in the low- and high-$q^2$ effective theories (see for example Section VI of Ref.~\cite{tensors}).\\[-2mm]

\section{Form Factors}
\label{sec:ffs}
In this section we discuss in detail our approach to form factors in both kinematic regions, as their behaviour is important for the construction of clean observables. We will see that the low-recoil region requires a specific treatment, as the extrapolation of current results on light-cone sum rules, the lattice determinations of the form factors, and the effective theory relationships are not fully compatible among each other.

\subsection{Large recoil}
\label{largerr}

\begin{figure}\centering
\includegraphics[width=5.2cm]{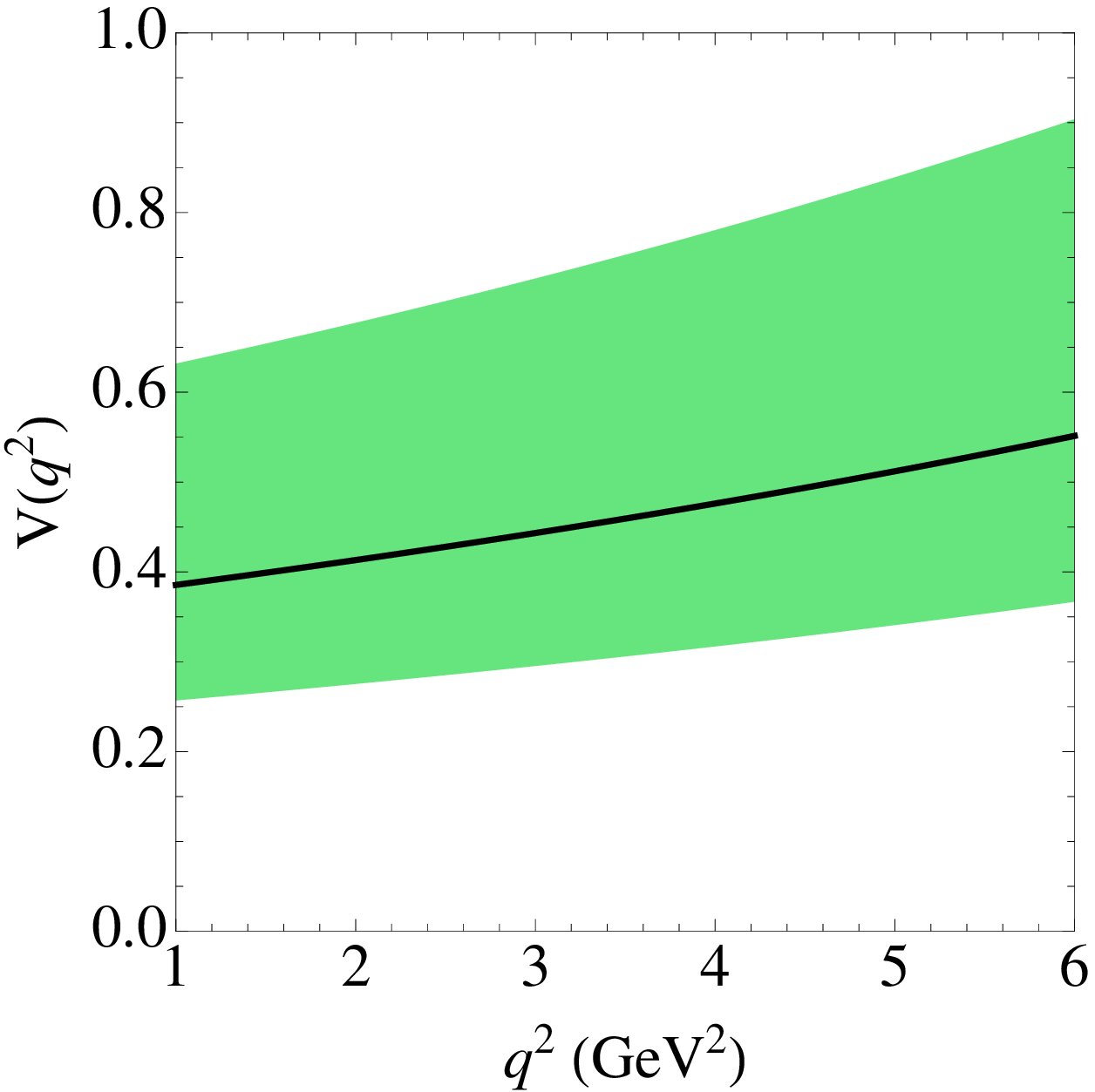}
\includegraphics[width=5.2cm]{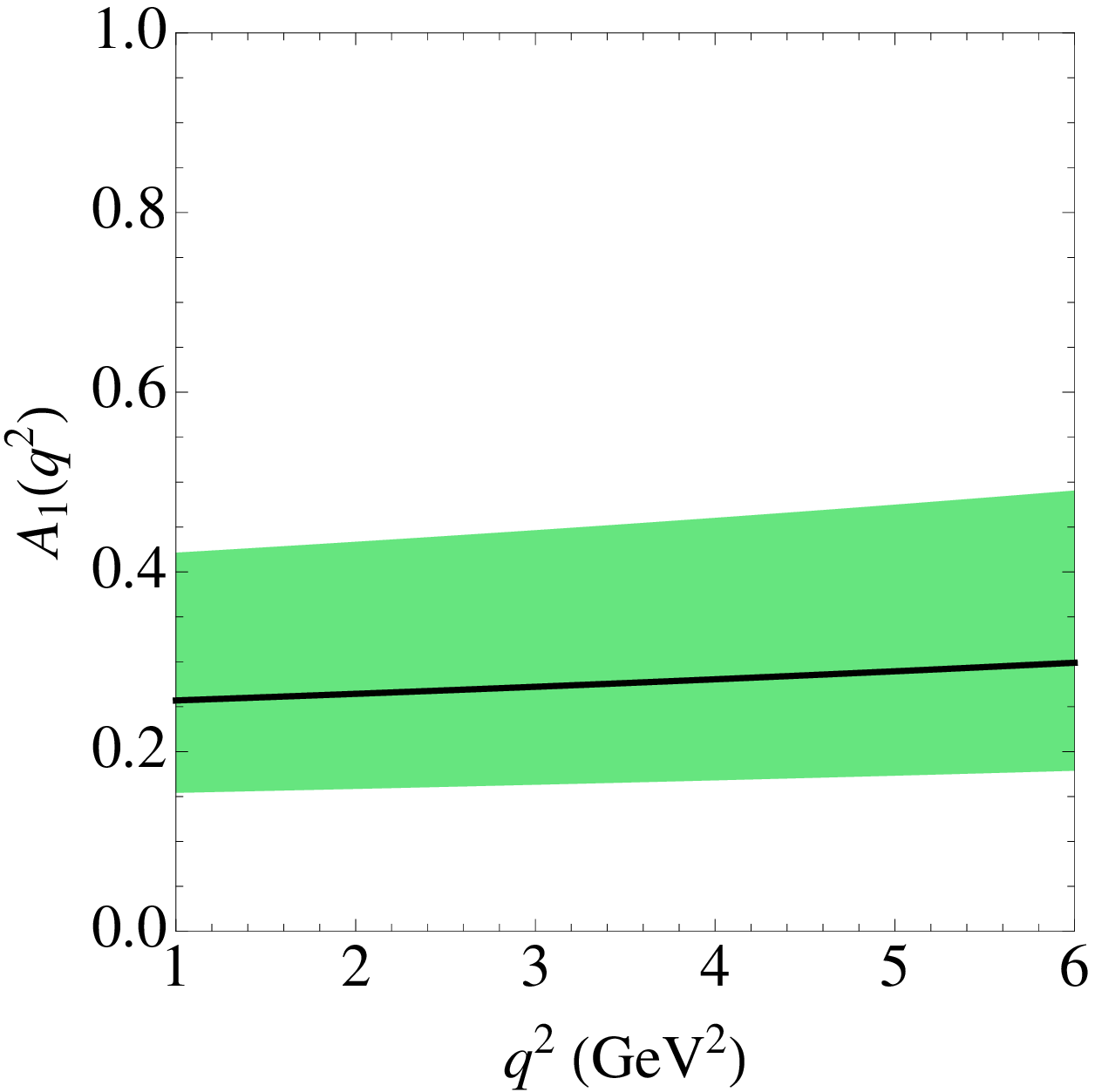}
\includegraphics[width=5.2cm]{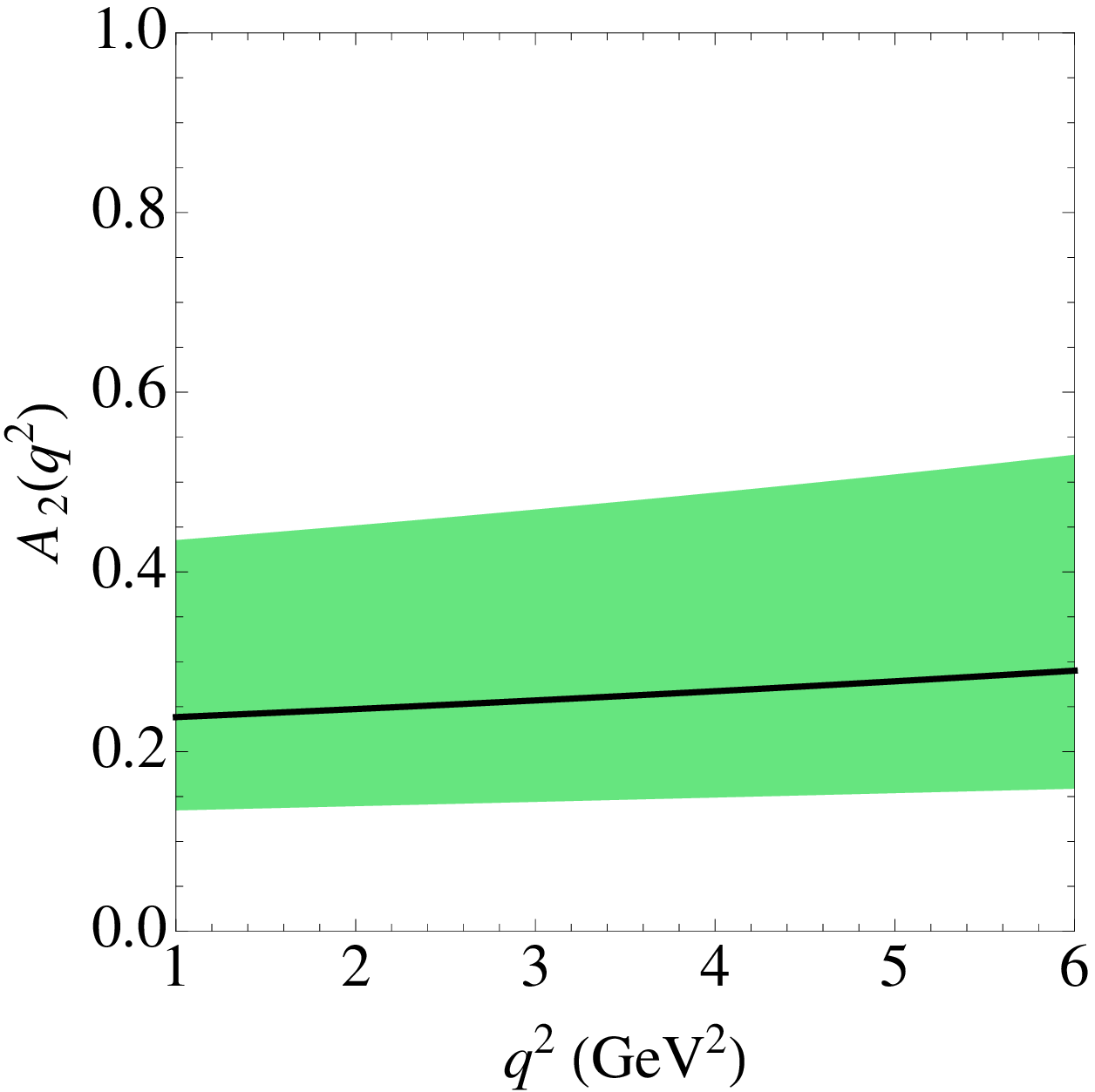}
\caption{Input form factors (from Ref.~\cite{1006.4945}) used to obtain the soft form factors $\xi_{\perp,\|}(q^2)$:  $V(q^2)$ (left), $A_1(q^2)$ (center), $A_2(q^2)$ (right). All errors are added in quadrature.
} 
\label{VA1A2}
\end{figure}

In the large recoil region, the amplitudes are expressed in terms of two ``soft" form factors $\xi_\perp(q^2)$ and $\xi_\|(q^2)$ \cite{0106067}. These are defined in terms of the QCD form factors $V(q^2)$, $A_1(q^2)$ and $A_2(q^2)$. Here we follow the prescription of Ref.~\cite{0412400} with a factorization scheme defining the soft form factors by the conventions
\eqa{
\xi_\perp(q^2)&=&\frac{m_B}{m_B+m_{K^*}} V(q^2) \ , \label{xiperp}\\
\xi_{\|}(q^2)&=&\frac{m_B+m_{K^*}}{2 E} A_1(q^2) - \frac{m_B - m_{K^*}}{m_B} A_2(q^2) \ .\label{xipar}
} 
The $q^2$ dependence of all form factors can be reproduced using a parametrization based on the Series Expansion with a single pole replacing the Blaschke factor (see for example Refs.~\cite{0807.2722} for discussions of the advantages and limits of the conformal mapping of the cut singularities onto the unit circle) 
\eq{
F(s)=\frac{F(0)\, m^2_{F}}{m^2_{F}-s} \left\{1+b_F \left(z(s,\tau_0)-z(0,\tau_0)+\frac{1}{2} \left(z[s,\tau_0]^2-z[0,\tau_0]^2\right) \right) \right\}\ ,  \label{khod}
}
where $F$ represents the form factor and 
\eq
{ z(s,\tau_0)=\frac{\sqrt{\tau_+ - s}-\sqrt{\tau_+ - \tau_0}}{\sqrt{\tau_+ -s}+\sqrt{\tau_+ - \tau_0}}\ , \ \tau_{\pm}=(m_B \pm m_{K*})^2\ , \  \tau_0=\tau_+ - \sqrt{\tau_+ - \tau_-} \sqrt{\tau_+}\ .
}
The form factors at $q^2=0$ and the slope parameters $b_F$ are computed via light-cone sum rules with $B$-meson distribution amplitudes in Ref.~\cite{1006.4945} (these values for the form factors will be called KMPW). The results for $V(q^2)$, $A_{1,2}(q^2)$ are shown in Figure~\ref{VA1A2}. An earlier and commonly quoted source for these form factors, computed using light-meson light-cone sum rules, is Ref.~\cite{0412079}.
Even though the size of the uncertainties in Ref.~\cite{0412079} is considerably smaller than in KMPW, we prefer to use KMPW for the following reasons. The size of the error in light-cone sum-rules computations does not only depend on the particular method used (for example light vs. heavy meson wavefunctions), but also depends on a delicate estimation of ``systematic'' errors associated to the built in assumptions of each procedure. There is in fact a wide spread of quoted uncertainties for $B\to K^*$ form factors in the recent literature, that range from a $\sim$ 10\% to a $\sim$ 40\% error for the same form factor \cite{buras,1006.4945}. For example, the values $A_0(0) = 0.33 \pm 0.03$ and $V (0) = 0.31 \pm 0.04$ given in Ref.~\cite{buras} should be compared to the values $A_0(0) = 0.29 \pm 0.10$ and $V (0) = 0.36 \pm 0.17$ as quoted in KMPW. Even central values have shifted significantly, see for instance the value $V (0) = 0.41 \pm 0.05$ from Ref.~\cite{0412079} before its update of Ref.~\cite{buras} (also consistent with Ref.~\cite{1004.3249}).
Given that all the values of the form factors  $V(q^2),A_{1,2}(q^2)$ always fall inside the error bars of KMPW,  we choose KMPW  in our numerical analyses in order to obtain more conservative results. This choice has a marginal impact on clean observables, but can have an important effect on other form-factor-dependent observables ($S_i,...$), as illustrated in Ref.~\cite{1207.2753} (see also Sec.~\ref{sec:const}). From now on we will always refer to KMPW when discussing the numerics of all form factors.

In principle, due to the large-recoil symmetry relations among the form factors that are valid up to corrections of order $\alpha_s$ and $\Lambda/m_b$, one is entitled to define $\xi_\|(q^2)$ also in terms of $T_{2,3}(q^2)$  (see Eq.~(24) of Ref.~\cite{beneke+feldmann}).
The resulting soft form factor is in very good agreement with the one obtained from Eq.(\ref{xipar}). 

The values of the soft form factors  at zero are determined by
\eq{
\xi_\perp(0)=\frac{m_B}{m_B+m_V} V(0) \qquad
\xi_{\|}(0)=2 \frac{m_V}{m_B} A_0(0)}
which corresponds to  $\xi_\perp(0)=0.31 ^{+0.20}_{-0.10}$ and $\xi_\|(0)=0.10^{+0.03}_{-0.02}$. Notice that $\xi_{\|}(0)$ can be determined also through $A_{1,2}(0)$ (see Eq.~(\ref{xipar}))  due to the large recoil relation
$2 m_V A_0(0)= (m_B+m_V) A_1(0) - (m_B-m_V) A_2(0)$.
 In order to correctly account for the correlation between the errors of $A_1(0)$ and $A_2(0)$ we determine $\xi_{\|}(0)$ using $A_0(0)=0.29^{+0.10}_{-0.07}$ from KMPW. 
In Ref.~\cite{0412400} a slightly different normalization for $\xi_\perp(0)$ is used,  that is obtained from $T_1(0)$ and not $V(0)$ after including an $\alpha_s$ correction.

\begin{figure}\centering
\includegraphics[width=7.2cm]{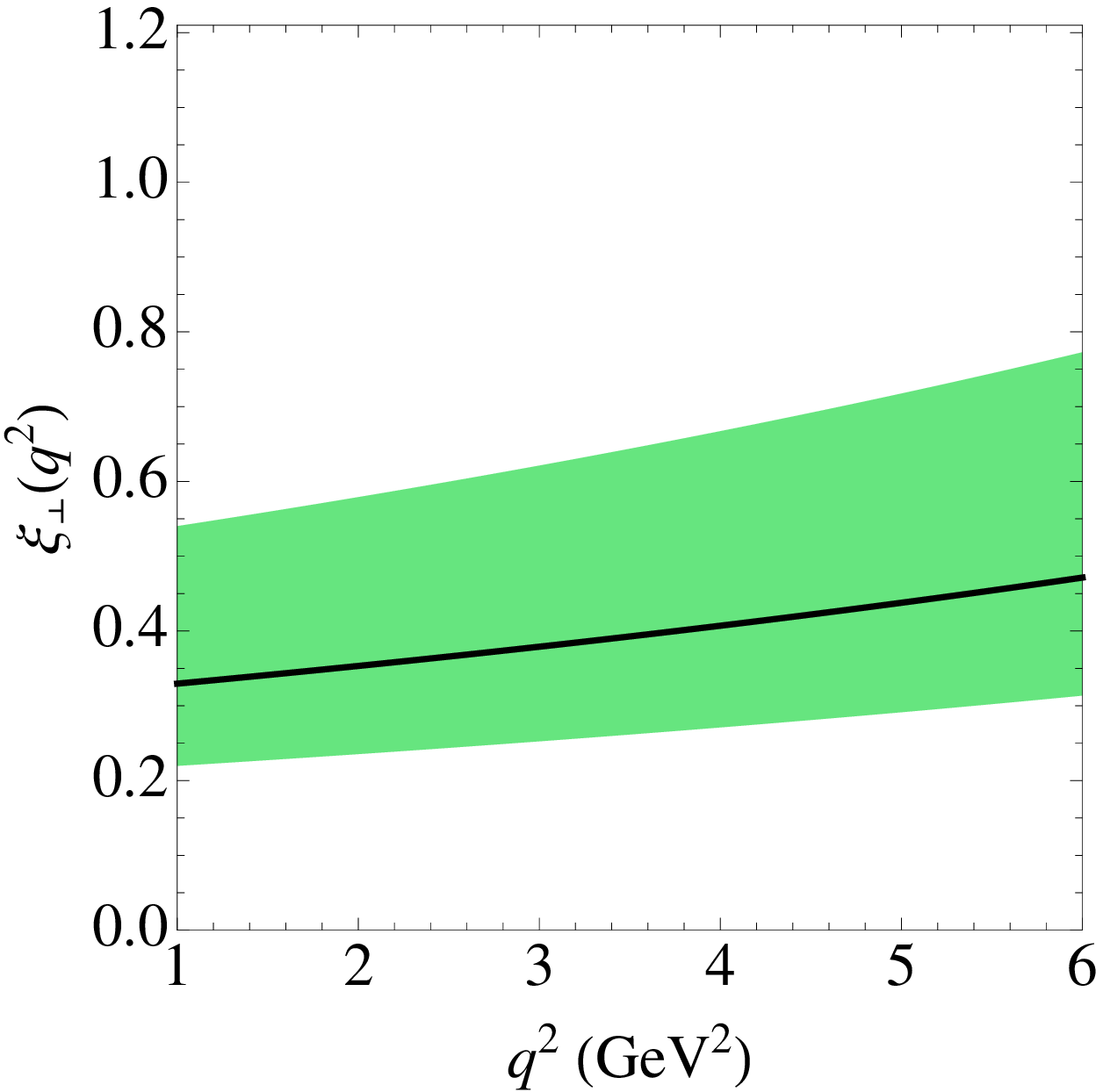}\hspace{1cm}
\includegraphics[width=7.2cm]{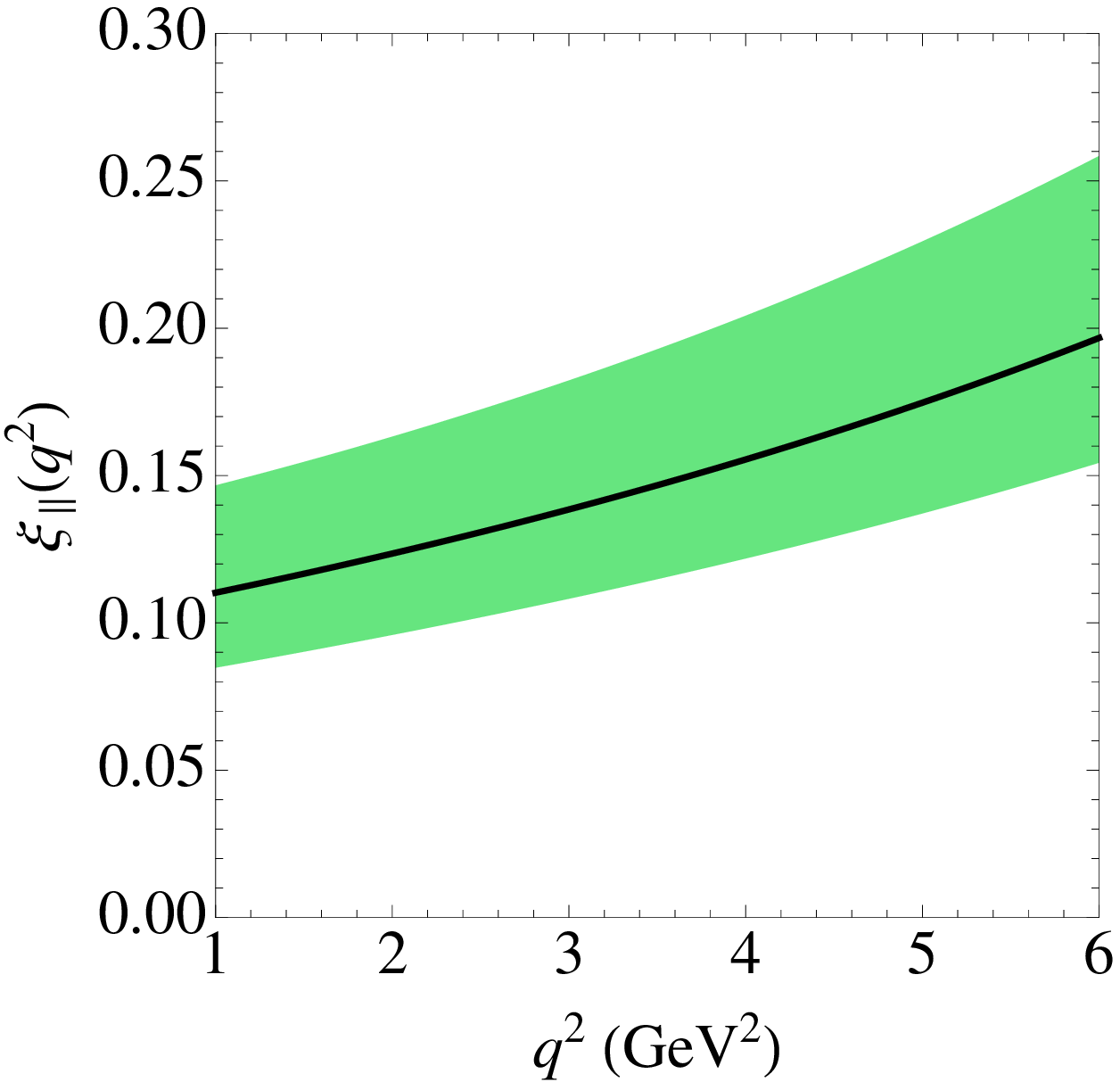}
\caption{Soft form factors $\xi_\perp(q^2)$ (left) and $\xi_\| (q^2)$ (right). These are obtained as described in the text from the results of Ref.~\cite{1006.4945}.} \label{softformfactors}
\end{figure}

Once $\xi_\perp(q^2)$ and $\xi_\|(q^2)$ are defined
using Eq.~(\ref{khod}), with $F=\xi_{\|,\perp}$, and the input values given in Table~\ref{TableFFs}
 (see Fig. \ref{softformfactors}), all form factors follow  using \cite{beneke+feldmann}
\eqa{
A_1(q^2)&=& \frac{2 E}{m_B+m_{K^*}} \xi_\perp(q^2) + \Delta A_1 + {\cal O}(\Lambda/m_b) \nonumber \\
 A_2(q^2)&=& \frac{m_B}{m_B-m_{K^*}} [\xi_\perp(q^2) -  \xi_\| (q^2)] 
      +\frac{m_B}{2E} \frac{m_B+m_{K^*}}{m_B-m_{K^*}}\Delta A_1 + {\cal O}(\Lambda/m_b) \nonumber \\
 A_0(q^2)&=& \frac{E}{m_{K^*}} \frac{\xi_\| (q^2)}{\Delta_{\|}(q^2)} + {\cal O}(\Lambda/m_b) \label{a0ff}
}
 where the first relation has no $\alpha_s$ corrections at first order ($\Delta A_1=O(\alpha_s^2)$).  The second relation comes from the definition of the prescription, while the third one includes an $\alpha_s$ correction explicitly inside the factor $\Delta_{\|}(q^2)=1+{\cal O}(\alpha_s) f(q^2)$, where $f(q^2) \to  0 $ when $q^2 \to 0$ (see \cite{beneke+feldmann} for the explicit definition of $\Delta_\|(q^2)$). 
 
\begin{figure}\centering
\includegraphics[height=7.2cm,width=12.0cm]{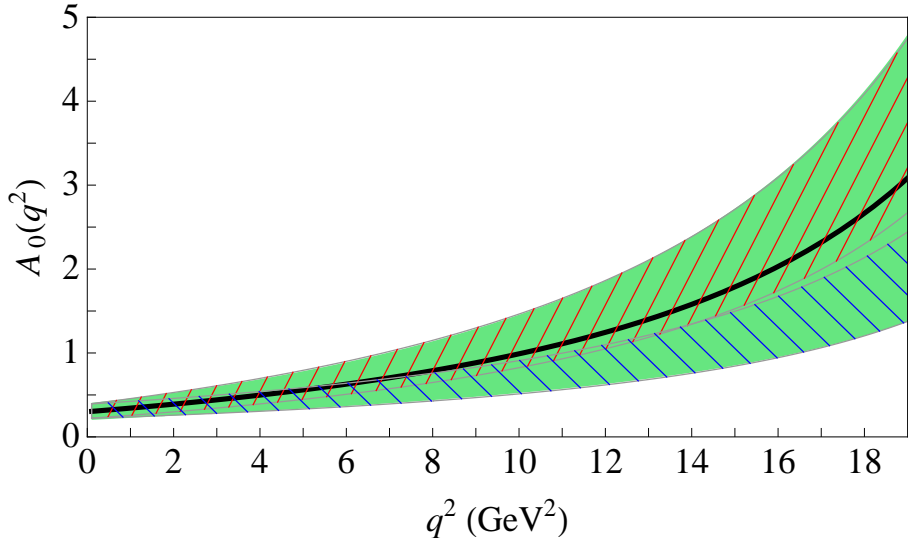}
\caption{For numerical estimates we define an enlarged $A_0(q^2)$ form factor (entering the amplitude $A_t$) in the full-q$^2$ region with enlarged error bars covering the two different determinations: directly from KMPW (red mesh) or from Eq.~(\ref{a0ff}) (blue mesh). This enlarged $A_0(q^2)$  can be obtained from Eq.~(\ref{khod}) with a normalisation $F_{A_0}(0)=0.3\pm0.1$ very similar to KMPW, but with a substantially larger error associated to  the slope $b_{A_0}=-14.5\pm 9.0$.} 
\label{A0At}
\end{figure}

One can compare the axial form factors defined from Eq.~(\ref{a0ff}) with the values obtained from light-cone sum rules in the case of KPMW. We have checked explicitly that $A_1(q^2)$ obtained from Eq.~(\ref{a0ff}) exhibits a very good compatibility with the value computed by KMPW, which was expected as their results for $A_1$ and $V$ fulfilled
the large recoil relations in Eq.~(\ref{effectiverelationships}) satisfactorily within errors.
The same holds for $A_2(q^2)$ with a similar small deviation. On the contrary the last relation of Eq.~(\ref{a0ff}) exhibits a  sizeable difference in the comparison between  the $A_0(q^2)$ obtained from KMPW (red-meshed region in Fig.~\ref{A0At}) and the $A_0(q^2)$ from Eq.~(\ref{a0ff}) (blue-meshed region in Fig.~\ref{A0At}), pointing to  non-negligible  ${\cal O}(\Lambda/m_b)$ corrections. Consequently,  we have enlarged the error size of $A_0(q^2)$ to cover both determinations, as shown  in Fig.~\ref{A0At}.  Notice that the form factor $A_0(q^2)$ only enters in the amplitude $A_t$ which is always suppressed by $m_\ell^2/q^2$, so that this choice will have only a limited impact on our discussion.

\begin{table}
\ra{1.5}
\small
\begin{center}
\begin{tabular}{@{}cccc@{}}
\toprule
Form factor & $F(0)$ & $b_F$ & $m_F$ (GeV)\\ [1mm]
\hline
$\xi_\perp(q^2)$ & $0.31^{+0.20}_{-0.10}$ &  $-4.8^{+0.8}_{- 0.4}$ & $5.412$   \\[1mm]
$\xi_\|(q^2)$ & $0.10^{+0.03}_{-0.02}$ & $-11.8^{+0.8}_{-1.9}$ & $5.366$ \\[1 mm]
\hline
$V(q^2)$&  $0.36^{+0.23}_{- 0.12}$ & $-4.8^{+0.8}_{- 0.4}$ & $5.412$   \\[1mm]
$A_1(q^2)$ & $0.25^{+0.16}_{- 0.10}$ & $0.34^{+0.86}_{-0.80}$ & $5.829$ \\[1mm]
$A_2(q^2)$  & $0.23^{+0.19}_{- 0.10}$ & $-0.85^{+2.88}_{-1.35}$ & $5.829$ \\[1mm]
$T_3(q^2)$  & $0.22^{+0.17}_{- 0.10}$ & $-10.3^{+2.5}_{-3.1}$ & $5.829$ \\[1mm]
\bottomrule
\end{tabular}
\caption{
$B\to K^*$ soft form factors at large recoil (upper cell) and form factors used in the low-recoil region (lower cell). 
Inputs are taken from Ref.~\cite{1006.4945}.
}
\label{TableFFs}
\end{center}
\end{table}

In conclusion, in the large recoil region, once we determine $\xi_\perp(q^2)$ and $\xi_\|(q^2)$ using Eq.~(\ref{khod}) and the numerical inputs of Table~\ref{TableFFs}, we obtain the  form factors  $A_1(q^2)$ and $A_2(q^2)$ from Eqs.~(\ref{a0ff}). The form factor $A_0(q^2)$ appearing in the amplitude $A_t$ is shown in Fig.~(\ref{A0At}). Finally, the tensor form factors ${\cal T}_{1,2,3}$  (or ${\cal T}_{\perp,\|}$), required in the QCDF expression of $B\to K^*\ell\ell$ amplitudes, are computed following
Ref.~\cite{0412400}. The results are  extrapolated up to 8.68 GeV$^2$ to allow a complete comparison with experimental data.
The resulting $\xi_{\|}(q^2)$ and $\xi_\perp(q^2)$  are inserted in the QCDF-corrected  form factors ${\cal T}_{\perp,\|}$ \cite{0412400} required to compute the transversity amplitudes for $B\to K^*\ell^+\ell^-$, including all factorizable and non-factorizable corrections at one loop.

\subsection{Low recoil}
\label{lowr}

In the low-recoil region, the form factors cannot be determined in the same manner. The light-cone sum rule approach is valid at low-$q^2$, and the results in KMPW are presented as valid only up to $14$~GeV$^2$. However, following Refs.~\cite{bobeth,tensors} we will extrapolate them up to 19~GeV$^2$ and check the consistency with the lattice QCD results which can be obtained at high-$q^2$~\cite{mescia}.
As for the large recoil region we will use KMPW form factors in order to remain conservative.

\begin{figure}\centering
\includegraphics[width=7.2cm]{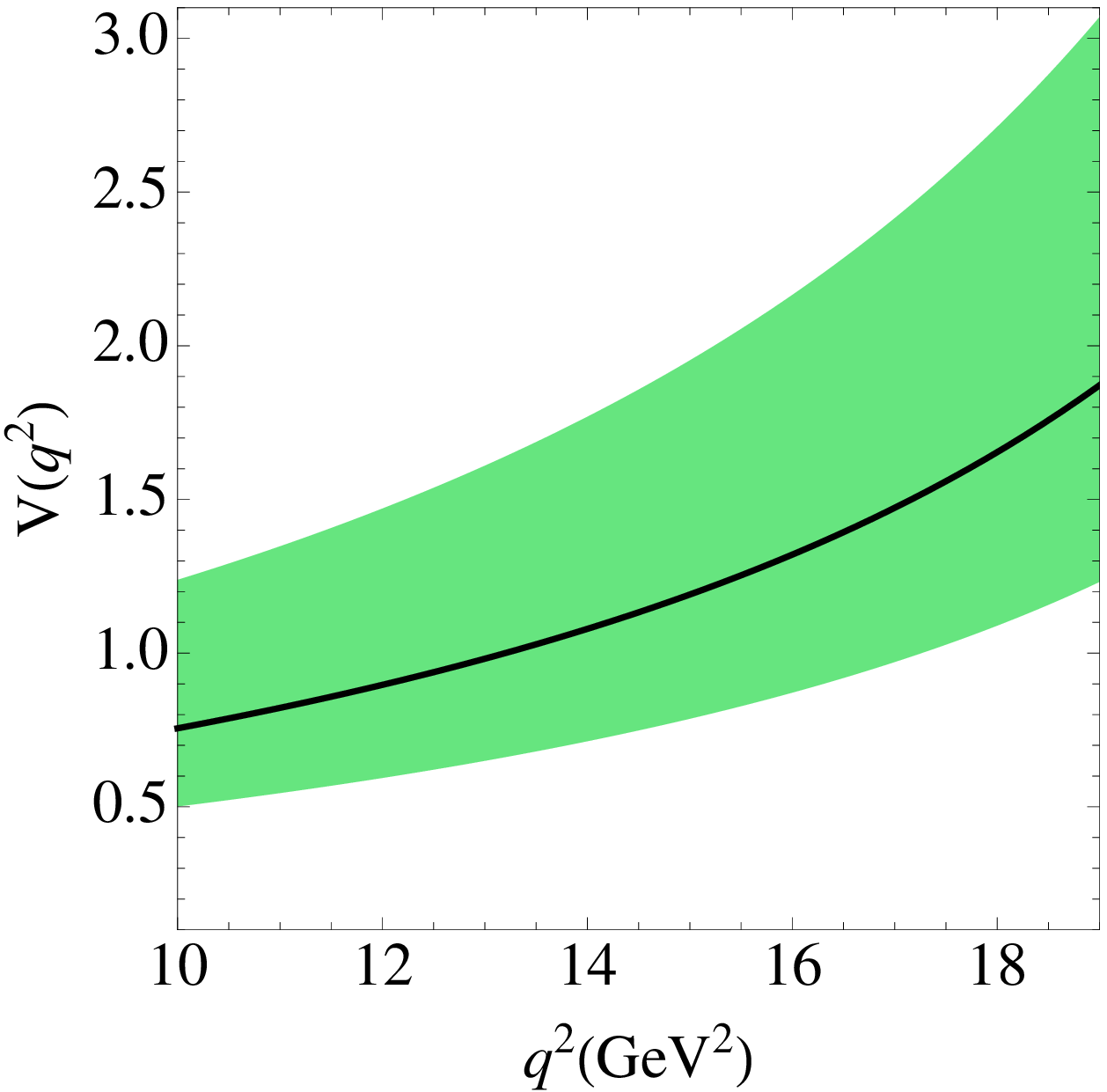}\hspace{1cm}
\includegraphics[width=7.2cm]{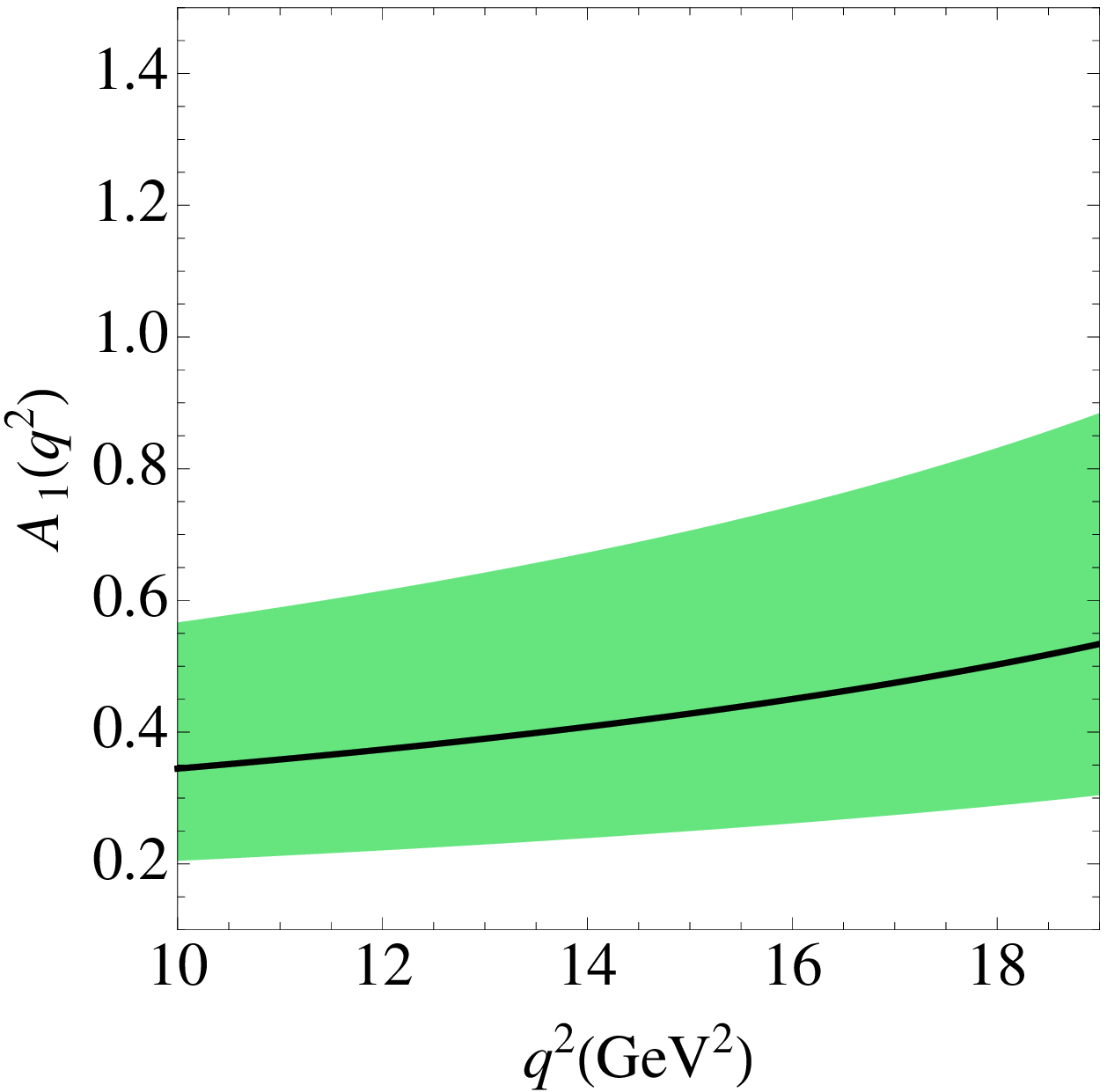}\\[5mm]
\includegraphics[width=7.2cm]{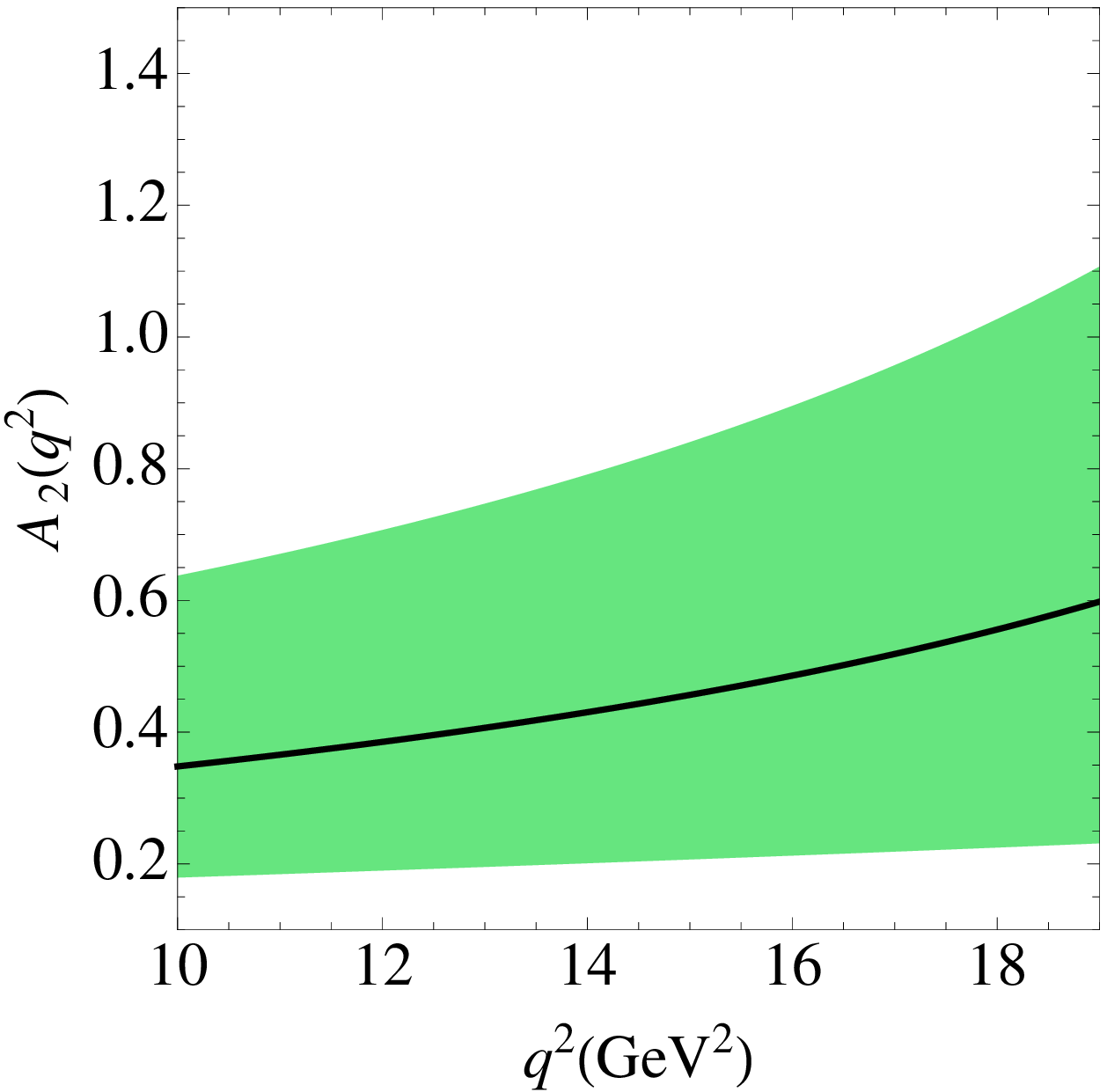}\hspace{1cm}
\includegraphics[width=7.2cm]{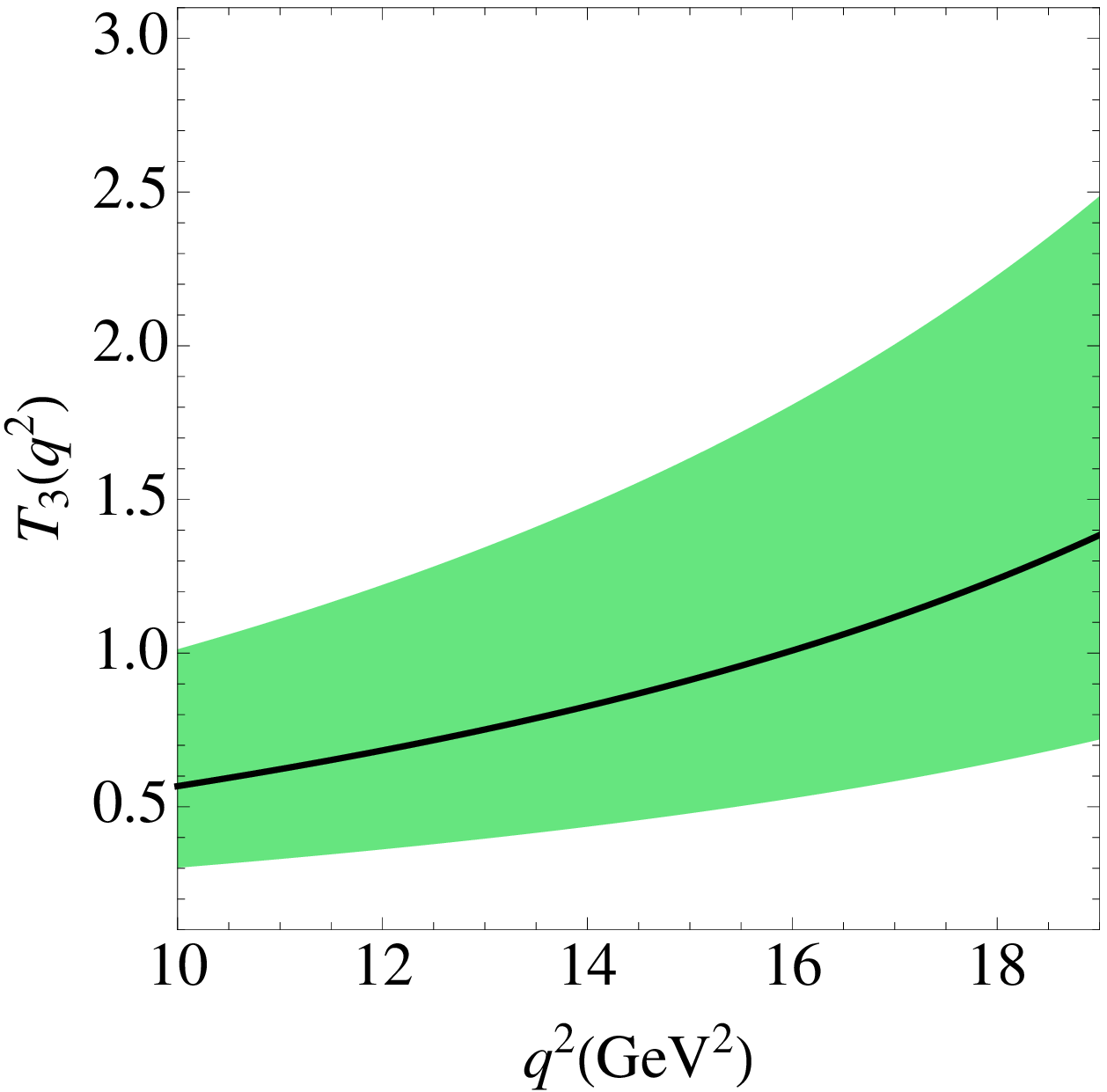}
\caption{KMPW form factors in the low recoil region.  All errors are added in quadrature.
} 
\label{restofFF}
\end{figure} 

\begin{figure}\centering
\includegraphics[width=7.2cm]{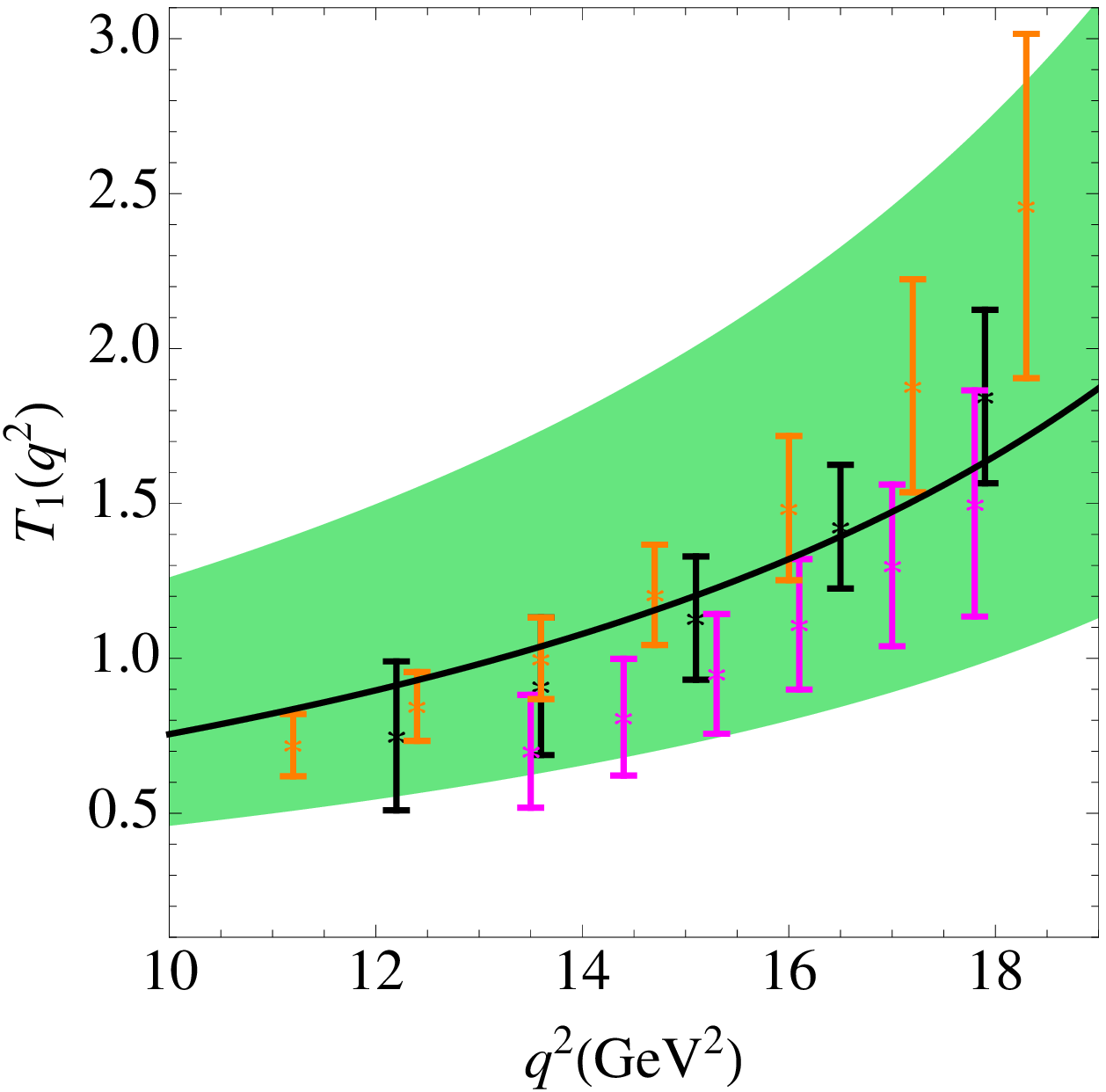}\hspace{1cm}
\includegraphics[width=7.2cm]{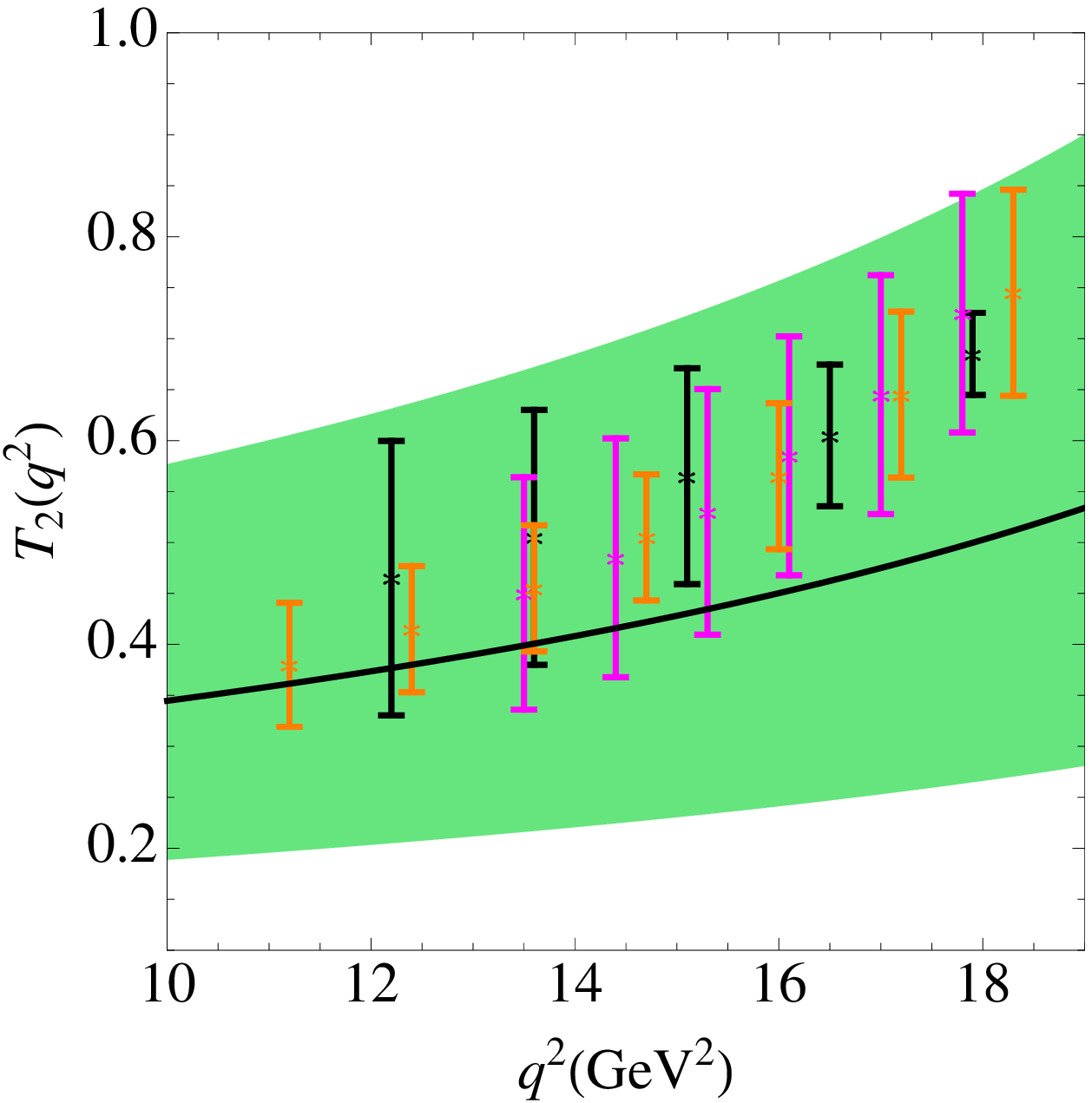}
\caption{Tensor form factors $T_1$ (left) and $T_2$ (right) at low recoil obtained imposing the relations $R_{1,2}$ including a $20\%$ $\Lambda/m_b$ correction, compared with lattice QCD results. The three sets of lattice data points correspond to the three sets of results presented in Table~1 of Ref.~\cite{mescia}.}
\label{T12}
\end{figure}

In this region where all degrees of freedom are soft, we expect the heavy-quark expansion to be a good approximation. Using this approach, in Ref.~\cite{grinstein+pirjol} a set of ratios were found that are expected to approach one in the exact symmetry limit, but away from this limit are broken by $\alpha_s$  and $1/m_b$ corrections. This idea was reconsidered in Ref.~\cite{bobeth}, where three ratios were introduced:
\eq{\label{bob}
R_1= \frac{T_1(q^2)}{V(q^2)}\ , \quad \quad R_2=\frac{T_2(q^2)}{A_1(q^2)}\ , \quad \quad R_3=\frac{q^2}{m_B^2} \frac{T_3(q^2)}{A_2(q^2)}\ .
}
In Ref.~\cite{bobeth}  the first two ratios were found  to be near one, but the third one was found to be around 0.4. 

This uncomfortable situation suggests one to reconsider these ratios.
Indeed the first two of those ratios can be also found in Ref.~\cite{grinstein+pirjol} (see Eq.~(A35) and Eq.~(A36) in that reference), but the last ratio, corresponding to Eq.~(A37), exhibits a more complicated structure:
\eq{
{\hat R_3}=\frac{q^2}{m_B^2} \frac{T_3}{2 \frac{m_V}{m_B} A_0(q^2) - \left(1+\frac{m_V}{m_B} \right) A_1(q^2) + \left(1 - \frac{m_V}{m_B} \right)  A_2 (q^2)}\ .
\label{correct}
}
We have checked that in the KMPW case the ratio ${\hat R_3}$  is indeed in the correct ballpark. The discrepancy between the two versions of $R_3$ is rooted in the scaling laws of the form factors according to HQET power counting (see Eq.~(A4) of Ref.~\cite{grinstein+pirjol}). According to these rules, the three terms in the denominator of ${\hat R_3}$ are of different order in $m_b$, so that $R_3$ and ${\hat R_3}$ are equivalent according to this power counting. However the central values of the form factors extrapolated from light-cone sum rule results do not seem to obey the HQET power counting
numerically, so that the three terms in the denominator of Eq.~(\ref{correct}) are numerically competing and yield a poorly determined value for ${\hat R_3}$ once uncertainties are taken into account.

In order to ensure the robustness of our results, given the problem with the definition of $R_3$, we choose to proceed as follows.  We 
determine $T_1$ and $T_2$ by exploiting the first two ratios ($R_{1,2}$) allowing for a 20\% breaking, i.e., $R_{1,2}=1+\delta_{1,2}$ (with $-0.2\leq \delta_{1,2}\leq 0.2$). 
For all other form factors (including  $T_3$) we  use KMPW extrapolated in the high-$q^2$ region, as shown in Fig.~\ref{restofFF}. 
We then compare with available lattice data~\cite{mescia} to validate the tensor form factors thus obtained.
As can be seen in Fig.~\ref{T12}, we find an excellent agreement between our determination of the tensor form factors $T_{1,2}$ using $R_{1,2}$  and lattice data. This also serves as a test of the validity of the extrapolation for $V(q^2)$ and $A_1(q^2)$.

One may be worried that we drop one of the HQET relationships to use a value for $T_3$ that does not fulfill the expected HQET relation, especially to 
discuss clean observables that have been built based on the existence of these HQET relationships. The problem is actually less acute than it may seem at first sight. Indeed, $T_3$ occurs only in $A_0^{L,R}$, multiplied by a factor $\lambda(q^2)$ that vanishes at the no-recoil endpoint $q^2\to (m_B-m_V)^2$ (with a fairly small derivative). The other terms contributing to the longitudinal transversity amplitudes are suppressed in the large-$m_B$ limit where $m_V/m_B\to 0$, but they are sizeable for the physical values of the mesons. Indeed, in the range of the low-recoil region, one finds the following relative contributions:
\eqa{
A_0^{L,R}[q^2=14\ {\rm GeV}^2] = 
  {\cal N}_0(q^2) \Big[&\C{9\mp10}^{-}& [ 80.8 \cdot A_1(q^2) - 20.5 \cdot A_2(q^2)] \nn\\ + &\C{7}^{-}&\!\!\!\![ 100.4 \cdot T_2(q^2)-28.9 \cdot T_3(q^2)\Big] }
  \eqa{
A_0^{L,R}[q^2=(m_B-m_V)^2] =
  {\cal N}_0(q^2) \Big[&\C{9\mp10}^{-}&  [ 48.3 \cdot A_1(q^2) - 0 \cdot A_2(q^2)] \nn\\ + &\C{7}^{-}&\!\!\!\! [ 68.0 \cdot T_2(q^2)-0 \cdot T_3(q^2)]\Big] 
}
where $ {\cal N}_0(q^2)$ is just a normalization and all form factors are numbers of order 1 in this kinematic region. Therefore, contrary to e.g., $T_2$, the numerical impact of $T_3$ is very small for the low-recoil values of the $B\to K^*\ell^+\ell^-$ observables.
Since $T_3$ plays only a marginal role in the discussion, we will keep the discussion on the construction of clean observables at low recoil assuming for simplicity that the relationship for $R_3$ in Eq.~(\ref{bob}) holds, but for numerical estimates, we will use the extrapolation of $T_3$ according to KMPW.
On the long term, an accurate lattice estimate for $T_3$ would be the best way to settle this uneasy situation and check the validity of the ratio $R_3$, exactly as for $T_{1,2}$.

In summary, the main two differences in our treatment of form factors in this region with respect to Ref.~\cite{bobeth} is that we use a more conservative approach to form factors  and that we do not use all the relations between $T_{1,2,3}$ and $V$, $A_{0,1,2,3}$ implied by the heavy quark symmetry, but only the two ratios ($R_{1,2}$) validated by a comparison to lattice data.

\section{Definition of clean CP conserving and CP violating observables in $q^2$-bins}
\label{sec:2}

Following Refs.~\cite{primary,1207.2753}, we have to consider a further experimental effect:
the various observables are obtained by fitting $q^2$-binned angular distributions, so that
the experimental results for the various observables 
must be compared to (functions 
of) the angular coefficients $J$ and $\bar{J}$ integrated over the relevant kinematic range. Therefore we
define the CP-averaged and CP-violating observables $\intbin{P_i}$ and $\intbin{P_i^{\cp}}$, integrated in $q^2$ bins, as:
\begin{align}
\av{P_1}_{\rm bin}&= \frac12 \frac{\int_{{\rm bin}} dq^2 [J_3+\bar J_3]}{\int_{{\rm bin}} dq^2 [J_{2s}+\bar J_{2s}]}\ ,
& \av{P_1^\cp }_{\rm bin}&= \frac12 \frac{\int_{{\rm bin}} dq^2 [J_3-\bar J_3]}{\int_{{\rm bin}} dq^2 [J_{2s}+\bar J_{2s}]}\ ,
\label{p1}\\
\av{P_2}_{\rm bin} &= \frac18 \frac{\int_{{\rm bin}} dq^2 [J_{6s}+\bar J_{6s}]}{\int_{{\rm bin}} dq^2 [J_{2s}+\bar J_{2s}]}\ ,
& \av{P_2^\cp }_{\rm bin} &= \frac18 \frac{\int_{{\rm bin}} dq^2 [J_{6s}-\bar J_{6s}]}{\int_{{\rm bin}} dq^2 [J_{2s}+\bar J_{2s}]}\ ,\\
\av{P_3}_{\rm bin} &= -\frac14 \frac{\int_{{\rm bin}} dq^2 [J_9+\bar J_9]}{\int_{{\rm bin}} dq^2 [J_{2s}+\bar J_{2s}]}\ ,
& \av{P_3^\cp }_{\rm bin} &= -\frac14 \frac{\int_{{\rm bin}} dq^2 [J_9-\bar J_9]}{\int_{{\rm bin}} dq^2 [J_{2s}+\bar J_{2s}]}\ ,
\end{align}
\begin{align}
\av{P'_4}_{\rm bin} &= \frac1{{\cal N}_\bin^\prime} \int_{{\rm bin}} dq^2 [J_4+\bar J_4]\ ,
& \av{{P'_4}^\cp }_{\rm bin} &= \frac1{{\cal N}_\bin^\prime} \int_{{\rm bin}} dq^2 [J_4-\bar J_4]\ ,\\
\av{P'_5}_{\rm bin} &= \frac1{2{\cal N}_\bin^\prime} \int_{{\rm bin}} dq^2 [J_5+\bar J_5]\ ,
& \av{{P'_5}^\cp }_{\rm bin} &= \frac1{2{\cal N}_\bin^\prime} \int_{{\rm bin}} dq^2 [J_5-\bar J_5]\ ,\\
\av{P'_6}_{\rm bin} &= \frac{-1}{2{\cal N}_\bin^\prime} \int_{{\rm bin}} dq^2 [J_7+\bar J_7]\ ,
& \av{{P'_6}^\cp }_{\rm bin} &= \frac{-1}{2{\cal N}_\bin^\prime} \int_{{\rm bin}} dq^2 [J_7-\bar J_7]\ ,
\end{align}
where the normalization ${\cal N}_\bin^\prime$ is defined as
\eq{{\cal N}_\bin^\prime = {\textstyle \sqrt{-\int_\bin dq^2 [J_{2s}+\bar J_{2s}] \int_{{\rm bin}} dq^2 [J_{2c}+\bar J_{2c}]}}\ .}
We also introduce the redundant\footnote{As was noted in Ref.~\cite{1207.2753}  the symmetries of the distribution show that $Q$ can be expressed in terms of all other observables (up to a sign, see appendix in Ref.~\cite{1207.2753}) and in this sense it is redundant. However, the binning procedure (or scalar contributions) will break this redundancy, recovered only in the limit of vanishing bin widths and in the absence of scalars.}    quantity $P'_8=Q'$ defined in Ref.~\cite{1207.2753}, and its CP-conjugated counterpart:
\begin{align}
\av{P'_8}_{\rm bin} &= \frac{-1}{{\cal N}_\bin^\prime} \int_{{\rm bin}} dq^2 [J_8+\bar J_8]\ ,
& \av{{P'_8}^\cp }_{\rm bin} &= \frac{-1}{{\cal N}_\bin^\prime} \int_{{\rm bin}} dq^2 [J_8-\bar J_8]\ .
\end{align}

These definitions are general: they hold for $m_\ell\ne 0$ and in the presence of scalar and tensor operators. In the limit of infinitesimal binning the definitions of $P_{1,2,3}$ coincide with the definitions in Ref.~\cite{primary} except for a factor of $\beta_\ell(q^2) \equiv \sqrt{1-4m_\ell^2/q^2}$ in $P_2$. This factor was introduced in Ref.~\cite{primary} in the differential definition of $P_2$ precisely to cancel an explicit $\beta_{\ell}$ dependence in the numerator and make the observable insensitive to the lepton mass. However, in defining a binned observable (as noted in Ref.~\cite{1209.1525}) this cancellation takes place only approximately and there is no more compelling reason to remove this factor. 

Using the arguments in Refs.~\cite{primary} and \cite{bobeth}, it is not difficult to check the status of these observables at low $q^2$ and large $q^2$ in relation with Table~\ref{TableObs}. All these observables are indeed built by considering a particular angular coefficient and normalizing it to cancel form factors in the appropriate region, as explained in Sec.~\ref{sec:clean}.
Besides these clean observables, we consider also quantities often discussed: the differential (CP-averaged) branching ratio $d{\rm BR}/dq^2$, the CP asymmetry $A_\cp$, the CP-averaged forward-backward asymmetry $\afb$ and longitudinal polarization fraction $F_L$, and the corresponding CP asymmetries $\afb^\cp$ and $F_L^\cp$. The definitions for the binned observables in terms of the integrated angular coefficients are:
\begin{align}
\av{\afb}&=-\frac34 \frac{\int dq^2 [J_{6s}+\bar J_{6s}]}{\av{d \Gamma/dq^2}+\av{d \bar \Gamma/dq^2}} \ ,
&\av{\afb^\cp}&=-\frac34 \frac{\int dq^2 [J_{6s}-\bar J_{6s}]}{\av{d \Gamma/dq^2}+\av{d \bar \Gamma/dq^2}} \ ,\\[2mm]
\av{F_L}&=-\frac{\int dq^2 [J_{2c}+\bar J_{2c}]}{\av{d \Gamma/dq^2}+\av{d \bar \Gamma/dq^2}} \ ,
&\av{F_L^\cp}&=-\frac{\int dq^2 [J_{2c}-\bar J_{2c}]}{\av{d \Gamma/dq^2}+\av{d \bar \Gamma/dq^2}} \ ,\\[2mm]
\av{\frac{d{\rm BR}}{dq^2}}&=\tau_B \frac{\av{d \Gamma/dq^2}+\av{d \bar \Gamma/dq^2}}2\ ,
&\av{A_\cp}&=\frac{\av{d \Gamma/dq^2}-\av{d \bar \Gamma/dq^2}}{\av{d \Gamma/dq^2}+\av{d \bar \Gamma/dq^2}}\ ,
\label{acp}
\end{align}
where
\eq{
\av{d \Gamma/dq^2} =\frac14 \int dq^2 [3 J_{1c} + 6 J_{1s} - J_{2c} -2 J_{2s}]
}
and analogously for $\bar\Gamma$.

Some of these observables are related to others that have been defined in the literature. For example (see \cite{primary}) $P_1=A_T^{(2)}$ \cite{kruger}, $2 P_2=A_T^{(\rm re)}$, $2 P_3=-A_T^{(im)}$ \cite{becirevic}, $P_{4,5,8}=H_T^{(1,2,4)}$ \cite{bobeth,tensors}, as well as
\eq{
H_T^{(3)} = \frac{2 P_2}{\sqrt{1-P_1^2}}\ ,\qquad H_T^{(5)} = \frac{2 P_3}{\sqrt{1-P_1^2}}\ ,
\label{rels}}
where in terms of the angular coefficients, the observables $H_T^{(3,5)}$ are given by\footnote{We drop a factor of $\beta_\ell$ in $H_T^{(3)}$ with respect to Ref.~\cite{tensors}. The arguments are the same as the ones given above for $P_{1,2,3}$.} \cite{tensors}
\eqa{
\intbin{H_T^{(3)}} &=& \frac{\int_\bin dq^2[J_{6s} + \bar J_{6s}]}{2 \sqrt{4 (\int_\bin dq^2 [J_{2s} + \bar J_{2s}])^2 - (\int_\bin dq^2 [J_3 + \bar J_3])^2}}\ ,\\
\intbin{H_T^{(5)}} &=& \frac{-\int_\bin dq^2 [J_{9} + \bar J_{9}]}{\sqrt{4 (\int_\bin dq^2 [J_{2s} + \bar J_{2s}])^2 - (\int_\bin dq^2 [J_3 + \bar J_3])^2}}\ .
}
The definitions for the integrated unprimed observables $\av{P_{4,5,6,8}^{(\cp)}}$ are given in App.~\ref{app:otherobs}. 

The CP asymmetry $P_{2}^\cp$  is related (but not equal) to the low-recoil observable $a_\cp^{(3)}$ \cite{bobeth2}, which is the CP-violating partner of  $H_T^{(3)}$. 
At low recoil (in the absence of scalar or tensor operators  \cite{tensors})  $a_\cp^{(3)}$ is also equal to the CP-violating partner of $H_T^{(2)}$ which is related (but not equal) to $P_5^\cp$, defined in  App.~\ref{app:otherobs}. 
Analogously, the asymmetry $P_8^\cp$ is related at low recoil to $a_\cp^{(4)}$ \cite{tensors} (CP-violating partner of $H_T^{(4)}$). At low recoil and in the absence of tensor operators this asymmetry is equal to the CP-violating partner of $H_T^{(5)}$, related to $P_3^\cp$ (cf. Eq.~(\ref{rels})). Again this equivalence is not true at large recoil. Besides $P_3^\cp$, we will consider this CP asymmetry related to $H_T^{(5)}$ in the full $q^2$ region, which we shall call $H_T^{(5)\cp}$. The exact definitions for $H_T^{(3)\cp}$ and $H_T^{(5)\cp}$ are given in App.~\ref{app:otherobs}.
 Finally, the asymmetries $A_\cp$ and $\afb^\cp$ are related to $a_\cp^{(1,2)}$ of Ref.~\cite{bobeth2}.
We recall that Table~\ref{TableObs} provides a summary of the equivalence of the observables and their experimental and theoretical status.

In the following sections we will study these integrated observables in detail, giving predictions within the SM and studying their sensitivity to hadronic uncertainties and also to New Physics.

\section{SM predictions for integrated observables}
\label{sec:sm}

In this section we provide SM predictions for the set of integrated observables defined in Section~\ref{sec:2}. We focus on the binning most commonly used by experimental collaborations (see Refs.~\cite{1205.3422,LHCbinned,0904.0770,1108.0695,1204.3933,1208.3987}): [0.1,2], [2,4.3], [4.3,8.68] and [1,6] GeV$^2$ at large recoil, [14.18,16] and [16,19] GeV$^2$ at low recoil, and a bin between the two narrow $c\bar c$ resonances, [10.09,12.89] GeV$^2$. Some of these bins contain $q^2$ regions outside the strict range of application of the corresponding theoretical frameworks. First, the region of very large recoil $q^2\sim 0.1-1$ GeV$^2$ contains contributions from light resonances which are not accounted for in QCDF. A thorough analysis of this region has been performed in Ref.~\cite{camalich}, and some of its features are recalled in Section~\ref{comparison}. However we will not include the effect of these light resonances in our results, as their impact is small on integrated quantities considered here. Second, the region $q^2\sim 6-8.68$ GeV$^2$ can be affected by non-negligible charm-loop effects (see Ref.~\cite{1006.4945}). Within the middle bin [10.09,12.89] GeV$^2$, in between the $J/\Psi$ and $\Psi(2s)$ peaks, the charm-loop contribution leads to a destructive interference, leading to a suppression of the decay rate in this region. However, the predictions within this region should be considered as crude estimates \cite{1006.4945}. In this paper, this region will be treated by interpolating central values and errors between the large and low recoil regions.

\begin{table}
\ra{1.4}
\small
\centering
\begin{tabular}{@{}cccccccccc}
\toprule
$\!\C1(\mu_b)\!$ &   $\!\C2(\mu_b)\!$ &  $\!\C3(\mu_b)\!$ &  $\!\C4(\mu_b)\!$
& $\!\C5(\mu_b)\!$ & $\!\C6(\mu_b)\!$ & $\!\C7^{\rm eff}(\mu_b)\!$ & $\!\C8^{\rm eff}(\mu_b)\!$
& $\!\C9(\mu_b)\!$ & $\! \C{10}(\mu_b)\!$ \\[1mm]
\hline
-0.2632 & 1.0111 & -0.0055 & -0.0806 & 0.0004 &
0.0009 &  -0.2923 & -0.1663 & 4.0749 & -4.3085\\
\bottomrule
\end{tabular}
\caption{NNLO Wilson coefficients at the scale $\mu_b$.}
\label{tabWCs}
\end{table}

Our SM predictions are obtained in the usual way. The integrated observables are defined in Eqs.~(\ref{p1})-(\ref{acp})  in terms of the coefficients $J_i(q^2)$, which are simple functions of the transversity amplitudes (see for example Ref.~\cite{primary}). The transversity amplitudes can be written in terms of Wilson coefficients and $B\to K^*$ form factors following Refs.~\cite{0106067,0412400,bobeth}. Concerning the Wilson coefficients and the treatment of uncertainties, we proceed as in Refs.~\cite{primary,1207.2753}. In particular, the SM Wilson coefficients are computed at the matching scale $\mu_0=2 M_W$, and run down to the hadronic scale $\mu_b=4.8\,{\rm GeV}$ following Refs.~\cite{0512066,0306079,0411071,0312090,0609241}~\footnote{The slightly different values of $\C9(\mu_b)$ and $\C{10}(\mu_b)$ compared to the usual values encountered in the literature stems from the fact that we include higher-order electromagnetic corrections in the evaluation of these coefficients following the formulae gathered in Ref~\cite{0512066}. In particular, Table 5 in that reference shows that $\C9(\mu_b)$ and $\C{10}(\mu_b)$ are affected by subleading but not negligible corrections in $\alpha_{em}$, denoted 
$\C9^{(12)}$ and $\C{10}^{(12)}$. This yields a shift compared to analyses not including higher-order electromagnetic corrections in the evaluation of their coefficients. (See also Ref.~\cite{0712.3009})}. The evolution of couplings and quark masses proceeds analogously. All relevant input parameters used in the numerical analysis, including the values of the SM Wilson coefficients at the scale $\mu_b$, are collected in Tables~\ref{tabWCs} and \ref{TabInputs}. 

\begin{table}
\centering
\begin{tabular}{@{}lrrl@{}}
\toprule[1.1pt]
$\mu_b=4.8 \ {\rm GeV}$ &   & $\mu_0=2M_W$ &  \cite{Misiak:2006zs} \\
\hline
$m_B=5.27950   \ {\rm GeV}$ & \cite{pdg} & $m_{K^*}=0.89594   \ {\rm GeV}$  & \cite{pdg} \\
\hline
$m_{B_s} = 5.3663  \ {\rm{GeV}}$ & \cite{pdg} & $m_{\mu} = 0.105658367  \ {\rm GeV}$ & \cite{pdg} \\
\hline
$ \sin^2 \theta_W                      = 0.2313$ & \cite{pdg}         &  & \\
$M_W=80.399\pm 0.023  \ {\rm GeV}$ &  \cite{pdg} &
$M_Z=91.1876 \ {\rm GeV}$ &  \cite{pdg} \\
\hline
$ \alpha_{em}(M_Z)                     =1/128.940 $ & \cite{Misiak:2006zs}&
$ \alpha_s(M_Z)                        = 0.1184 \pm 0.0007  $ & \cite{pdg} \\
\hline
$ m_t^{\rm pole} = 173.3\pm 1.1  \ {\rm GeV}      $ & \cite{Alcaraz:2009jr}
 & $ m_b^{1S}                   = 4.68 \pm 0.03   \ {\rm GeV}   $  &\cite{Bauer:2004ve} \\
$ m_c^{\overline{MS}}(m_c)                   = 1.27 \pm 0.09  \ {\rm GeV}    $ & \cite{pdg} &
$ m_s^{\overline{MS}}(2\ {\rm GeV})=0.101 \pm 0.029 \ {\rm GeV}$ &   \cite{pdg} \\
\hline
$\lambda_{CKM}=0.22543\pm 0.0008$ &  \cite{ckmfitter} &
$A_{CKM}=0.805\pm 0.020$ & \cite{ckmfitter}  \\
$\bar\rho=0.144\pm 0.025$ & \cite{ckmfitter} &
$\bar\eta=0.342\pm 0.016$ & \cite{ckmfitter}  \\
\hline
$ \Lambda_h=0.5\ {\rm GeV}$ & \cite{Kagan:2001zk} & $ f_B = 0.190 \pm 0.004\ {\rm GeV}    $ & \cite{1203.3862}\\
$f_{K^*,||}=0.220 \pm 0.005$\ {\rm GeV} & \cite{buras}
 & $f_{K^*,\perp}(2\ {\rm GeV})=0.163(8)\ {\rm GeV}$
 & \cite{buras}\\
$ a_{1,||,\perp}(2\ {\rm GeV})=0.03\pm 0.03$ & \cite{buras}& $ a_{2,||,\perp}(2\ {\rm GeV})=0.08\pm 0.06$ & \cite{buras}\\
$\lambda_B(\mu_h)=0.51\pm 0.12\ {\rm GeV}$& \cite{buras}&  $\tau_B=2.307\cdot 10^{12}$ GeV  &  \cite{pdg}\\
\bottomrule[1.1pt]
\end{tabular}
\caption{Input parameters used in the analysis.}
\label{TabInputs}
\end{table}

Concerning uncertainties related to $\Lambda/m_b$ corrections, we proceed as follows. In the large recoil region we include a 10\% multiplicative error in each amplitude, with an arbitrary strong phase, implemented as described in Refs.~\cite{matias2,primary}. In the low recoil region, we allow for a 20\% correction to the ratios $R_1$ and $R_2$, as described in Section~\ref{lowr}, so that $R_{1,2}\sim1\pm 20\%$. We note that this correction is suppressed by a factor $\sim \C7/\C9$ with respect to the dominant part of the amplitude [see Eqs.~(\ref{asd1}) and (\ref{asd2})]. In the SM $\C7^{\rm SM}/\C9^{\rm SM}\sim 0.1$, so the total correction is a few percent, as noticed in Ref.~\cite{bobeth2}. However, in some NP scenarios this suppression might not be so effective.

The results are collected in Tables~\ref{tabSM1} and \ref{tabSM2}, and in App.~\ref{appA}, and they exhibit some expected behaviours.  CP asymmetries are very small. $P_1$ and $P_3$, related to $J_3$ and $J_9$ which involve suppressed helicity form factors in the low $q^2$ region~\cite{camalich} are null tests of the Standard Model for the first bins~\cite{kruger,becirevic}. In addition,  $P_4$ and $P_5$ involve combinations of form factors which become equal in the low-recoil limit, and are thus very close to 1 and -1, respectively, for the last bins~\cite{bobeth}.


\begin{table}
\ra{1.35}
\rb{6mm}
\refstepcounter{table}
\label{tabSM1}
\footnotesize
{\ \ \textsf{\small Table \arabic{table}. Standard Model Predictions for the CP-averaged optimized basis.}}
\begin{center}
\rowcolors{1}{}{lgris}
\begin{tabular}{@{}lcrrr@{}}
\toprule[1.1pt]
Bin (GeV$^2$) & & \cen{$\av{P_1}=\av{A_T^{(2)}}$} & \cen{$\av{P_2}=\textstyle{\frac12} \av{A_T^{\rm (re)}}$} & $\av{P_3}=-\textstyle{\frac12} \av{A_T^{\rm (im)}}$ \\ [1mm]
\hline
%
[\,1\,,\,2\,]  & &
$0.007_{-0.005 - 0.051}^{+0.008 + 0.054}$ & 
$0.399_{-0.023 - 0.008}^{+0.022 + 0.006}$ &
$-0.003_{-0.002 - 0.024}^{+0.001 + 0.027}$\\ [1mm]
[\,0.1\,,\,2\,]  & &
$0.007_{-0.004 - 0.044}^{+0.007 + 0.043}$ & 
$0.172_{-0.009 - 0.018}^{+0.009 + 0.018}$ &
$-0.002_{-0.001 - 0.023}^{+0.001 + 0.02}$\\ [1mm]
[\,2\,,\,4.3\,]  & &
$-0.051_{-0.009 - 0.045}^{+0.010 + 0.045}$ & 
$0.234_{-0.085 - 0.016}^{+0.058 + 0.015}$ &
$-0.004_{-0.003 - 0.022}^{+0.001 + 0.022}$\\ [1mm]
[\,4.3\,,\,8.68\,]  & &
$-0.117_{-0.002 - 0.052}^{+0.002 + 0.056}$ & 
$-0.407_{-0.037 - 0.006}^{+0.048 + 0.008}$ &
$-0.001_{-0.001 - 0.027}^{+0.000+ 0.027}$\\ [1mm]
[\,10.09\,,\,12.89\,]  & &
$-0.181_{-0.361 - 0.029}^{+0.278 + 0.032}$ & 
$-0.481_{-0.005 - 0.002}^{+0.08 + 0.003}$ &
$0.003_{-0.001 - 0.015}^{+0.000+ 0.014}$\\ [1mm]
[\,14.18\,,\,16\,]  & &
$-0.352_{-0.467 - 0.015}^{+0.696 + 0.014}$ & 
$-0.449_{-0.041 - 0.004}^{+0.136 + 0.004}$ &
$0.004_{-0.001 - 0.002}^{+0.000+ 0.002}$\\ [1mm]
[\,16\,,\,19\,]  & &
$-0.603_{-0.315 - 0.009}^{+0.589 + 0.009}$ & 
$-0.374_{-0.126 - 0.004}^{+0.151 + 0.004}$ &
$0.003_{-0.001 - 0.001}^{+0.001 + 0.001}$\\ [1mm]
[\,1\,,\,6\,]  & &
$-0.055_{-0.008 - 0.042}^{+0.009 + 0.040}$ & 
$0.084_{-0.076 - 0.019}^{+0.057 + 0.019}$ &
$-0.003_{-0.002 - 0.022}^{+0.001 + 0.020}$\\ [1mm]
%
%
\midrule[1.1pt]
%
%
\rowcolor{white} & & \cen{$\av{P'_4}$} & \cen{$\av{P'_5}$} & \cen{$\av{P'_6}$} \\ [1mm]
\hline
%
[\,1\,,\,2\,]  & &
$-0.160_{-0.031 - 0.013}^{+0.040 + 0.013}$ & 
$0.387_{-0.063 - 0.015}^{+0.047 + 0.014}$ & 
$-0.104_{-0.042 - 0.016}^{+0.025 + 0.016}$\\ [1mm]
[\,0.1\,,\,2\,]  & &
$-0.342_{-0.019 - 0.017}^{+0.026 + 0.018}$ & 
$0.533_{-0.036 - 0.020}^{+0.028 + 0.017}$ & 
$-0.084_{-0.035 - 0.026}^{+0.021 + 0.026}$\\ [1mm]
[\,2\,,\,4.3\,]  & &
$0.569_{-0.059 - 0.021}^{+0.070 + 0.020}$ & 
$-0.334_{-0.111 - 0.019}^{+0.095 + 0.02}$ & 
$-0.098_{-0.046 - 0.031}^{+0.03 + 0.031}$\\ [1mm]
[\,4.3\,,\,8.68\,]  & &
$1.003_{-0.015 - 0.029}^{+0.014 + 0.024}$ & 
$-0.872_{-0.029 - 0.029}^{+0.043 + 0.03}$ & 
$-0.027_{-0.021 - 0.059}^{+0.012 + 0.059}$\\ [1mm]
[\,10.09\,,\,12.89\,]  & &
$1.082_{-0.144 - 0.017}^{+0.140 + 0.014}$ & 
$-0.893_{-0.110 - 0.017}^{+0.223 + 0.018}$ & 
$0.001_{-0.004 - 0.034}^{+0.003 + 0.034}$\\ [1mm]
[\,14.18\,,\,16\,]  & &
$1.161_{-0.332 - 0.007}^{+0.190 + 0.007}$ & 
$-0.779_{-0.363 - 0.009}^{+0.328 + 0.010}$ & 
$0.000_{-0.000 - 0.000}^{+0.000 + 0.000}$\\ [1mm]
[\,16\,,\,19\,]  & &
$1.263_{-0.248 - 0.004}^{+0.119 + 0.004}$ & 
$-0.601_{-0.367 - 0.007}^{+0.282 + 0.008}$ & 
$0.000_{-0.000- 0.000}^{+0.000+ 0.000}$\\ [1mm]
[\,1\,,\,6\,]  & &
$0.555_{-0.055 - 0.019}^{+0.065 + 0.018}$ & 
$-0.349_{-0.098 - 0.017}^{+0.086 + 0.019}$ & 
$-0.089_{-0.043 - 0.03}^{+0.028 + 0.031}$\\ [1mm]
%
%
\midrule[1.1pt]
%

%
\rowcolor{white} &  & \cen{$10^7 \times \av{d{\rm BR}/dq^2}$} & \cen{$\av{\afb}$} & \cen{$\av{F_L}$} \\ [1mm]
\hline
%
[\,1\,,\,2\,]  & &
$0.437_{-0.148 - 0.023}^{+0.345 + 0.026}$ &
$-0.212_{-0.144 - 0.015}^{+0.11 + 0.014}$ & 
$0.605_{-0.229 - 0.024}^{+0.179 + 0.021}$\\ [1mm]
[\,0.1\,,\,2\,]  & &
$1.446_{-0.561 - 0.054}^{+1.537 + 0.057}$ &
$-0.136_{-0.045 - 0.016}^{+0.048 + 0.016}$ & 
$0.323_{-0.178 - 0.020}^{+0.198 + 0.019}$\\ [1mm]
[\,2\,,\,4.3\,]  & &
$0.904_{-0.314 - 0.055}^{+0.664 + 0.061}$ &
$-0.081_{-0.068 - 0.009}^{+0.054 + 0.008}$ & 
$0.754_{-0.198 - 0.018}^{+0.128 + 0.015}$\\ [1mm]
[\,4.3\,,\,8.68\,]  & &
$2.674_{-0.973 - 0.145}^{+2.326 + 0.156}$ &
$0.220_{-0.112 - 0.016}^{+0.138 + 0.014}$ & 
$0.634_{-0.216 - 0.022}^{+0.175 + 0.022}$\\ [1mm]
[\,10.09\,,\,12.89\,]  & &
$2.344_{-1.100 - 0.063}^{+2.814 + 0.069}$ &
$0.371_{-0.164 - 0.011}^{+0.150 + 0.010}$ & 
$0.482_{-0.208 - 0.013}^{+0.163 + 0.014}$\\ [1mm]
[\,14.18\,,\,16\,]  & &
$1.290_{-0.815 - 0.013}^{+2.122 + 0.013}$ &
$0.404_{-0.191 - 0.005}^{+0.199 + 0.005}$ & 
$0.396_{-0.241 - 0.004}^{+0.141 + 0.004}$\\ [1mm]
[\,16\,,\,19\,]  & &
$1.450_{-0.922 - 0.015}^{+2.333 + 0.015}$ &
$0.360_{-0.172 - 0.005}^{+0.205 + 0.004}$ & 
$0.357_{-0.133 - 0.003}^{+0.074 + 0.003}$\\ [1mm]
[\,1\,,\,6\,]  & &
$2.155_{-0.742 - 0.123}^{+1.646 + 0.138}$ &
$-0.035_{-0.033 - 0.009}^{+0.036 + 0.008}$ & 
$0.703_{-0.212 - 0.019}^{+0.149 + 0.017}$\\ [1mm]
%
%
\bottomrule[1.1pt]
\end{tabular} 
\end{center} 
\end{table}  



\begin{table}
\ra{1.35}
\rb{6mm}
\refstepcounter{table}
\label{tabSM2}
\footnotesize
{\ \ \textsf{\small Table \arabic{table}. Standard Model Predictions for the CP-violating optimized basis.}}
\begin{center}
\rowcolors{1}{}{lgris}
\begin{tabular}{@{}lcrrr@{}}
\toprule[1.1pt]
Bin (GeV$^2$) & & \cen{$10^2 \times \av{P_1^\cp}$} & \cen{$10^2 \times \av{P_2^\cp}$} & \cen{$10^2 \times \av{P_3^\cp}$} \\ [1mm]
\hline
%
[\,1\,,\,2\,]  & &
$-0.010_{-0.004 - 0.155}^{+0.002 + 0.150}$ & 
$-0.403_{-0.074 - 0.031}^{+0.008 + 0.036}$ &
$-0.044_{-0.009 - 0.077}^{+0.016 + 0.074}$\\ [1mm]
[\,0.1\,,\,2\,]  & &
$0.001_{-0.001 - 0.128}^{+0.001 + 0.133}$ & 
$-0.133_{-0.034 - 0.061}^{+0.004 + 0.059}$ &
$-0.028_{-0.004 - 0.062}^{+0.008 + 0.069}$\\ [1mm]
[\,2\,,\,4.3\,]  & &
$-0.061_{-0.009 - 0.152}^{+0.004 + 0.14}$ & 
$-1.018_{-0.120 - 0.013}^{+0.033 + 0.018}$ &
$-0.047_{-0.007 - 0.073}^{+0.020 + 0.066}$\\ [1mm]
[\,4.3\,,\,8.68\,]  & &
$-0.088_{-0.009 - 0.074}^{+0.003 + 0.079}$ & 
$-0.650_{-0.127 - 0.009}^{+0.060 + 0.012}$ &
$-0.008_{-0.001 - 0.037}^{+0.007 + 0.035}$\\ [1mm]
[\,10.09\,,\,12.89\,]  & &
$-0.053_{-0.015 - 0.026}^{+0.017 + 0.028}$ & 
$-0.208_{-0.095 - 0.007}^{+0.095 + 0.007}$ &
$0.001_{-0.001 - 0.013}^{+0.002 + 0.013}$\\ [1mm]
[\,14.18\,,\,16\,]  & &
$-0.004_{-0.006 - 0.000}^{+0.009 + 0.000}$ & 
$0.000_{-0.000- 0.000}^{+0.000+ 0.000}$ &
$0.001_{-0.000- 0.000}^{+0.000+ 0.000}$\\ [1mm]
[\,16\,,\,19\,]  & &
$-0.006_{-0.004 - 0.000}^{+0.007 + 0.000}$ & 
$0.000_{-0.000- 0.000}^{+0.000+ 0.000}$ &
$0.001_{-0.000- 0.000}^{+0.000+ 0.000}$\\ [1mm]
[\,1\,,\,6\,]  & &
$-0.060_{-0.007 - 0.119}^{+0.004 + 0.110}$ & 
$-0.828_{-0.097 - 0.007}^{+0.028 + 0.012}$ &
$-0.036_{-0.004 - 0.057}^{+0.015 + 0.051}$\\ [1mm]
%
%
\midrule[1.1pt]
%
%
\rowcolor{white} & & \cen{$10^2 \times \av{{P'_4}^\cp}$} & \cen{$10^2 \times \av{{P'_5}^\cp}$} & \cen{$10^2 \times \av{{P'_6}^\cp}$} \\ [1mm]
\hline
%
[\,1\,,\,2\,]  & &
$0.144_{-0.040 - 0.153}^{+0.139 + 0.138}$ & 
$-0.891_{-0.151 - 0.128}^{+0.013 + 0.141}$ & 
$-1.007_{-0.214 - 0.134}^{+0.402 + 0.129}$\\ [1mm]
[\,0.1\,,\,2\,]  & &
$-0.04_{-0.054 - 0.149}^{+0.129 + 0.139}$ & 
$-0.582_{-0.157 - 0.137}^{+0.036 + 0.148}$ & 
$-0.874_{-0.165 - 0.138}^{+0.328 + 0.137}$\\ [1mm]
[\,2\,,\,4.3\,]  & &
$0.631_{-0.041 - 0.119}^{+0.091 + 0.111}$ & 
$-1.277_{-0.102 - 0.097}^{+0.03 + 0.106}$ & 
$-0.805_{-0.139 - 0.127}^{+0.336 + 0.122}$\\ [1mm]
[\,4.3\,,\,8.68\,]  & &
$0.782_{-0.023 - 0.043}^{+0.054 + 0.040}$ & 
$-0.896_{-0.099 - 0.040}^{+0.045 + 0.045}$ & 
$-0.255_{-0.027 - 0.072}^{+0.109 + 0.073}$\\ [1mm]
[\,10.09\,,\,12.89\,]  & &
$0.399_{-0.155 - 0.022}^{+0.162 + 0.020}$ & 
$-0.339_{-0.167 - 0.019}^{+0.145 + 0.020}$ & 
$-0.051_{-0.016 - 0.026}^{+0.029 + 0.024}$\\ [1mm]
[\,14.18\,,\,16\,]  & &
$0.013_{-0.008 - 0.000}^{+0.006 + 0.000}$ & 
$0.000_{-0.000- 0.000}^{+0.000+ 0.000}$ & 
$0.000_{-0.000- 0.000}^{+0.000+ 0.000}$\\ [1mm]
[\,16\,,\,19\,]  & &
$0.013_{-0.007 - 0.000}^{+0.006 + 0.000}$ & 
$0.000_{-0.000- 0.000}^{+0.000+ 0.000}$ & 
$0.000_{-0.000- 0.000}^{+0.000+ 0.000}$\\ [1mm]
[\,1\,,\,6\,]  & &
$0.597_{-0.036 - 0.104}^{+0.080 + 0.095}$ & 
$-1.140_{-0.096 - 0.082}^{+0.028 + 0.091}$ & 
$-0.691_{-0.111 - 0.114}^{+0.284 + 0.106}$\\ [1mm]
%
%
\midrule[1.1pt]
%

%
\rowcolor{white} &  & \cen{$10^2 \times \av{A_{\rm CP}}$} & \cen{$10^2 \times \av{\afb^\cp}$} & \cen{$10^2 \times \av{F_L^\cp}$} \\ [1mm]
\hline
%
[\,1\,,\,2\,]  & &
$0.005_{-0.518 - 0.133}^{+0.373 + 0.113}$ &
$0.214_{-0.108 - 0.040}^{+0.152 + 0.041}$ & 
$0.387_{-0.163 - 0.056}^{+0.142 + 0.048}$\\ [1mm]
[\,0.1\,,\,2\,]  & &
$-0.29_{-0.469 - 0.103}^{+0.370 + 0.100}$ &
$0.105_{-0.036 - 0.049}^{+0.045 + 0.052}$ & 
$0.208_{-0.121 - 0.038}^{+0.141 + 0.035}$\\ [1mm]
[\,2\,,\,4.3\,]  & &
$0.424_{-0.260 - 0.067}^{+0.186 + 0.056}$ &
$0.351_{-0.196 - 0.039}^{+0.301 + 0.043}$ & 
$0.479_{-0.150 - 0.042}^{+0.115 + 0.034}$\\ [1mm]
[\,4.3\,,\,8.68\,]  & &
$0.673_{-0.060 - 0.013}^{+0.071 + 0.011}$ &
$0.350_{-0.169 - 0.025}^{+0.213 + 0.025}$ & 
$0.402_{-0.147 - 0.023}^{+0.121 + 0.023}$\\ [1mm]
[\,10.09\,,\,12.89\,]  & &
$0.366_{-0.145 - 0.008}^{+0.150 + 0.008}$ &
$0.160_{-0.107 - 0.009}^{+0.139 + 0.008}$ & 
$0.169_{-0.050 - 0.008}^{+0.033 + 0.008}$\\ [1mm]
[\,14.18\,,\,16\,]  & &
$0.012_{-0.006 - 0.000}^{+0.005 + 0.000}$ &
$0.000_{-0.000 - 0.000}^{+0.000 + 0.000}$ & 
$0.004_{-0.003 - 0.000}^{+0.002 + 0.000}$\\ [1mm]
[\,16\,,\,19\,]  & &
$0.010_{-0.005 - 0.000}^{+0.005 + 0.000}$ &
$0.000_{-0.000 - 0.000}^{+0.000 + 0.000}$ & 
$0.004_{-0.002 - 0.000}^{+0.002 + 0.000}$\\ [1mm]
[\,1\,,\,6\,]  & &
$0.422_{-0.249 - 0.066}^{+0.184 + 0.054}$ &
$0.346_{-0.183 - 0.035}^{+0.261 + 0.038}$ & 
$0.446_{-0.155 - 0.040}^{+0.123 + 0.035}$\\ [1mm]
%
%
\bottomrule[1.1pt]
\end{tabular} 
\end{center} 
\end{table}  


\begin{figure}
\begin{center}
\includegraphics[height=5cm,width=16cm]{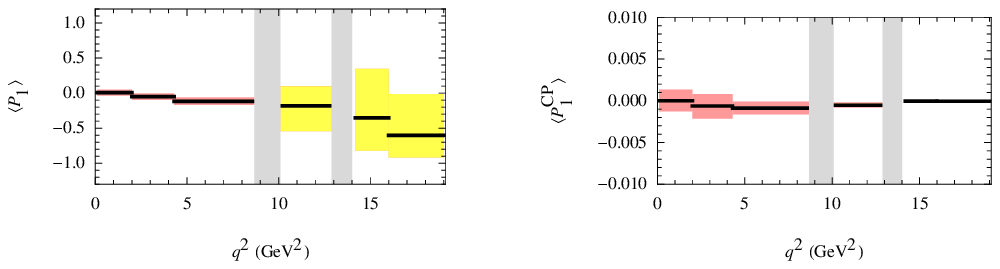}
\includegraphics[height=5cm,width=16cm]{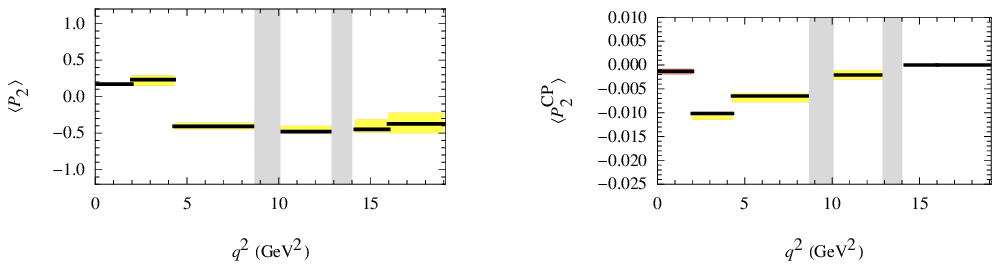}
\includegraphics[height=5cm,width=16cm]{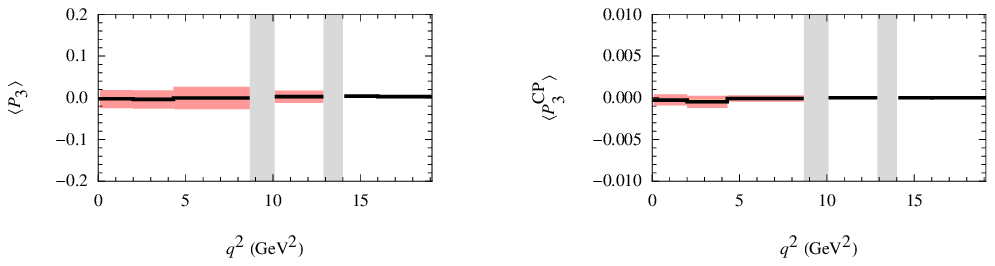}
\end{center}
\vspace{-0.4cm}
\caption{Binned Standard Model predictions for the observables $\av{P_{1,2,3}^{\sss\rm (CP)}}$, corresponding to the bins measured experimentally (see Tables \ref{tabSM1} and \ref{tabSM2}). The red (dark gray) error bar correspond to the $\Lambda/m_b$ corrections, the yellow one (light gray) to the other sources of uncertainties. If one of the two bands is missing, it means that the associated uncertainty is negligible compared to the dominant one.
}
\label{SMplotsPs1}
\end{figure}

\begin{figure}
\begin{center}
\includegraphics[height=5cm,width=16cm]{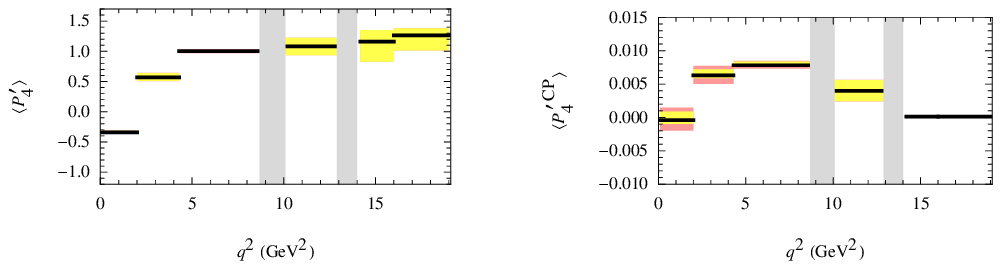}
\includegraphics[height=5cm,width=16cm]{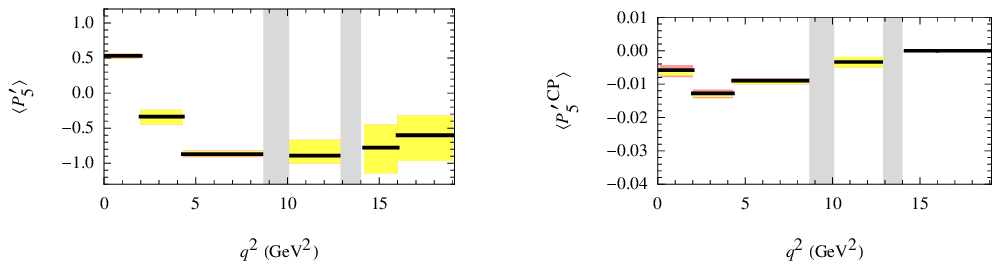}
\includegraphics[height=5cm,width=16cm]{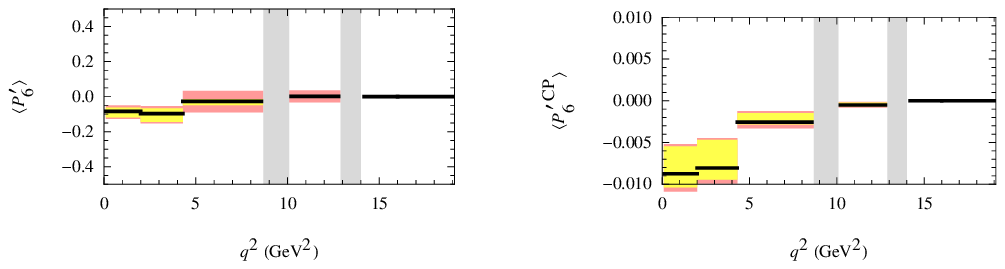}
\includegraphics[height=5cm,width=16cm]{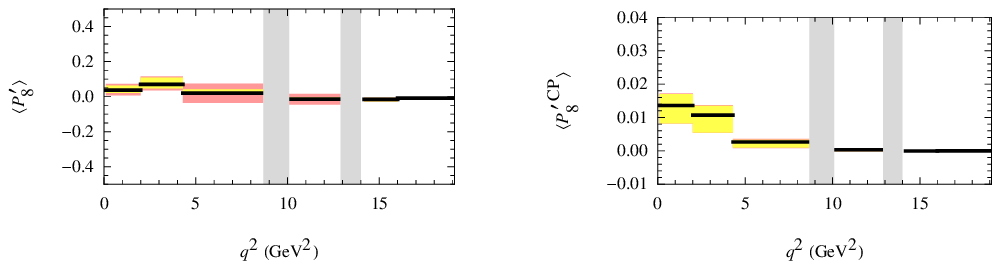}
\end{center}
\vspace{-0.4cm}
\caption{Binned Standard Model predictions for the observables $\av{P_{4,5,6,8}^{\prime \sss\rm (CP)}}$,  with the same conventions as in Fig.~\ref{SMplotsPs1}.}
\label{SMplotsPs2}
\end{figure}

\section{New Physics opportunities }
\label{sec:const}

Our main motivation has consisted in finding an optimal basis for the analysis of $B\to K^*\ell\ell$ data,
with a balance between the theoretical control on hadronic pollution and the experimental accessibility.
The importance of finding clean observables for NP searches has been emphasized in Ref.~\cite{1207.2753}. It has been shown that, while a NP contribution to an angular coefficient $J_i$ can be substantial, a non-clean observable sensitive to the coefficient $J_i$ (such as $J_i$ itself, or $S_i$) might not  be able to detect such NP because of large theoretical uncertainties, even if the SM prediction for this observable is accurate. On the contrary, the corresponding clean observable $P_i$ might show a clear distinction between the SM and the NP scenario even once theoretical uncertainties are included.

\begin{figure}
\begin{center}
\includegraphics[width=7cm]{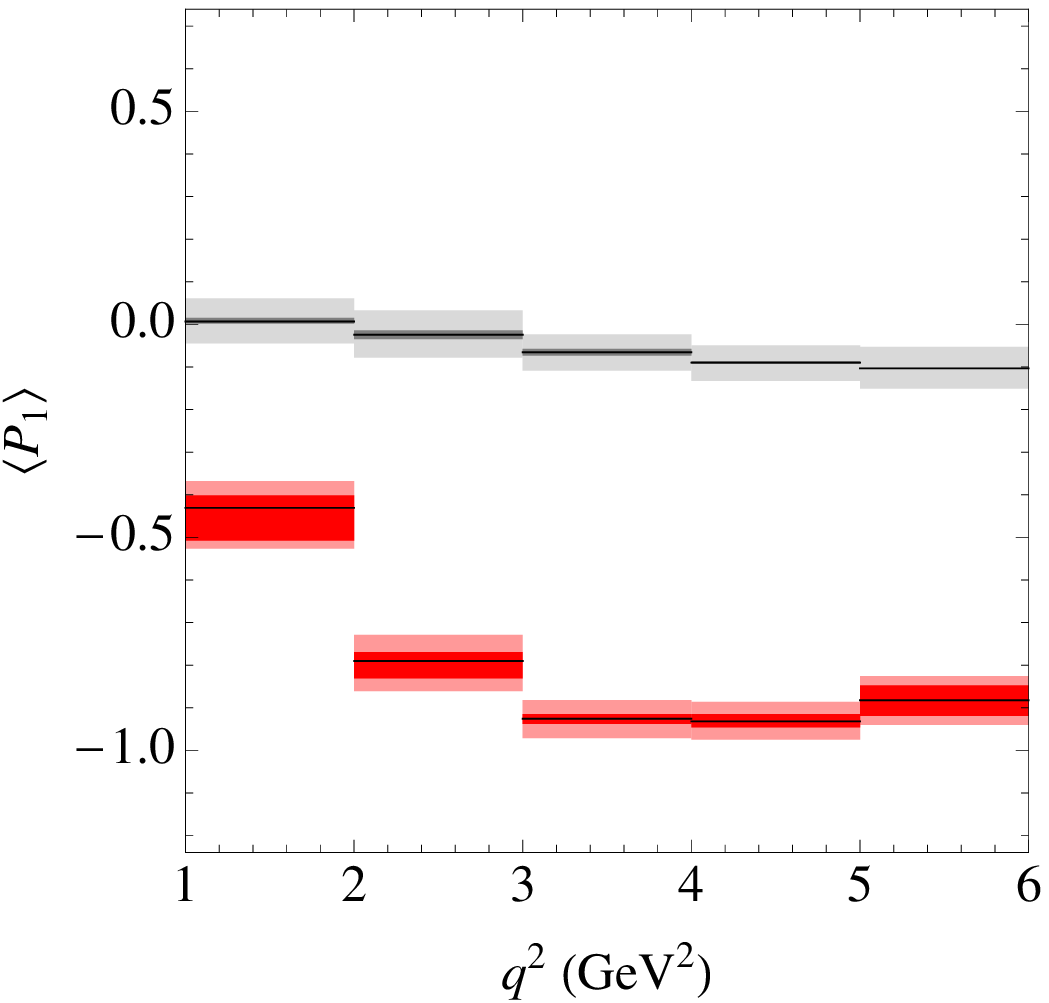} \hspace{1cm} \includegraphics[width=7cm]{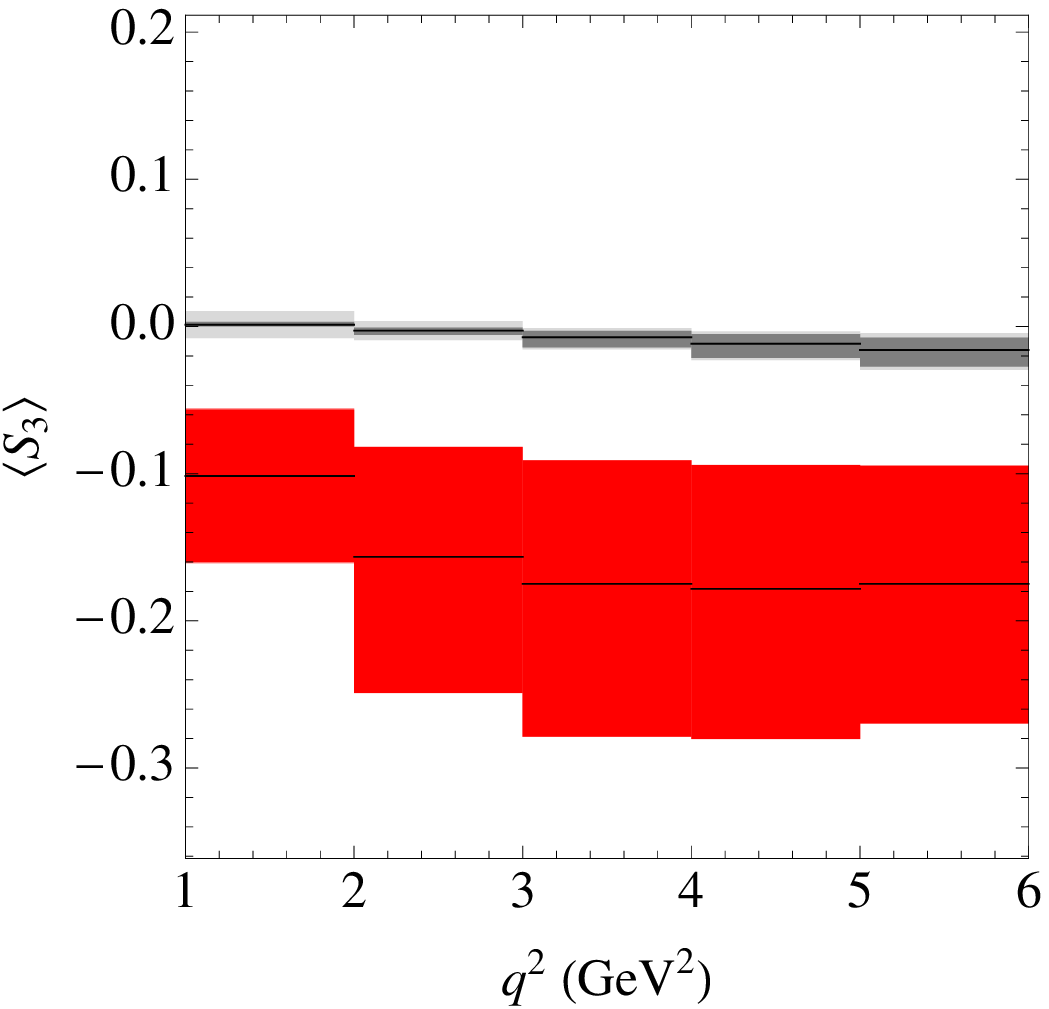}
\end{center}
\vspace{-0.4cm}
\caption{The sensitivity to New Physics for two CP-averaged observables related to the coefficient $J_3$: $P_1$ (left), and $S_3$ (right).
SM predictions are shown in gray, and the NP point $(\delta \C7,\delta \Cp7,\delta\C9,\delta\C{10})=(0.3,-0.4,1,6)$ is shown in red. Our estimate of power corrections is included in light gray (SM) and light red (NP).
$P_1$ is much more sensitive to New Physics than $S_3$, due to its reduced hadronic uncertainties.
}
\label{P1vsS3}
\end{figure}

This feature has been exemplified in Ref~\cite{1207.2753} studying the observables $P_1$ and $S_3$ both in the SM and within a NP benchmark point compatible with other constraints from rare $B$ decays. In Fig.~\ref{P1vsS3} we update this discussion by showing binned predictions at large recoil for $\av{P_1}$ and $\av{S_3}$ in the SM and in the NP scenario given by $(\delta \C7,\delta \Cp7,\delta\C9,\delta\C{10})=(0.3,-0.4,1,6)$ (where $\delta \C{i}=\C{i}(\mu_b)-\C{i}^{SM}(\mu_b)$). Clearly, $P_1$ is much more sensitive to New Physics than $S_3$, due to its reduced hadronic uncertainties.

\begin{figure}
\begin{center}
\includegraphics[width=3.8cm]{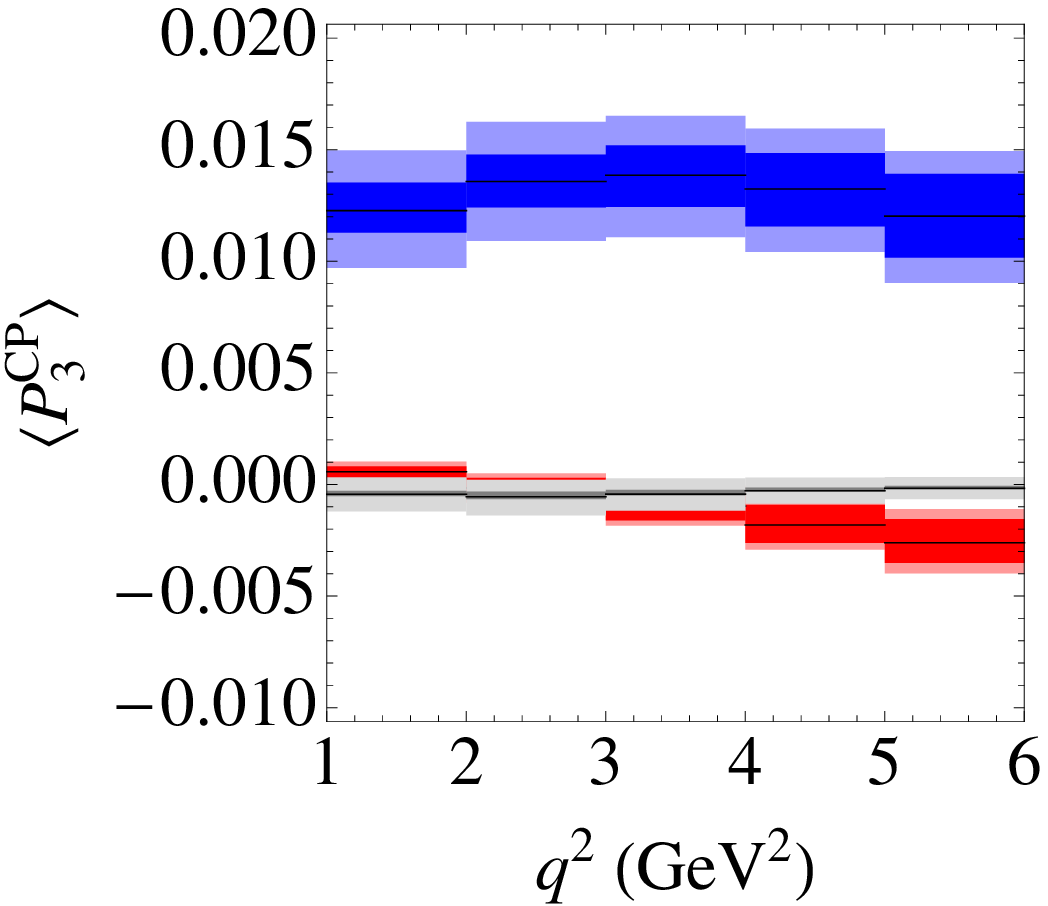} \hspace{1mm} \includegraphics[width=3.9cm]{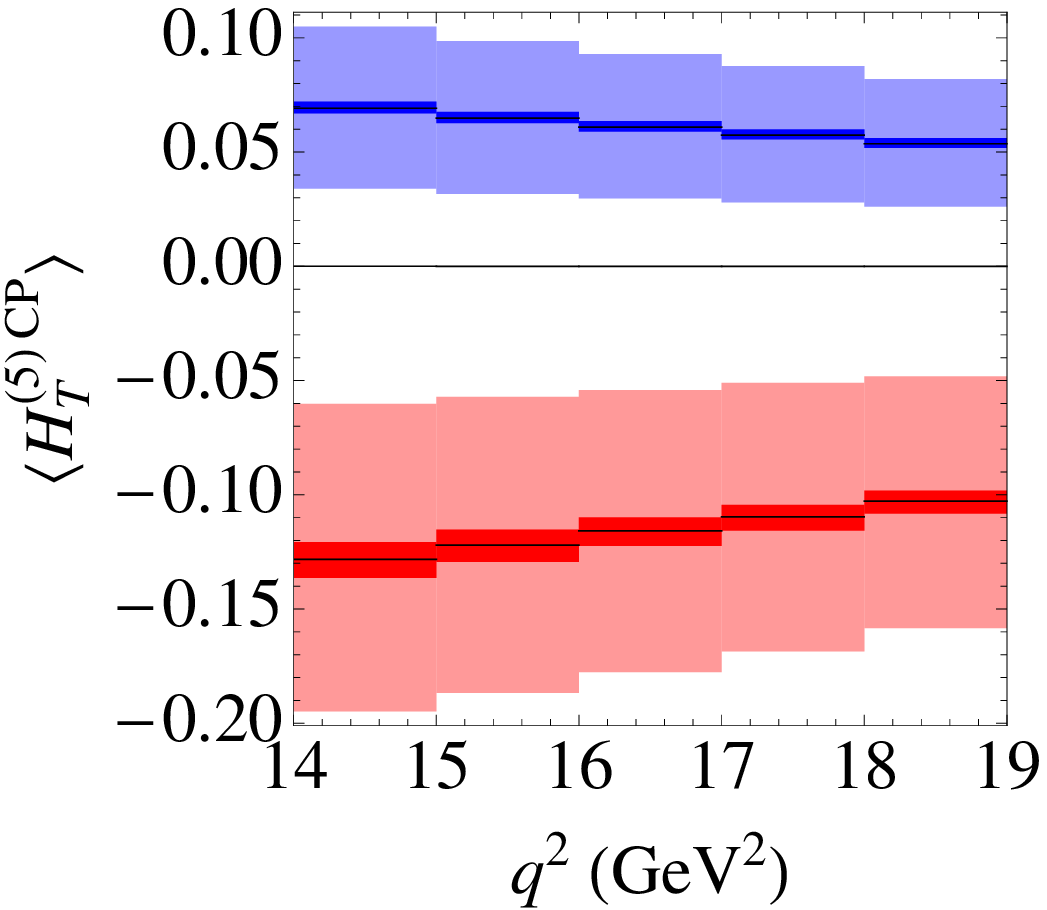}\hspace{1mm} 
\includegraphics[width=3.7cm]{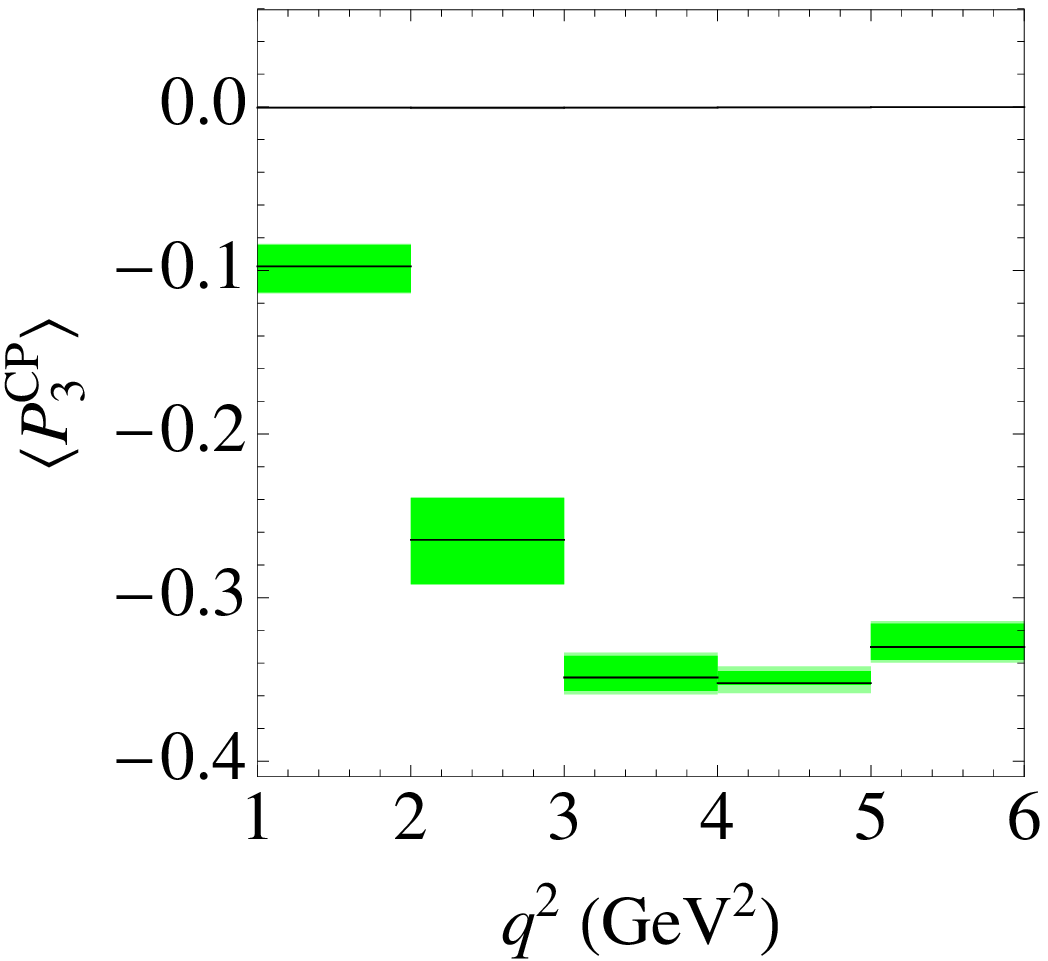}\hspace{1mm} \includegraphics[width=3.8cm]{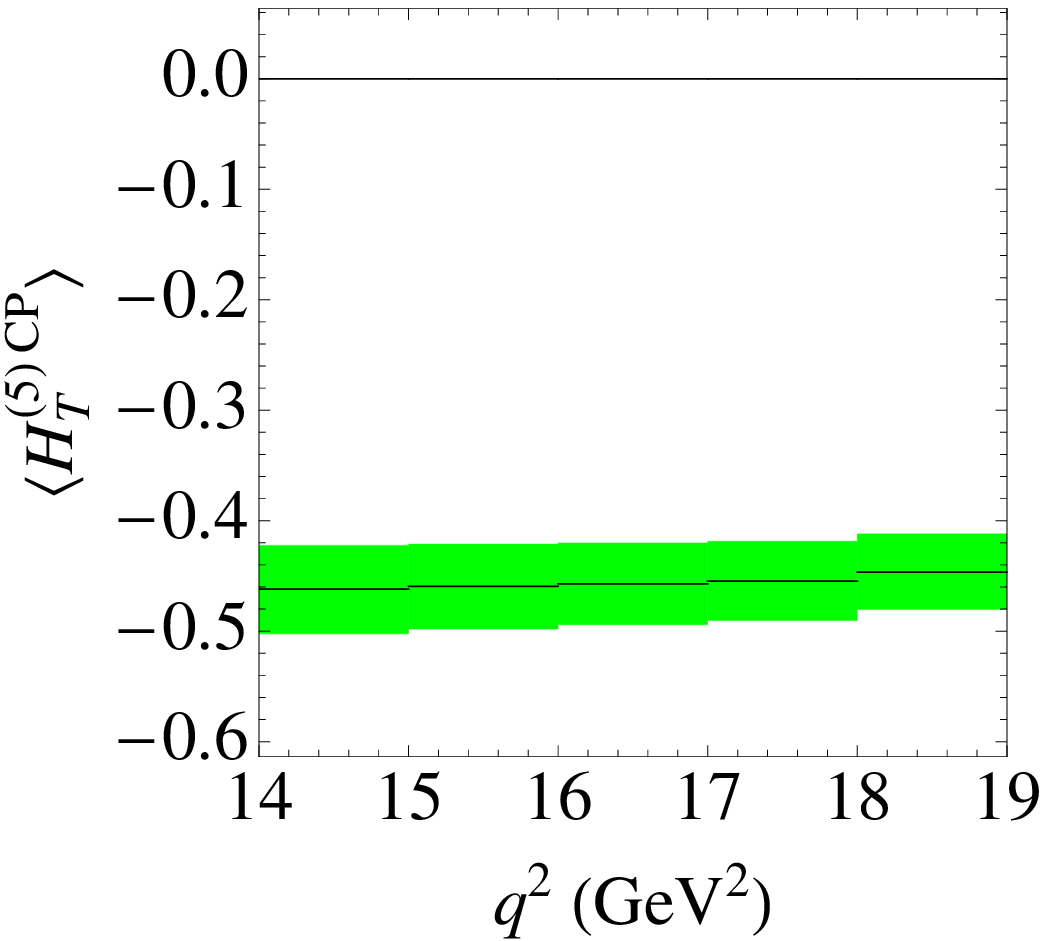}\\[6mm]
\includegraphics[width=3.8cm]{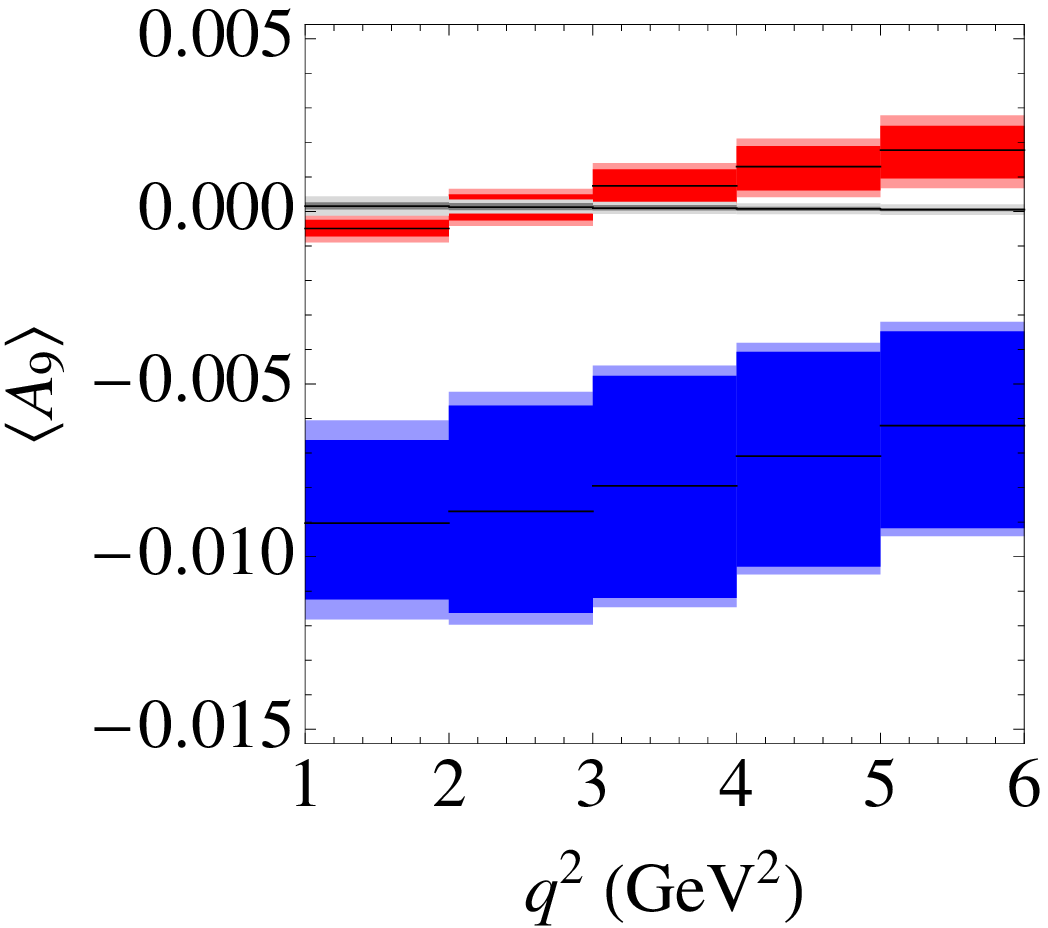}\hspace{1mm} \includegraphics[width=3.85cm]{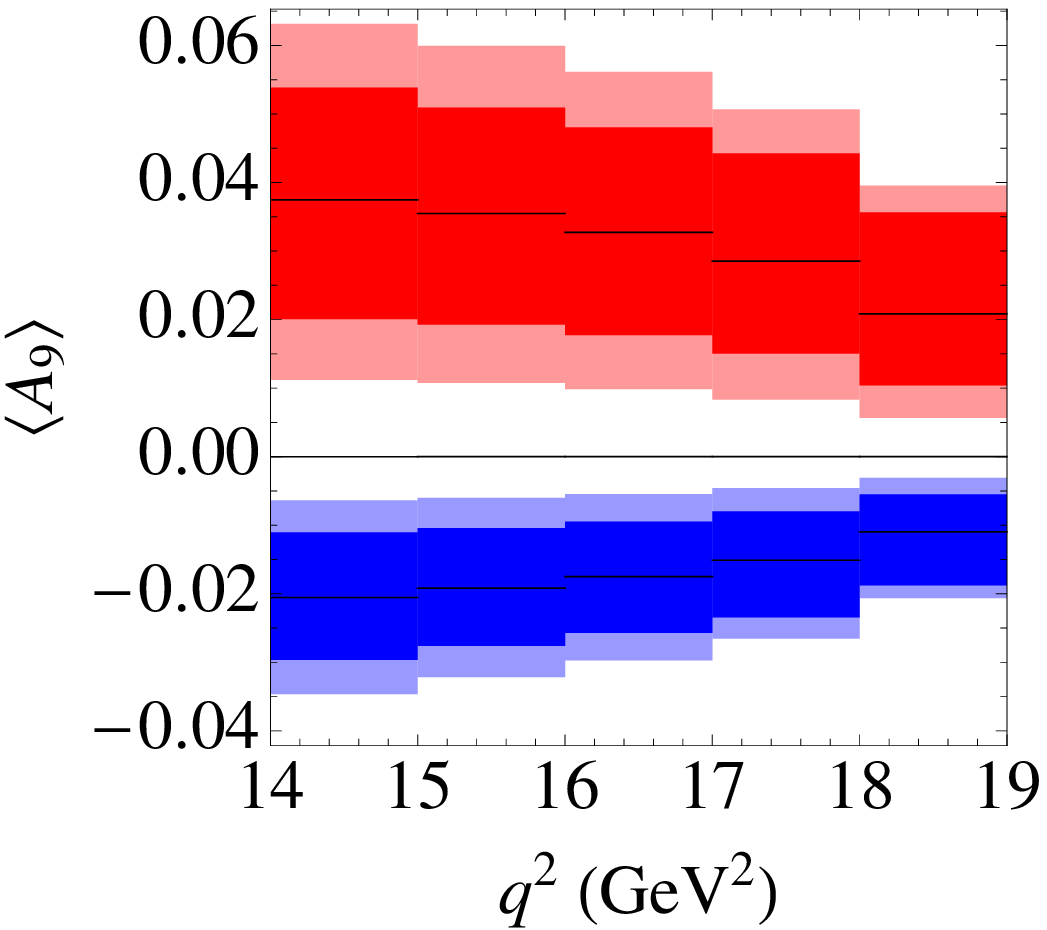}\hspace{1mm} 
\includegraphics[width=3.75cm]{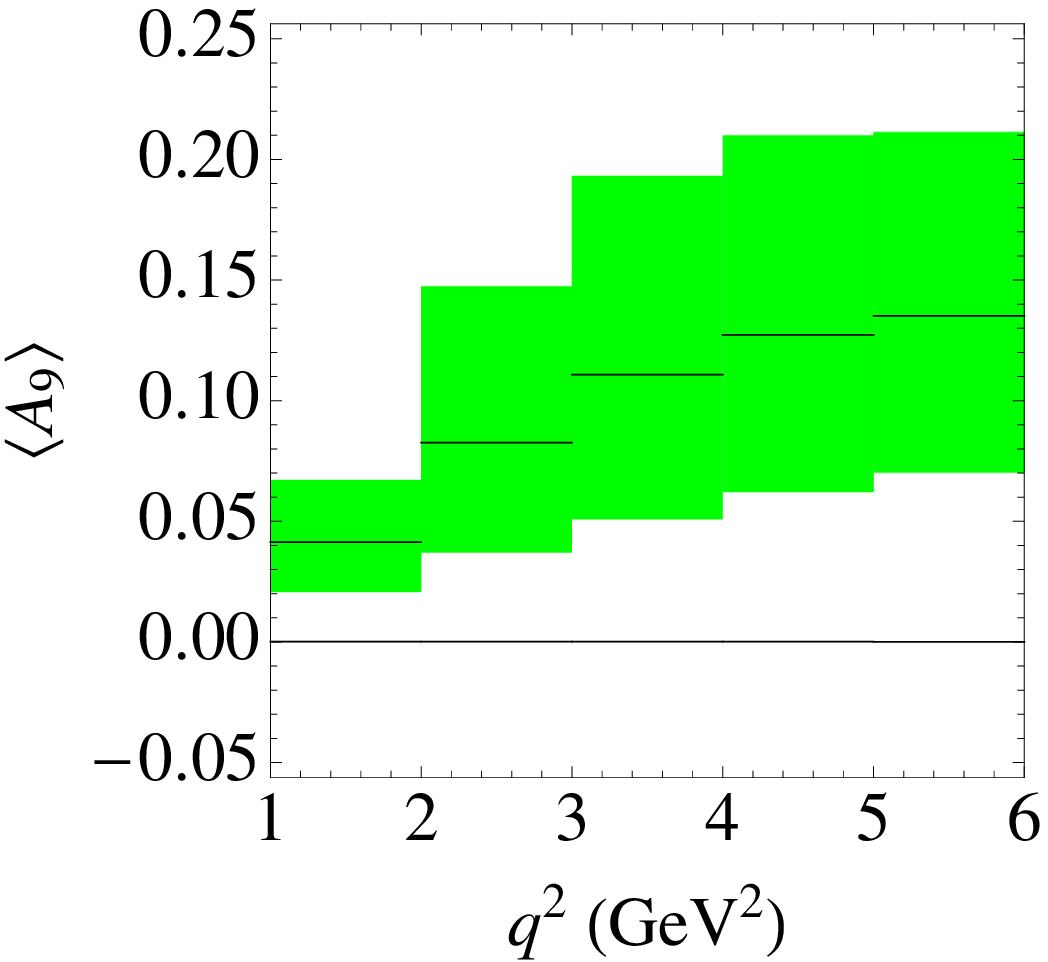}\hspace{1mm} \includegraphics[width=3.8cm]{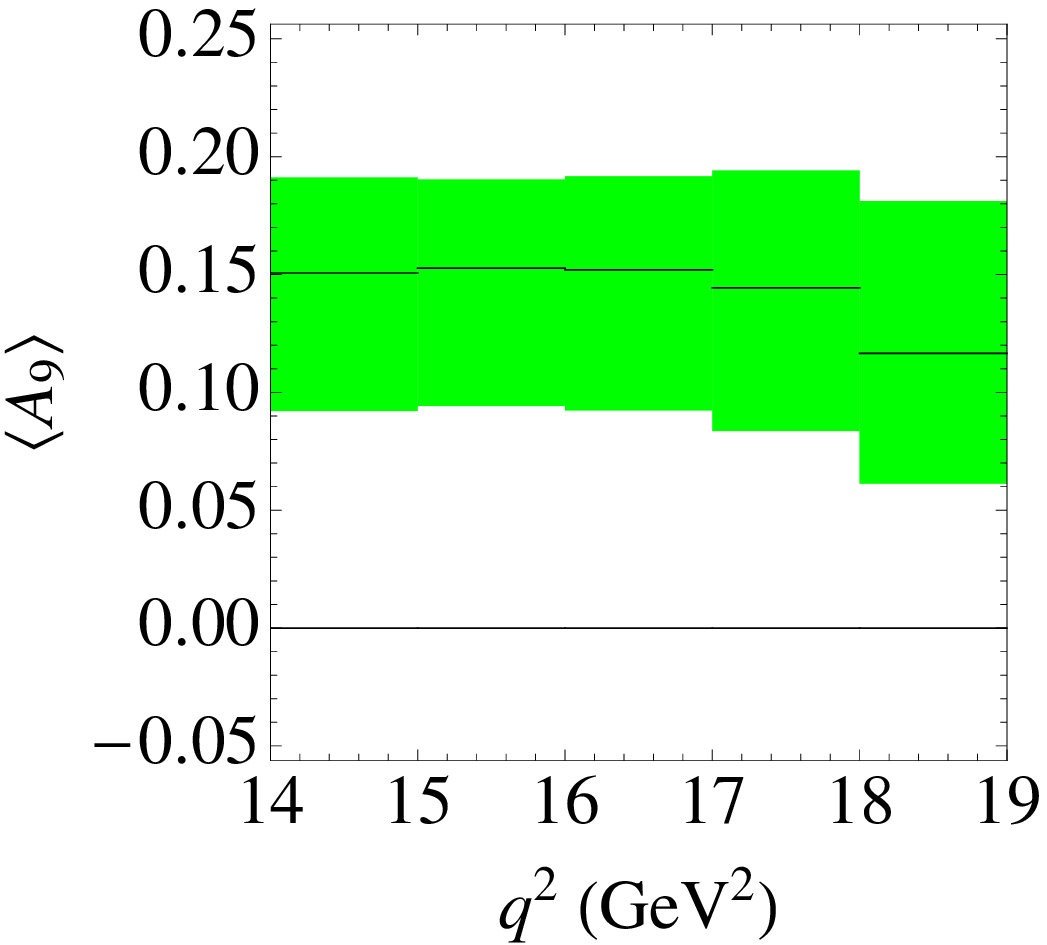}\\[6mm]
\end{center}
\vspace{-0.4cm}
\caption{The case of New Physics in three observables related to the coefficient $J_9$: $P_3^\cp$, $H_T^{\sss\rm (5) CP}$ and $A_9$. Red and blue binned curves (left plots) correspond to predictions for non-standard complex left-handed currents: $(\delta \C7,\delta\C9,\delta\C{10})=(0.1+0.5i,-1.4,1-1.5i)$ (blue) and $(\delta \C7,\delta\C9,\delta\C{10})=(1.5+0.3i,-8+2i,8-2i)$ (red). Green  binned curves (right plots) corresponds to $\delta \Cp{10}=-1.5+2i$. The SM predictions (with errors) correspond to the narrow grey bins around zero. $H_T^{\sss\rm (5) CP}$ and $P_3^\cp$ are much more sensitive to New Physics than $A_9$, due to their reduced hadronic uncertainties.
}
\label{P3vsA9}
\end{figure}

A similar conclusion can be reached in the case of CP-violating observables. For illustration, we consider  the case of CP-violating observables related to the angular coefficient $J_9$. The observable $A_9$ is not a clean observable, while $P_3^\cp$ is the corresponding clean observable in the large recoil region. At low recoil, the clean observable $H_T^{\sss\rm (5) CP}$ is also considered.
In Fig.~\ref{P3vsA9} we show binned predictions for $\av{P_3^\cp}$, $\av{H_T^{\sss\rm (5) CP}}$ and $\av{A_9}$ in the SM and in three NP scenarios: two scenarios with complex left-handed currents given by $(\delta \C7,\delta\C9,\delta\C{10})=(0.1+0.5i,-1.4,1-1.5i)$ and $(\delta \C7,\delta\C9,\delta\C{10})=(1.5+0.3i,-8+2i,8-2i)$, and a scenario with a complex contribution to $\Cp{10}$:  $\delta \Cp{10}=-1.5+2i$. These scenarios are consistent with all current constraints \cite{Altmannshofer:2011gn}.
The SM prediction is very close to zero for all these observables. But a departure from the SM point has a dramatic effect in the hadronic uncertainties in the prediction of $A_9$. On the other hand, the clean observables $P_3^\cp$ and $H_T^{\sss\rm (5) CP}$, designed for low and high-$q^2$ regions respectively, are much more robust. 

These examples  show how  $P_3^\cp$ (at large recoil) and $H_T^{\sss\rm (5) CP}$ (at low recoil) present unique opportunities to discover or constrain New Physics, and should be given priority over the non-clean observable $A_9$. More generally, they illustrate the interest of choosing clean observables to distinguish between the SM case and NP scenarios from a binned angular analysis of the $B\to K^*\ell \ell$ decay.

\section{Impact of the S-wave pollution}
\label{sec:Swave}

Another source of uncertainties come for the S-wave contribution on the angular distribution $B \to K^+\pi^- l^+ l^-$, which will correspond to a pollution of the angular observables extracted under the assumption that the process is only mediated through P-wave $K^{*0}$ decaying into $K^+\pi^-$.

In Ref.~\cite{1207.4004} the S-wave pollution to the decay $B \to K^*(\to K\pi) l^+l^-$ coming from the companion decay $B \to K_0^* (\to K\pi) l^+l^-$ was computed. The model used there assumed that both $P$ and $S$ waves were correctly described by $q^2$-dependent  $B\to K^*$ or $B\to K^*_0$ form factors, multiplied by
a Breit-Wigner function depending on the $K\pi$ invariant mass (possibly distorted by non-resonant effects such as the elusive $K_0^*(800)$ or  $\kappa$ resonance~\cite{DescotesGenon:2006uk})\footnote{This model has the advantage of simplicity, as it factorises the dependence of the amplitudes on the dilepton and the dihadron masses into two different pieces. However, it should be understood as a rough description of off-shell effects, as it hinges on the assumptions that the $B\to K^*$ or $B\to K^*_0$  form factors are not significantly altered once the light strange meson is not on shell any more, and that the $K^*_{(0)} K \pi$ coupling is essentially independent of the dihadron mass $m^2_{K\pi}$ from threshold up to around 2 GeV.}.
It was then claimed that transverse asymmetries, obtained from uni-angular distributions, suffer unavoidably the S-wave contamination. Soon after and in the same framework, it was shown in Ref. \cite{1209.1525} that using folded distributions instead of uni-angular distributions it should be possible to extract those asymmetries free from this pollution. Indeed, due to the distinct angular dependence of the S-wave terms one can disentangle the interesting signal of the P-wave from the S-wave polluting terms. However an additional  problem arises due to the normalization used for the distribution, that we will discuss in the following.

The angular distribution that describes the four-body decay $B\to K^*(\to K\pi) l^+l^-$ including  the S-wave pollution from the companion decay $B\to K_0^*(\to K\pi) l^+l^-$ is \cite{1207.4004,1111.1513}
 \eqa{\label{distWS}
\frac{d^4\Gamma}{dq^2\,d\!\cos\theta_K\,d\!\cos\theta_l\,d\phi}&=&\frac9{32\pi} \bigg[
J_{1s} \sin^2\theta_K + J_{1c} \cos^2\theta_K + (J_{2s} \sin^2\theta_K + J_{2c} \cos^2\theta_K) \cos 2\theta_l\nn\\[1.5mm]
&&\hspace{-2.7cm}+ J_3 \sin^2\theta_K \sin^2\theta_l \cos 2\phi + J_4 \sin 2\theta_K \sin 2\theta_l \cos\phi  + J_5 \sin 2\theta_K \sin\theta_l \cos\phi \nn\\[1.5mm]
&&\hspace{-2.7cm}+ (J_{6s} \sin^2\theta_K +  {J_{6c} \cos^2\theta_K})  \cos\theta_l    
+ J_7 \sin 2\theta_K \sin\theta_l \sin\phi  + J_8 \sin 2\theta_K \sin 2\theta_l \sin\phi \nn\\[1.5mm]
&&\hspace{-2.7cm}+ J_9 \sin^2\theta_K \sin^2\theta_l \sin 2\phi \bigg]\,X+ W_S
}
where new angular coefficients arise (including a Breit-Wigner function in their definition)
\eqa{\label{wsfunc}
W_S &=&\frac{1}{4\pi} \left[ {\tilde J}_{1a}^c+{\tilde J}_{1b}^c \cos \theta_K+ ({\tilde J}_{2a}^c+{\tilde J}_{2b}^c \cos\theta_K ) \cos 2\theta_\ell + {\tilde J}_{4} \sin \theta_K \sin 2 \theta_\ell \cos \phi \right.\nn\\ &&\left.+
{\tilde J}_{5} \sin \theta_K \sin \theta_\ell \cos \phi + {\tilde J}_{7} \sin \theta_K \sin \theta_\ell \sin \phi+
{\tilde J}_{8} \sin \theta_K \sin 2\theta_\ell \sin \phi\right]
}
as well as a factor included to take into account the width of the resonance:
\eqa{X=\int dm_{K\pi}^2 |BW_{K^*}(m_{K\pi}^2)|^2} 

One can  consider for the normalization of the angular distribution  not  the P-wave component alone ($\Gamma^\prime_{K^*}$) but  the sum of S and P wave amplitudes (including both the $K^* $ and $K_0^*$ components) defined by
\eq{\Gamma^\prime_{{\it full}}=\Gamma_{K^*}^\prime+\Gamma_{S}^\prime}
where we  denote $\Gamma'_x=d\Gamma_x/dq^2$ ($x=K^*,S$) with $\Gamma^\prime_S$ the distribution of the $K_0^*$.
Their expression in terms of the angular coefficients are (see Refs. \cite{1207.4004,1209.1525} for detailed definitions of the new coefficients ${\tilde J_i}$) 
\eq{\Gamma_{K^*}^\prime=\frac{1}{4} (3 J_{1c} + 6 J_{1s} - J_{2c}Ê-2 J_{2s} ) X, \quad \Gamma_S^\prime=2 {\tilde J}_{1a}^c - \frac{2}{3} {\tilde J}_{2a}^c}
and  the  longitudinal  polarization fraction associated to the $\Gamma_S^\prime$ distribution is
\eq{F_S=\frac{\Gamma'_{S}}{\Gamma'_{{\it full}}} \quad \quad {\rm and} \qquad \qquad  1-F_S=\frac{\Gamma'_{K^*}}{\Gamma'_{{\it full}}}
}
As pointed out in Ref. \cite{1209.1525} (see Eq. (23)),
the use of the $\Gamma^\prime_{{\it full}}$ normalization for the angular distribution induces  a polluting factor  (called $C$ in Ref.~\cite{1209.1525} or equivalently  
$1- F_S$ here) that multiplies the P-wave component distribution. For simplicity, in order to obtain the bounds on the polluting terms entering $W_S$ we will work with the distribution in the massless lepton limit  (the distribution in the massive case is discussed in Ref.~\cite{1209.1525})
 \eqa{\label{distfull}
\frac{1}{\Gamma'_{{\it full}}}\frac{d^4\Gamma}{dq^2\,d\!\cos\theta_K\,d\!\cos\theta_l\,d\phi}=&&
\nn\\[1.5mm]
&&\hspace{-5.2cm}\frac9{32\pi} \bigg[
\frac{3}{4} F_T \sin^2\theta_K + F_L \cos^2\theta_K  
+ (\frac{1}{4} F_T \sin^2\theta_K - F_L \cos^2\theta_K) \cos 2\theta_l\nn\\[1.5mm]
&&\hspace{-5.2cm}+ \frac{1}{2} P_1 F_T \sin^2\theta_K \sin^2\theta_l \cos 2\phi + \sqrt{F_T F_L} \left(\frac{1}{2} P_4'  \sin 2\theta_K \sin 2\theta_l \cos\phi  + P_5'  \sin 2\theta_K \sin\theta_l \cos\phi  \right)\nn\\[1.5mm]
&&\hspace{-5.2cm}+ 2 P_2 F_T \sin^2\theta_K   \cos\theta_l    
- \sqrt{F_T F_L} \left( P_6'  \sin 2\theta_K \sin\theta_l \sin\phi  -\frac{1}{2} Q'  \sin 2\theta_K \sin 2\theta_l \sin\phi \right) \nn\\[1.5mm]
&&\hspace{-2.7cm}- P_3 F_T \sin^2\theta_K \sin^2\theta_l \sin 2\phi \bigg]\,(1-F_S)+ \frac{1}{\Gamma'_{{\it full}}}W_S
}
The coefficients of the polluting term can be parametrized as
\eqa{ \label{polut}\frac{W_S}{\Gamma'_{{\it full}}}&=& \frac{3}{16 \pi} \left[ F_S  \sin^2\theta_\ell + A_S \sin^2\theta_\ell \cos\theta_K  + {A_S^4} \sin \theta_K \sin 2 \theta_\ell \cos \phi \right.\nn\\ &&\left.+
{A_S^5} \sin \theta_K \sin \theta_\ell \cos \phi + {A_S^7} \sin \theta_K \sin \theta_\ell \sin \phi+
{A_S^8} \sin \theta_K \sin 2\theta_\ell \sin \phi\right]
}
where we have used the equalities in the massless limit ${\tilde J_{1a}^c}=-{\tilde J_{2a}^c}$ and ${\tilde J_{1b}^c}=-{\tilde J_{2b}^c}$.

We will now estimate the size of the S-wave polluting terms ($\tilde J_i$) normalized to $\Gamma'_{{\it full}}$. 
Identifying the coefficients in Eq.(\ref{polut}) with Eq.(\ref{wsfunc}) we find:
\eqa{
A_S=\frac{8}{3}\frac{\tilde J_{1b}^c}{\Gamma'_{{\it full}}} \quad {\rm and} \quad 
A_S^i=\frac{4}{3} \frac{\tilde J_{i}}{\Gamma'_{{\it full}}}
}
with $i=4,5,7,8$. From the explicit expressions of the ${\tilde J_i}$ (see Ref.~\cite{1209.1525} for definitions) one finds for $F_S$\footnote{The  amplitudes used here are proportional to those introduced in Ref.~\cite{1207.4004}:  ${\cal M}_{0,\perp,\|}^{L,R}=-i \sqrt{\frac{3}{8}} A_{0,\perp,\|}^{L,R}$ and  ${\cal M'}_{0}^{L,R}=-i \sqrt{\frac{3}{8}} A_{0}'^{L,R}$. Notice that here $F_L=(|A_{0}^{L}|^2+|A_{0}^{R}|^2) X/\Gamma'_{K^*}$,  where $X$ cancels between numerator and denominator.
}
\eq{F_S= \frac{8}{3} \frac{\tilde J_{1a}^c}{\Gamma'_{{\it full}}}=\frac{|A_0'^L|^2+|A_0'^R|^2}{\Gamma'_{{\it full}}} Y\qquad 
Y=\int dm_{K\pi}^2 |BW_{K_0^*}(m_{K\pi}^2)|^2} 
with $Y$  a factor included to take into account the scalar component including the $K_0^*$ resonance. 
The corresponding lineshape is denoted $BW_{K_0^*}$, even though it is likely not to be a simple Breit-Wigner shape, due to the possibility of a non-trivial scalar continuum. The contribution from the $S$-wave is expected to be small compared to the $P$-wave one.

The  other terms in Eq.(\ref{polut}) comes from the $S$- and $P$-wave interference and are
\eqa{\frac{\tilde J_{1b}^c}{\Gamma'_{{\it full}}}&=& \frac{3}{4}\sqrt{3} \frac{1}{\Gamma'_{{\it full}}} \int  {\rm Re} \left[(
A_0'^L A_0^{L*} + A_0'^R A_0^{R*})
 BW_{K_0^*}(m_{K\pi}^2) BW_{K^*}^\dagger (m_{K\pi}^2)\right]dm_{K\pi}^2\nn \\
\frac{\tilde J_{4}}{\Gamma'_{{\it full}}}&=& \frac{3}{4}\sqrt{\frac{3}{2}} \frac{1}{\Gamma'_{{\it full}}} \int  {\rm Re} \left[(
A_0'^L A_\|^{L*} + A_0'^R A_\|^{R*})
 BW_{K_0^*}(m_{K\pi}^2) BW_{K^*}^\dagger (m_{K\pi}^2)\right]dm_{K\pi}^2 \nn \\
\frac{\tilde J_{5}}{\Gamma'_{{\it full}}}&=&
\frac{3}{2}\sqrt{\frac{3}{2}} \frac{1}{\Gamma'_{{\it full}}} \int  {\rm Re} \left[(
A_0'^L A_\perp^{L*} - A_0'^R A_\perp^{R*})
 BW_{K_0^*}(m_{K\pi}^2) BW_{K^*}^\dagger (m_{K\pi}^2)\right]dm_{K\pi}^2
 \nn \\
\frac{\tilde J_{7}}{\Gamma'_{{\it full}}}&=& \frac{3}{2}\sqrt{\frac{3}{2}} \frac{1}{\Gamma'_{{\it full}}} \int  {\rm Im} \left[(
A_0'^L A_\|^{L*} - A_0'^R A_\|^{R*})
 BW_{K_0^*}(m_{K\pi}^2) BW_{K^*}^\dagger (m_{K\pi}^2)\right]dm_{K\pi}^2\nn \\
\frac{\tilde J_{8}}{\Gamma'_{{\it full}}}&=&\frac{3}{4}\sqrt{\frac{3}{2}} \frac{1}{\Gamma'_{{\it full}}} \int  {\rm Im} \left[(
A_0'^L A_\perp^{L*} + A_0'^R A_\perp^{R*})
 BW_{K_0^*}(m_{K\pi}^2) BW_{K^*}^\dagger (m_{K\pi}^2)\right]dm_{K\pi}^2}
A bound on these ratios is obtained once we define the $S-P$ interference integral
\eq{
Z=\int  \left| 
 BW_{K_0^*}(m_{K\pi}^2) BW_{K^*}^\dagger (m_{K\pi}^2)\right| dm_{K\pi}^2
}
 and use the bound from the Cauchy-Schwartz inequality
\eqa{
&&\left|\int ({\rm Re}, {\rm Im}) \left[(A_0'^L A_j^{L*} \pm A_0'^R A_j^{R*})   BW_{K_0^*}(m_{K\pi}^2) BW_{K^*}^\dagger (m_{K\pi}^2)\right]
dm_{K\pi}^2 \right|\nonumber\\
&&\qquad\qquad\qquad \leq Z \times \sqrt{[|A_0'^L|^2+|A_0'^R|^2][|A_j^L|^2+|A_j^R|^2]}
}

\begin{table}
\ra{1.5}
\small
\begin{center}
\begin{tabular}{@{}ccccc@{}}
\toprule[1.1pt]
Coefficient  & \mpa{2cm}{Large recoil\\ $\infty$ Range} & \mpa{3cm}{Low recoil\\$\infty$ Range} & \mpa{3cm}{Large Recoil \\Finite Range} & \mpa{3cm}{Low Recoil\\Finite Range}  \\ [2mm]
\midrule
$|A_S|$ & $0.33$ & $0.25$ & $0.67$ & $0.49$ \\[1mm]
$|A_S^4|$ & $0.05$ & $0.10$ & $0.11$ & $0.19$  \\[1 mm]
$|A_S^5|$ & $0.11$ & $0.11$ & $0.22$ & $0.23$ \\[1mm]
$|A_S^7|$ & $0.11$ & $0.19$ & $0.22$  & $0.38$  \\[1mm]
$|A_S^8|$ & $0.05$& $0.06$ & $0.11$ & $0.11$  \\[1 mm]
\bottomrule[1.1pt]
\end{tabular}
\caption{
Illustrative values of the size of the bounds for the choices of $F_S, F_L, P_1$ and $F$ described in the text.
}
\label{swave}
\end{center}
\end{table}

 The definitions of $F_S$ and $F_L$  yield the following bound:
 \eq{|A_S| \leq 2\sqrt{3} \sqrt{ F_S (1-F_S) F_L} \,\frac{Z}{\sqrt{XY}}}
where the factor $(1-F_S)$ arises due to the fact that $F_L$ is defined with respect to $\Gamma'_{K^*}$ rather than $\Gamma'_{{\it full}}$.
Using the definition of $P_1$ in terms of $|A_{\perp,||}^{L,R}|^2$ one finds for the other terms in Eq.(\ref{polut}) the following bounds
\eqa{\label{boundss}
|A_S^4| &\leq& \sqrt{\frac{3}{2}} \sqrt{ F_S (1-F_S) (1-F_L) \left(\frac{1-P_1}{2} \right)} \,\frac{Z}{\sqrt{XY}} \nn \\
|A_S^5| &\leq& 2 \sqrt{\frac{3}{2}} \sqrt{ F_S (1-F_S) (1-F_L) \left(\frac{1+P_1}{2} \right)} \,\frac{Z}{\sqrt{XY}} \nn \\
|A_S^7| &\leq& 2 \sqrt{\frac{3}{2}} \sqrt{ F_S (1-F_S) (1-F_L) \left(\frac{1-P_1}{2} \right)} \,\frac{Z}{\sqrt{XY}} \nn \\
|A_S^8| &\leq& \sqrt{\frac{3}{2}} \sqrt{ F_S (1-F_S) (1-F_L) \left(\frac{1+P_1}{2} \right)}  \,\frac{Z}{\sqrt{XY}}
}

 In order to assign a numerical value to these bounds we have to evaluate the integrals $X$, $Y$ and $Z$ that enter the factor 
 $F=\frac{Z}{\sqrt{XY}}$. Here we can consider two options: a first option that we call ``infinite range" where we take the integrals in the whole $m_{K\pi}$ range. In this case, we get $X=Y=1$, $0.37 \leq Z \leq 0.45$ and $0.37 \leq F^{\infty} \leq 0.45$. And a second option where we consider a ``finite range" for the integrals around $m_{K^*} \pm 0.1$ GeV (corresponding to the constraints put on the invariant dihadron mass for the experimental analysis),  we use the parametrization given in \cite{becirevic}, and we vary the parameters of the $K_0^*$ Breit-Wigner ($0\leq g_{\kappa} \leq 0.2$ and ${\rm arg} (g_{\kappa})$ inside $\pi/2,\pi$ range)  to obtain
$0.113 \leq Z \leq 0.176$.   Similarly for the other integrals one gets in this case $0.019 \leq Y \leq 0.045$ and $X=0.848$. And the corresponding factor $F^{(m_{K^*} \pm 0.1)}$ is now inside the range $0.89 \leq F^{(m_{K^*} \pm 0.1)} \leq 0.90$.

We can provide  two illustrative examples for the large- and low-recoil regions. We take in  the large-recoil region the following values  
  $F_S \sim 7\%$ \cite{expFS} (assuming that the scalar contribution is similar to that in the decay  $B^0 \to J/\psi K^+ \pi^-$), $F_L\sim 0.7$
  and $P_1 \sim 0$, and for the low recoil region, the same value of $F_S$, $F_L \sim 0.38$ and $P_1 \sim -0.48$, where the  values for $F_L$ and $P_1$ are the average of the central values of the SM predictions in the last two bins. For the $F$ factor  we take the maximal possible values. The corresponding bounds are gathered in Table~\ref{swave}.
These estimates  can be more precise once we have a direct measurement of $F_S$.

\section{Comparison with other works}
\label{comparison}

In addition to  general global fits to radiative $B$-decays~\cite{1207.2753,Altmannshofer:2011gn,bobeth4} (see also Refs.~\cite{bobeth3,1104.3342,1206.1502,1207.0688,1209.0262,Muru}),
there has been a growing literature aiming at determining the best set of observables with a reduced sensitivity to hadronic inputs, and the ability of these observables to find New Physics.

Compared to Ref.~\cite{buras}, we obtain a fair agreement with the predictions of the $S_i$ and $A_i$ within errors, even if we take a different approach for the treatment of form factors and different input values for them. On top of this, there is also some difference on the Wilson coefficient values for dileptonic operators where we included extra electromagnetic corrections. In some cases like $S_{4,5}$ or the tiny asymmetry $A_{6s}$ the agreement with the central value is perfect. But as
these $S_i,A_i$ observables are not protected from hadronic uncertainties in general, there are cases
where the SM value is tiny (e.g., $10^{-2}$ for $S_3$) and basically driven by NLO contributions, where 
the result is more sensitive to the use of full form factors or of specific relationships derived from an effective theory approach,
and to the input values chosen.

The high-$q^2$ region has been the focus of 
a series of papers~\cite{bobeth,bobeth2,tensors}, relying
on relationships between form factors from the heavy-quark expansion derived in Ref.~\cite{grinstein+pirjol} (as discussed
in Section~\ref{sec:ffs}). In Refs.~\cite{bobeth,tensors}, a set of 5 clean (CP-averaged) observables called $H_T^{(i)}$ was introduced, which is equivalent to the set introduced here for clean observables at high-$q^2$. The corresponding CP-violating observables were also discussed, as well as the case of $B\to K\ell\ell$ transitions (with only 3 angular observables avalaible), and their potential for New Physics.  We confirm that CP-averaged observables (e.g., $H_T^{(1)}=P_4$) have a small sensitivity to hadronic uncertainties, but we stress that this needs not be the case for CP-violating observables in the presence of New Physics. We agree with the numerical results for the  SM predictions of $B\to K^*\ell\ell$, but we quote larger uncertainties in particular for the branching ratio. This is due most probably to a different choice of form factors (Ref.~\cite{0412079} (BZ) versus KMPW \cite{1006.4945}). As discussed in Ref.~\cite{bobeth} and discussed again in Section~\ref{sec:ffs}, there are some difficulties to accomodate the extrapolation of BZ form factors at high $q^2$ with the HQET relationships exploited to compute the angular coefficients. We have discussed the comparison with available lattice data and our choice of using KMPW together with HQET relationships until accurate lattice data are available for the low-recoil region.

The low-$q^2$ has been discussed in Ref.~\cite{camalich} recently, with an interpretation of angular observables in terms of helicity form factors~\cite{1004.3249}. The pattern of suppression of some form factors with respect to others, indicated by QCD factorisation/SCET analyses, was shown to hold even after the inclusion of radiative and power corrections as well as non-factorisable effects and to yield a strong suppression of the angular coefficients $J_3$ and $J_9$.
For the phenomenological analyses, several inputs for the form factors were considered, not only (rescaled) BZ and KMPW, but also QCD sum rules and truncated Dyson-Schwinger equations. The comparison of the various models confirmed $\Lambda/m_b$ corrections of a few percent to the QCD factorisation/SCET relationships, in the line of the estimates used in this paper. But if some of the angular coefficients are dominated by form factor uncertainties, sizeable contributions may also arise from charm-loop contributions. Compared to our own analysis, we have allowed for a larger range of uncertainties concerning form factors, but no charm-loop contributions. This explains the agreement concerning the central values for the low-$q^2$ observables, but significantly larger errors in Ref.~\cite{camalich} (especially for $P_2,P_4',P_5'$ and $P_6'$). The very low-$q^2$ region (below 1 GeV$^2$) was also analysed, using a phenomenological model to account for light resonances. The effect of the latter was shown to be very small once binning effects were including, and has not been considered in our own analysis.

The issue of long-distance charm loop effects was also considered in Ref.~\cite{1006.4945}. This paper was aimed at calculating one particular effect, the soft-gluon emission
from the charm loop, which is only one of the several nonlocal hadronic  effects for $B\to K^* \ell \ell$
caused by four-quark, quark-penguin and $O_8$ operators. The modification induced to $C_9$ was encoded
in a shift $\delta C_9$ where only the factorizable charm loop and nonfactorizable
soft gluon are taken into account, up to 20\% in the low-$q^2$ region (below 4 GeV$^2$).  It is interesting to notice that this particular effect is difficult to assess and can be large, casting some doubts on the possibility to exploit the bins between $J/\psi$ and $\psi(2S)$ for comparison with experiment.
We have not included the results of Refs.~\cite{1006.4945,camalich} waiting for a more comprehensive theory of charm-loop effects before including them in our analysis.


\section{Conclusions}
\label{sec:con}

Measurements on the angular distribution of the decay $B\to K^*(\to K\pi) \ell^+\ell^-$ are being performed intensively at flavour facilities. In the near future, these measurements will either reveal 
hints of NP in flavour physics, or set the strongest constraints so far on radiative and semileptonic $|\Delta B|=|\Delta S|=1$ operators. However, in order for these measurements to be effective, the focus has to be put on theoretically ``clean" observables.

In this paper we have studied and collected all relevant clean angular observables in both kinematic regions of interest (large and low recoils), giving a unified and comprehensive description of both regions, and a thorough reassessment of the form factor input. We have also considered a full set of CP-violating observables, $P_i^{\cp}$. We reviewed the various observables proposed in Table~\ref{TableObs}, and we can identify an optimal basis containing a maximal number of clean observables that constitutes a compromise between a clean theoretical prediction  and a simple experimental extraction. All SM predictions can be found in Tables~\ref{tabSM1}-\ref{tabSM2} in Sec.~\ref{sec:sm}
 and Tables~\ref{tabSM3}-\ref{tabSM7} in App.~\ref{appA}.

The relevance in focusing on clean observables can be seen more clearly by studying the NP sensitivity of different observables probing in principle the same angular coefficient, but with a normalisation enhancing or suppressing the sensitivity to form factors. By considering different NP scenarios (compatible with current bounds) we find that $\av{P_1}$ is much more sensitive than $\av{S_3}$ to NP effects due to it reduced hadronic uncertainties. The same is found for  the CP asymmetries $\av{P_3^\cp}$ (at large recoil) and $\av{H_T^{(5)}}$ (at low recoil), which are much finer probes of NP than $\av{A_9}$.

An important systematic effect that has to be fully understood is the S-wave component due to $B\to K^*_0(\to\pi K)\ell\ell$ events, which is not negligible even at $m_{K\pi}\sim m_{K^*}$. Disentangling the S- and P-wave contributions requires a more complete angular analysis, which if not performed leads to intrinsic systematics in the extraction of the $B\to K^*(\to K\pi) \ell^+\ell^-$ angular observables. We have 
determined bounds on the size of the interference terms in the angular distribution, which constitute an upper bound on these systematic uncertainties.

By providing an appropriate basis with a limited sensitivity to hadronic contributions and thus a better potential to identify New Physics contributions, and by giving SM predictions for these observables, we hope that the results and predictions presented in this paper will help the discussion of the next series of results on $B\to K^*\ell^+\ell^-$.


\subsubsection*{Acknowledgments}

It is a pleasure to thank Nicola Serra and Thorsten Feldmann for discussions and comments.
J.M. enjoys financial support from FPA2011-25948, SGR2009-00894. J.V. is supported in part by ICREA-Academia funds and FPA2011-25948.


\appendix
\section{Definitions of additional observables}\label{app:otherobs}

The integrated unprimed observables $\av{P_{4,5,6}^{(\cp)}}$ are given by
\begin{align}
\av{P_4}_{\rm bin} &= \frac{\sqrt{2}}{{\cal N}_\bin^-} \int_{{\rm bin}} dq^2 [J_4+\bar J_4]\ ,
& \av{{P_4}^\cp }_{\rm bin} &= \frac{\sqrt{2}}{{\cal N}_\bin^-} \int_{{\rm bin}} dq^2 [J_4-\bar J_4]\ ,\\
\av{P_5}_{\rm bin} &= \frac1{\sqrt{2}{\cal N}_\bin^+} \int_{{\rm bin}} dq^2 [J_5+\bar J_5]\ ,
& \av{{P_5}^\cp }_{\rm bin} &= \frac1{\sqrt{2}{\cal N}_\bin^+} \int_{{\rm bin}} dq^2 [J_5-\bar J_5]\ ,\\
\av{P_6}_{\rm bin} &= \frac1{\sqrt{2}{\cal N}_\bin^-} \int_{{\rm bin}} dq^2 [J_7+\bar J_7]\ ,
& \av{{P_6}^\cp }_{\rm bin} &= \frac1{\sqrt{2}{\cal N}_\bin^-} \int_{{\rm bin}} dq^2 [J_7-\bar J_7]\ ,\\
\av{P_8}_{\rm bin} &= \frac{-\sqrt{2}}{{\cal N}_\bin^+} \int_{{\rm bin}} dq^2 [J_8+\bar J_8]\ ,
& \av{{P_8}^\cp }_{\rm bin} &= \frac{-\sqrt{2}}{{\cal N}_\bin^+} \int_{{\rm bin}} dq^2 [J_8-\bar J_8]\ ,
\end{align}
where ${\cal N}_\bin^+$ and ${\cal N}_\bin^-$ are defined as
\eqa{
{\cal N}_\bin^+ = {\textstyle \sqrt{-(\int_\bin dq^2 [2(J_{2s}+\bar J_{2s}) + (J_{3}+\bar J_{3})])(\int_{{\rm bin}} dq^2 [J_{2c}+\bar J_{2c}]) }}\ ,\\[2mm]
{\cal N}_\bin^- = {\textstyle \sqrt{-(\int_\bin dq^2 [2(J_{2s}+\bar J_{2s}) - (J_{3}+\bar J_{3})])(\int_{{\rm bin}} dq^2 [J_{2c}+\bar J_{2c}]) }}\ .}

The CP-violating observables $\av{H_T^{(3,5)\cp}}$ are given by
\eqa{
\intbin{H_T^{(3)\cp}} &=& \frac{\int_\bin dq^2[J_{6s} - \bar J_{6s}]}{2 \sqrt{4 (\int_\bin dq^2 [J_{2s} + \bar J_{2s}])^2 - (\int_\bin dq^2 [J_3 + \bar J_3])^2}}\ ,\\
\intbin{H_T^{(5)\cp}} &=& \frac{-\int_\bin dq^2 [J_{9} - \bar J_{9}]}{\sqrt{4 (\int_\bin dq^2 [J_{2s} + \bar J_{2s}])^2 - (\int_\bin dq^2 [J_3 + \bar J_3])^2}}\ .
}

We also collect here the definitions of other observables \cite{buras}: 
\eq{
\intbin{S_i} = \frac{\int_\bin dq^2[J_{i} + \bar J_{i}]}{\intbin{d\Gamma/dq^2}+\intbin{d\bar\Gamma/dq^2}}\ ,\quad 
\intbin{A_i} = \frac{\int_\bin dq^2[J_{i} - \bar J_{i}]}{\intbin{d\Gamma/dq^2}+\intbin{d\bar\Gamma/dq^2}}\ .
}
We refer to Refs.~\cite{matias1,matias2,primary} for the definitions of $A_{T}^{(3,4,5)}$, taking into account that the substitution $J_i\to \int_{bin} dq^2 [J_i+\bar J_i]$ should be understood for all $J_i$.

\section{Compendium of SM results}
\label{appA}

\begin{figure}
\begin{center}
\includegraphics[height=5cm,width=16cm]{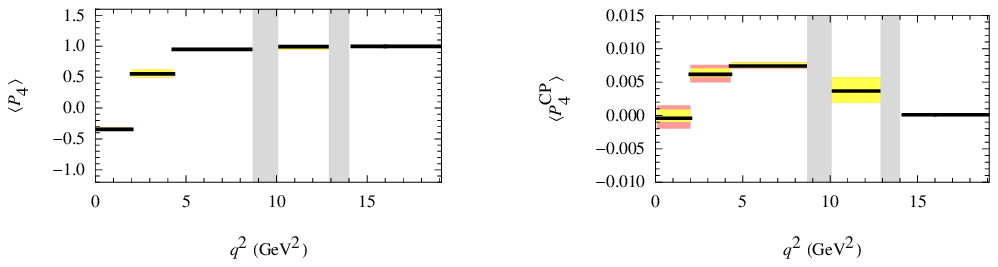}
\includegraphics[height=5cm,width=16cm]{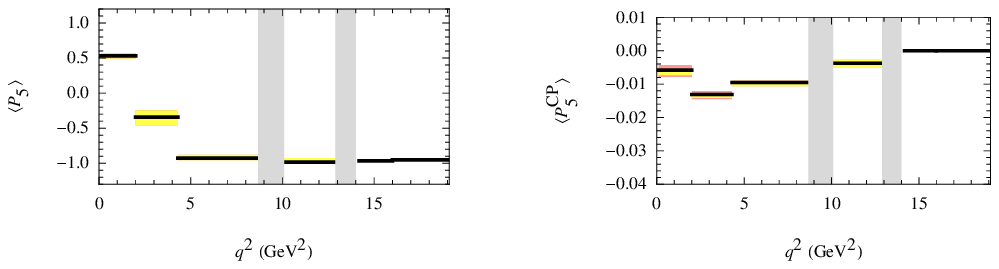}
\includegraphics[height=5cm,width=16cm]{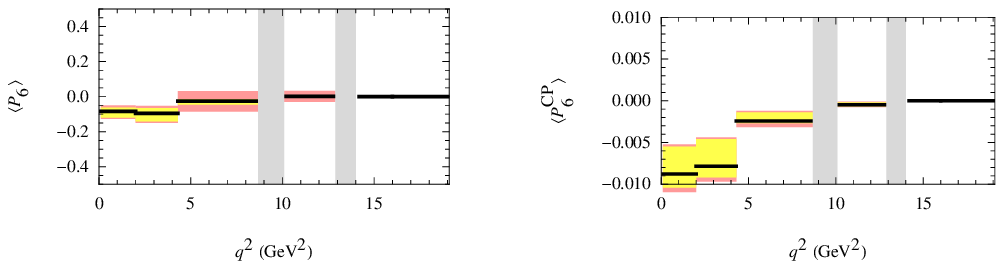}
\end{center}
\vspace{-0.4cm}
\caption{Binned Standard Model predictions for the observables $\av{P_{4,5,6}^{\sss\rm (CP)}}$,  with the same conventions as in Fig.~\ref{SMplotsPs1}.}
\label{SMplotsPs3}
\end{figure}

\begin{figure}
\begin{center}
\includegraphics[height=5cm,width=16cm]{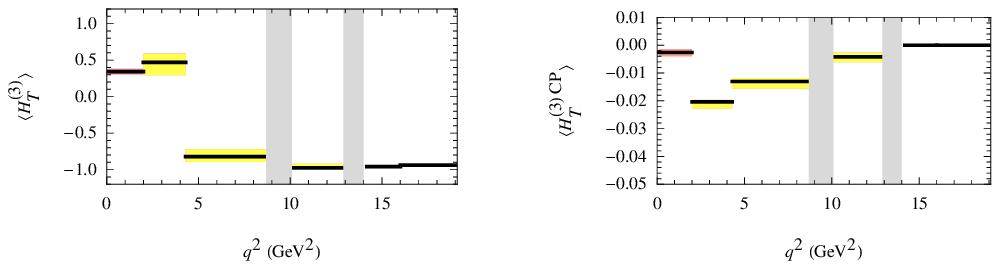}
\includegraphics[height=5cm,width=16cm]{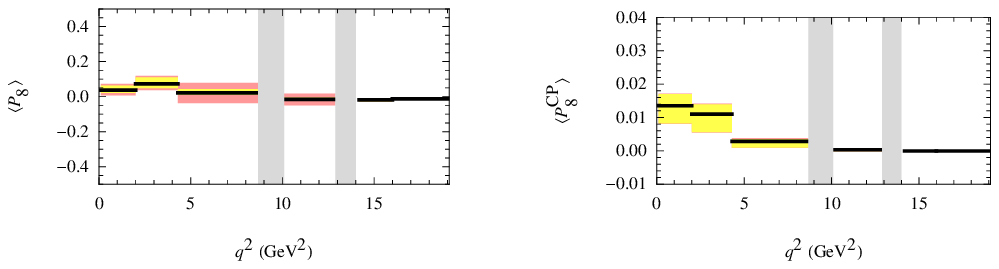}
\includegraphics[height=5cm,width=16cm]{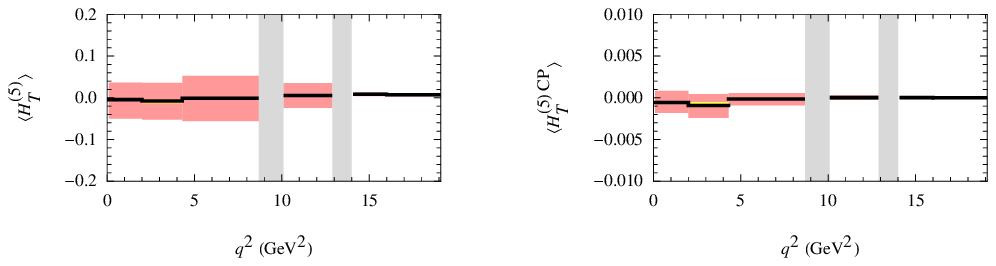}
\end{center}
\vspace{-0.4cm}
\caption{Binned Standard Model predictions for the observables $\av{H_T^{(3)\sss\rm (CP)}}$, $\av{P_{8}^{\sss\rm (CP)}}$, $\av{H_T^{(5)\sss\rm (CP)}}$,  with the same conventions as in Fig.~\ref{SMplotsPs1}.}
\label{SMplotsPs4}
\end{figure}

\begin{figure}\centering
\includegraphics[height=5cm,width=16cm]{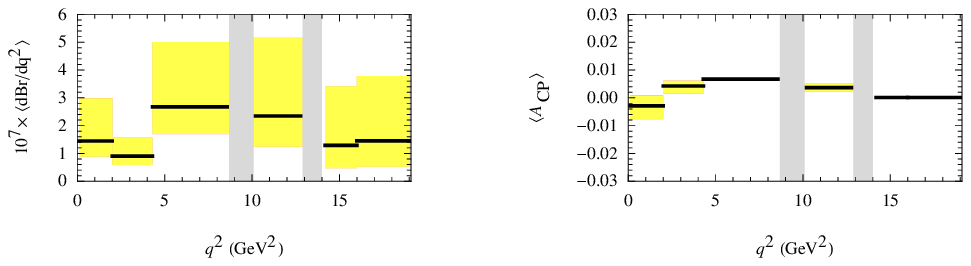}
\includegraphics[height=5cm,width=16cm]{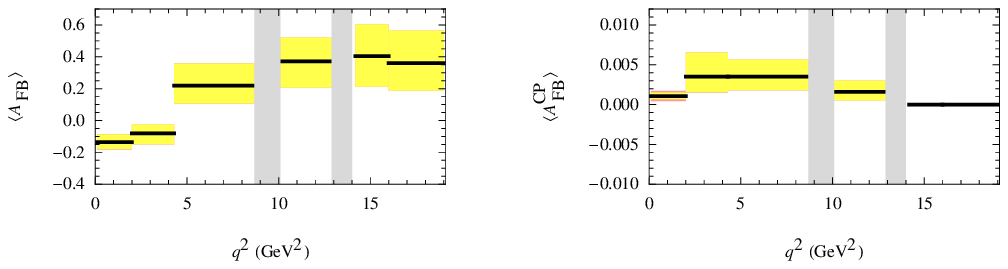}
\includegraphics[height=5cm,width=16cm]{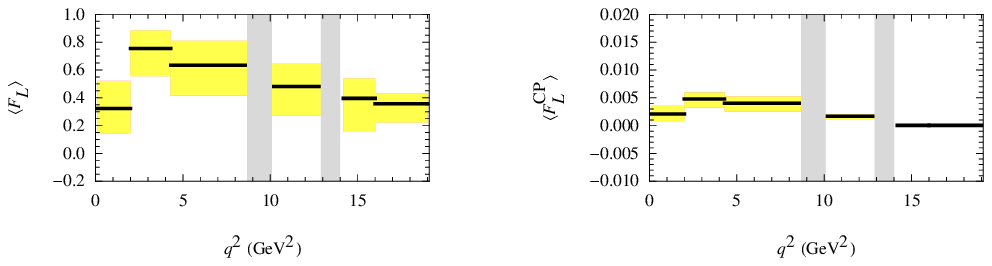}
\caption{Binned Standard Model predictions for the observables $\av{d\Gamma/dq^2}$, $\av{A_{\rm CP}}$, $\av{A_{\rm FB}^{\sss\rm (CP)}}$ and $\av{F_L^{\sss\rm (CP)}}$,  with the same conventions as in Fig.~\ref{SMplotsPs1}.}
\label{SMplotsAFBFL}
\end{figure}

In this Appendix we collect the SM predictions for the observables discussed in the paper
in addition to Tables~\ref{tabSM1} and \ref{tabSM2},  where the observables in the optimized CP-averaged and CP-violating bases are collected. All these results are also presented graphically in Figures~\ref{SMplotsPs1}-\ref{SMplotsAFBFL}. The binning is chosen to match the current experimental results. Predictions within different bins can be obtained as explained in the text, and are also available upon request.

The second series of errors quoted correspond to $\Lambda/m_b$ corrections, whereas the first series stem from all uncertainties in the inputs, and in particular in form factors. They have been obtained following the method presented in Refs.~\cite{matias1,matias2,primary}, as outlined in Section~\ref{sec:sm}.



\begin{table}
\ra{1.35}
\rb{6mm}
\refstepcounter{table}
\label{tabSM3}
\footnotesize
{\ \ \textsf{\small Table \arabic{table}. Standard Model Predictions for CP-averaged observables.}}
\begin{center}
\rowcolors{1}{}{lgris}
\begin{tabular}{@{}lcrrr@{}}
\toprule[1.1pt]
Bin (GeV$^2$) & & \cen{$\av{P_4}=\av{H_T^{(1)}}$} & \cen{$\av{P_5}=\av{H_T^{(2)}}$} & \cen{$\av{P_6}$}  \\ [1mm]
\hline
%
[\,1\,,\,2\,]  & &
$-0.160_{-0.032 - 0.015}^{+0.040 + 0.014}$ & 
$0.385_{-0.062 - 0.016}^{+0.045 + 0.016}$ & 
$-0.104_{-0.043 - 0.016}^{+0.026 + 0.016}$ \\ [1mm]
[\,0.1\,,\,2\,]  & &
$-0.344_{-0.019 - 0.018}^{+0.026 + 0.017}$ & 
$0.531_{-0.035 - 0.017}^{+0.026 + 0.017}$ & 
$-0.084_{-0.036 - 0.025}^{+0.021 + 0.026}$ \\ [1mm]
[\,2\,,\,4.3\,]  & &
$0.555_{-0.056 - 0.020}^{+0.066 + 0.020}$ & 
$-0.343_{-0.116 - 0.020}^{+0.098 + 0.021}$ & 
$-0.095_{-0.046 - 0.030}^{+0.030 + 0.030}$ \\ [1mm]
[\,4.3\,,\,8.68\,]  & &
$0.949_{-0.015 - 0.006}^{+0.014 + 0.004}$ & 
$-0.927_{-0.030 - 0.005}^{+0.046 + 0.007}$ & 
$-0.025_{-0.020 - 0.056}^{+0.011 + 0.056}$ \\ [1mm]
[\,10.09\,,\,12.89\,]  & &
$0.996_{-0.050 - 0.003}^{+0.007 + 0.001}$ & 
$-0.986_{-0.003 - 0.001}^{+0.058 + 0.003}$ & 
$0.001_{-0.004 - 0.032}^{+0.003 + 0.031}$ \\ [1mm]
[\,14.18\,,\,16\,]  & &
$0.998_{-0.002 - 0.001}^{+0.001 + 0.001}$ & 
$-0.968_{-0.004 - 0.002}^{+0.007 + 0.002}$ & 
$0.000_{-0.000- 0.000}^{+0.000+ 0.000}$ \\ [1mm]
[\,16\,,\,19\,]  & &
$0.997_{-0.003 - 0.001}^{+0.003 + 0.001}$ & 
$-0.954_{-0.006 - 0.001}^{+0.013 + 0.002}$ & 
$0.000_{-0.000- 0.000}^{+0.000+ 0.000}$ \\ [1mm]
[\,1\,,\,6\,]  & &
$0.540_{-0.052 - 0.015}^{+0.061 + 0.015}$ & 
$-0.359_{-0.103 - 0.017}^{+0.090 + 0.018}$ & 
$-0.087_{-0.042 - 0.029}^{+0.028 + 0.030}$ \\ [1mm]
%
%
\midrule[1.1pt]
%
%
\rowcolor{white} & & \cen{$\av{P_8}=\av{H_T^{(4)}}$} & \cen{$\av{P'_8}$} &  \\ [1mm]
\hline
%
[\,1\,,\,2\,]  & &
$0.059_{-0.019 - 0.018}^{+0.033 + 0.017}$ & 
$0.059_{-0.019 - 0.018}^{+0.033 + 0.017}$ & \\ [1mm]
[\,0.1\,,\,2\,]  & &
$0.037_{-0.015 - 0.025}^{+0.026 + 0.026}$ & 
$0.037_{-0.015 - 0.025}^{+0.026 + 0.026}$ & \\ [1mm]
[\,2\,,\,4.3\,]  & &
$0.072_{-0.024 - 0.026}^{+0.038 + 0.025}$ & 
$0.070_{-0.023 - 0.025}^{+0.038 + 0.024}$ & \\ [1mm]
[\,4.3\,,\,8.68\,]  & &
$0.021_{-0.011 - 0.058}^{+0.022 + 0.053}$ & 
$0.020_{-0.010 - 0.054}^{+0.021 + 0.050}$ & \\ [1mm]
[\,10.09\,,\,12.89\,]  & &
$-0.016_{-0.005 - 0.033}^{+0.010 + 0.031}$ & 
$-0.015_{-0.005 - 0.030}^{+0.010 + 0.028}$ & \\ [1mm]
[\,14.18\,,\,16\,]  & &
$-0.019_{-0.008 - 0.004}^{+0.006 + 0.004}$ & 
$-0.015_{-0.012 - 0.003}^{+0.009 + 0.003}$ & \\ [1mm]
[\,16\,,\,19\,]  & &
$-0.013_{-0.004 - 0.004}^{+0.004 + 0.003}$ & 
$-0.008_{-0.007 - 0.002}^{+0.005 + 0.002}$ & \\ [1mm]
[\,1\,,\,6\,]  & &
$0.065_{-0.022 - 0.026}^{+0.035 + 0.025}$ & 
$0.063_{-0.022 - 0.025}^{+0.034 + 0.024}$ & \\ [1mm]
%
%
\midrule[1.1pt]
%
%
\rowcolor{white} & & \cen{$\av{H_T^{(3)}}$} & \cen{$\av{H_T^{(5)}}$} &  \\ [1mm]
\hline
%
[\,1\,,\,2\,]  & &
$0.799_{-0.046 - 0.016}^{+0.043 + 0.014}$ & 
$-0.007_{-0.004 - 0.048}^{+0.002 + 0.055}$ & \\ [1mm]
[\,0.1\,,\,2\,]  & &
$0.343_{-0.018 - 0.037}^{+0.017 + 0.037}$ & 
$-0.004_{-0.003 - 0.045}^{+0.001 + 0.041}$ & \\ [1mm]
[\,2\,,\,4.3\,]  & &
$0.469_{-0.170 - 0.031}^{+0.115 + 0.031}$ & 
$-0.008_{-0.005 - 0.045}^{+0.003 + 0.043}$ & \\ [1mm]
[\,4.3\,,\,8.68\,]  & &
$-0.820_{-0.074 - 0.011}^{+0.097 + 0.014}$ & 
$-0.001_{-0.002 - 0.055}^{+0.001 + 0.054}$ & \\ [1mm]
[\,10.09\,,\,12.89\,]  & &
$-0.977_{-0.004 - 0.002}^{+0.057 + 0.004}$ & 
$0.006_{-0.002 - 0.030}^{+0.001 + 0.030}$ & \\ [1mm]
[\,14.18\,,\,16\,]  & &
$-0.959_{-0.000- 0.004}^{+0.007 + 0.004}$ & 
$0.008_{-0.001 - 0.004}^{+0.000+ 0.004}$ & \\ [1mm]
[\,16\,,\,19\,]  & &
$-0.938_{-0.002 - 0.003}^{+0.004 + 0.003}$ & 
$0.007_{-0.001 - 0.004}^{+0.000+ 0.004}$ & \\ [1mm]
[\,1\,,\,6\,]  & &
$0.168_{-0.152 - 0.037}^{+0.114 + 0.039}$ & 
$-0.006_{-0.004 - 0.043}^{+0.002 + 0.040}$ & \\ [1mm]
%
%
\bottomrule[1.1pt]
\end{tabular} 
\end{center} 
\end{table}  


\begin{table}
\ra{1.35}
\rb{6mm}
\refstepcounter{table}
\label{tabSM4}
\footnotesize
{\ \ \textsf{\small Table \arabic{table}. Standard Model Predictions for CP-violating observables.}}
\begin{center}
\rowcolors{1}{}{lgris}
\begin{tabular}{@{}lcrrr@{}}
\toprule[1.1pt]
Bin (GeV$^2$) & & \cen{$10^2 \times \av{P_4^\cp}$} & \cen{$10^2 \times \av{P_5^\cp}$} & \cen{$10^2 \times \av{P_6^\cp}$}  \\ [1mm]
\hline
%
[\,1\,,\,2\,]  & &
$0.144_{-0.041 - 0.154}^{+0.141 + 0.142}$ & 
$-0.888_{-0.146 - 0.134}^{+0.013 + 0.128}$ & 
$-1.011_{-0.212 - 0.137}^{+0.401 + 0.130}$ \\ [1mm]
[\,0.1\,,\,2\,]  & &
$-0.040_{-0.054 - 0.147}^{+0.130 + 0.141}$ & 
$-0.580_{-0.154 - 0.138}^{+0.035 + 0.143}$ & 
$-0.877_{-0.164 - 0.141}^{+0.327 + 0.137}$ \\ [1mm]
[\,2\,,\,4.3\,]  & &
$0.615_{-0.041 - 0.112}^{+0.089 + 0.111}$ & 
$-1.311_{-0.100 - 0.099}^{+0.030 + 0.094}$ & 
$-0.785_{-0.132 - 0.128}^{+0.326 + 0.117}$ \\ [1mm]
[\,4.3\,,\,8.68\,]  & &
$0.740_{-0.022 - 0.030}^{+0.052 + 0.026}$ & 
$-0.953_{-0.106 - 0.022}^{+0.049 + 0.026}$ & 
$-0.242_{-0.026 - 0.069}^{+0.103 + 0.064}$ \\ [1mm]
[\,10.09\,,\,12.89\,]  & &
$0.367_{-0.176 - 0.019}^{+0.205 + 0.017}$ & 
$-0.375_{-0.136 - 0.019}^{+0.096 + 0.019}$ & 
$-0.047_{-0.020 - 0.022}^{+0.029 + 0.022}$ \\ [1mm]
[\,14.18\,,\,16\,]  & &
$0.011_{-0.006 - 0.000}^{+0.005 + 0.000}$ & 
$0.000_{-0.000 - 0.000}^{+0.000 + 0.000}$ & 
$0.000_{-0.000 - 0.000}^{+0.000 + 0.000}$ \\ [1mm]
[\,16\,,\,19\,]  & &
$0.010_{-0.005 - 0.000}^{+0.004 + 0.000}$ & 
$0.000_{-0.000 - 0.000}^{+0.000 + 0.000}$ & 
$0.000_{-0.000 - 0.000}^{+0.000 + 0.000}$ \\ [1mm]
[\,1\,,\,6\,]  & &
$0.581_{-0.035 - 0.096}^{+0.078 + 0.090}$ & 
$-1.173_{-0.094 - 0.075}^{+0.027 + 0.076}$ & 
$-0.673_{-0.106 - 0.113}^{+0.275 + 0.101}$ \\ [1mm]
%
%
\midrule[1.1pt]
%
%
\rowcolor{white} & & \cen{$10^2 \times \av{P_8^\cp}$} & \cen{$10^2 \times \av{{P'_8}^\cp}$} &  \\ [1mm]
\hline
%
[\,1\,,\,2\,]  & &
$1.467_{-0.643 - 0.130}^{+0.425 + 0.131}$ & 
$1.472_{-0.642 - 0.125}^{+0.421 + 0.140}$ & \\ [1mm]
[\,0.1\,,\,2\,]  & &
$1.354_{-0.533 - 0.104}^{+0.343 + 0.112}$ & 
$1.359_{-0.532 - 0.123}^{+0.341 + 0.120}$ & \\ [1mm]
[\,2\,,\,4.3\,]  & &
$1.100_{-0.538 - 0.115}^{+0.291 + 0.120}$ & 
$1.071_{-0.521 - 0.117}^{+0.278 + 0.117}$ & \\ [1mm]
[\,4.3\,,\,8.68\,]  & &
$0.284_{-0.183 - 0.071}^{+0.069 + 0.067}$ & 
$0.267_{-0.172 - 0.069}^{+0.065 + 0.062}$ & \\ [1mm]
[\,10.09\,,\,12.89\,]  & &
$0.034_{-0.042 - 0.023}^{+0.017 + 0.024}$ & 
$0.031_{-0.039 - 0.022}^{+0.017 + 0.022}$ & \\ [1mm]
[\,14.18\,,\,16\,]  & &
$-0.004_{-0.002 - 0.001}^{+0.001 + 0.001}$ & 
$-0.003_{-0.003 - 0.001}^{+0.002 + 0.001}$ & \\ [1mm]
[\,16\,,\,19\,]  & &
$-0.002_{-0.001 - 0.001}^{+0.001 + 0.001}$ & 
$-0.002_{-0.001 - 0.000}^{+0.001 + 0.000}$ & \\ [1mm]
[\,1\,,\,6\,]  & &
$0.959_{-0.459 - 0.100}^{+0.239 + 0.103}$ & 
$0.932_{-0.444 - 0.101}^{+0.228 + 0.102}$ & \\ [1mm]
%
%
\midrule[1.1pt]
%
%
\rowcolor{white} & & \cen{$10^2 \times \av{H_T^{(3)\cp}}$} & \cen{$10^2 \times \av{H_T^{(5)\cp}}$} &  \\ [1mm]
\hline
%
[\,1\,,\,2\,]  & &
$-0.806_{-0.148 - 0.065}^{+0.016 + 0.069}$ & 
$-0.088_{-0.017 - 0.154}^{+0.031 + 0.147}$ & \\ [1mm]
[\,0.1\,,\,2\,]  & &
$-0.266_{-0.068 - 0.123}^{+0.007 + 0.12}$ & 
$-0.056_{-0.007 - 0.125}^{+0.016 + 0.139}$ & \\ [1mm]
[\,2\,,\,4.3\,]  & &
$-2.039_{-0.239 - 0.026}^{+0.066 + 0.034}$ & 
$-0.093_{-0.014 - 0.147}^{+0.040 + 0.131}$ & \\ [1mm]
[\,4.3\,,\,8.68\,]  & &
$-1.308_{-0.257 - 0.016}^{+0.121 + 0.018}$ & 
$-0.015_{-0.003 - 0.075}^{+0.014 + 0.071}$ & \\ [1mm]
[\,10.09\,,\,12.89\,]  & &
$-0.422_{-0.184 - 0.013}^{+0.158 + 0.013}$ & 
$0.002_{-0.001 - 0.027}^{+0.004 + 0.027}$ & \\ [1mm]
[\,14.18\,,\,16\,]  & &
$0.000_{-0.000- 0.000}^{+0.000+ 0.000}$ & 
$0.002_{-0.000- 0.001}^{+0.000+ 0.001}$ & \\ [1mm]
[\,16\,,\,19\,]  & &
$0.000_{-0.000- 0.000}^{+0.000+ 0.000}$ & 
$0.001_{-0.000- 0.001}^{+0.000+ 0.001}$ & \\ [1mm]
[\,1\,,\,6\,]  & &
$-1.658_{-0.193 - 0.014}^{+0.056 + 0.022}$ & 
$-0.071_{-0.009 - 0.113}^{+0.030 + 0.103}$ & \\ [1mm]
%
%
\bottomrule[1.1pt]
\end{tabular} 
\end{center} 
\end{table}  


\begin{table}
\ra{1.35}
\rb{6mm}
\refstepcounter{table}
\label{tabSM5}
\footnotesize
{\ \ \textsf{\small Table \arabic{table}. Standard Model Predictions for CP-averaged observables.}}
\begin{center}
\rowcolors{1}{}{lgris}
\begin{tabular}{@{}lcrrr@{}}
\toprule[1.1pt]
Bin (GeV$^2$) & & \cen{$\av{S_{2s}}$} & \cen{$\av{S_{2c}}$} & \cen{$\av{S_3}$}  \\ [1mm]
\hline
%
[\,1\,,\,2\,]  & &
$0.089_{-0.045 - 0.005}^{+0.058 + 0.006}$ & 
$-0.605_{-0.179 - 0.021}^{+0.229 + 0.024}$ & 
$0.001_{-0.001 - 0.009}^{+0.002 + 0.009}$ \\ [1mm]
[\,0.1\,,\,2\,]  & &
$0.132_{-0.045 - 0.004}^{+0.040 + 0.004}$ & 
$-0.323_{-0.198 - 0.019}^{+0.178 + 0.020}$ & 
$0.002_{-0.001 - 0.011}^{+0.002 + 0.011}$ \\ [1mm]
[\,2\,,\,4.3\,]  & &
$0.057_{-0.033 - 0.004}^{+0.051 + 0.004}$ & 
$-0.754_{-0.128 - 0.015}^{+0.198 + 0.018}$ & 
$-0.006_{-0.005 - 0.005}^{+0.004 + 0.005}$ \\ [1mm]
[\,4.3\,,\,8.68\,]  & &
$0.090_{-0.044 - 0.006}^{+0.055 + 0.005}$ & 
$-0.634_{-0.175 - 0.022}^{+0.216 + 0.022}$ & 
$-0.021_{-0.013 - 0.01}^{+0.010 + 0.010}$ \\ [1mm]
[\,10.09\,,\,12.89\,]  & &
$0.129_{-0.041 - 0.003}^{+0.052 + 0.003}$ & 
$-0.482_{-0.163 - 0.014}^{+0.208 + 0.013}$ & 
$-0.046_{-0.060 - 0.007}^{+0.073 + 0.008}$ \\ [1mm]
[\,14.18\,,\,16\,]  & &
$0.150_{-0.035 - 0.001}^{+0.060 + 0.001}$ & 
$-0.396_{-0.141 - 0.004}^{+0.241 + 0.004}$ & 
$-0.106_{-0.105 - 0.004}^{+0.222 + 0.004}$ \\ [1mm]
[\,16\,,\,19\,]  & &
$0.160_{-0.018 - 0.001}^{+0.033 + 0.001}$ & 
$-0.357_{-0.074 - 0.003}^{+0.133 + 0.003}$ & 
$-0.193_{-0.078 - 0.003}^{+0.177 + 0.003}$ \\ [1mm]
[\,1\,,\,6\,]  & &
$0.070_{-0.038 - 0.004}^{+0.054 + 0.005}$ & 
$-0.703_{-0.149 - 0.017}^{+0.212 + 0.019}$ & 
$-0.008_{-0.006 - 0.006}^{+0.004 + 0.005}$ \\ [1mm]
%
%
\midrule[1.1pt]
%
%
\rowcolor{white} & & \cen{$\av{S_4}$} & \cen{$\av{S_5}$} &  \cen{$\av{S_{6s}}$} \\ [1mm]
\hline
%
[\,1\,,\,2\,]  & &
$-0.037_{-0.003 - 0.003}^{+0.009 + 0.003}$ & 
$0.179_{-0.045 - 0.007}^{+0.028 + 0.007}$   & 
$0.283_{-0.146 - 0.019}^{+0.192 + 0.021}$  \\ [1mm]
[\,0.1\,,\,2\,]  & &
$-0.071_{-0.008 - 0.004}^{+0.018 + 0.004}$ & 
$0.220_{-0.042 - 0.009}^{+0.013 + 0.007}$   & 
$0.181_{-0.064 - 0.021}^{+0.06 + 0.022}$  \\ [1mm]
[\,2\,,\,4.3\,]  & &
$0.118_{-0.037 - 0.005}^{+0.034 + 0.005}$ & 
$-0.139_{-0.051 - 0.008}^{+0.053 + 0.008}$   & 
$0.107_{-0.072 - 0.011}^{+0.091 + 0.012}$  \\ [1mm]
[\,4.3\,,\,8.68\,]  & &
$0.239_{-0.042 - 0.008}^{+0.015 + 0.006}$ & 
$-0.416_{-0.027 - 0.015}^{+0.073 + 0.016}$   & 
$-0.293_{-0.184 - 0.019}^{+0.15 + 0.021}$  \\ [1mm]
[\,10.09\,,\,12.89\,]  & &
$0.269_{-0.053 - 0.004}^{+0.028 + 0.003}$ & 
$-0.444_{-0.024 - 0.009}^{+0.120 + 0.009}$   & 
$-0.494_{-0.199 - 0.013}^{+0.218 + 0.014}$  \\ [1mm]
[\,14.18\,,\,16\,]  & &
$0.283_{-0.120 - 0.002}^{+0.056 + 0.002}$ & 
$-0.380_{-0.104 - 0.005}^{+0.156 + 0.005}$   & 
$-0.539_{-0.265 - 0.007}^{+0.255 + 0.007}$  \\ [1mm]
[\,16\,,\,19\,]  & &
$0.302_{-0.086 - 0.001}^{+0.039 + 0.001}$ & 
$-0.287_{-0.136 - 0.004}^{+0.129 + 0.004}$   & 
$-0.48_{-0.273 - 0.006}^{+0.229 + 0.006}$  \\ [1mm]
[\,1\,,\,6\,]  & &
$0.123_{-0.034 - 0.005}^{+0.027 + 0.004}$ & 
$-0.154_{-0.047 - 0.008}^{+0.053 + 0.008}$   & 
$0.047_{-0.047 - 0.011}^{+0.044 + 0.012}$  \\ [1mm]
%
%
\midrule[1.1pt]
%
%
\rowcolor{white} & & \cen{$\av{S_7}$} & \cen{$\av{S_8}$} & \cen{$\av{S_9}$} \\ [1mm]
\hline
%
[\,1\,,\,2\,]  & &
$0.048_{-0.013 - 0.007}^{+0.018 + 0.007}$ & 
$-0.014_{-0.007 - 0.004}^{+0.005 + 0.004}$ & 
$0.001_{-0.001 - 0.01}^{+0.001 + 0.008}$  \\ [1mm]
[\,0.1\,,\,2\,]  & &
$0.034_{-0.013 - 0.011}^{+0.015 + 0.011}$ & 
$-0.008_{-0.006 - 0.005}^{+0.004 + 0.005}$ & 
$0.001_{-0.000- 0.010}^{+0.001 + 0.012}$  \\ [1mm]
[\,2\,,\,4.3\,]  & &
$0.041_{-0.012 - 0.013}^{+0.018 + 0.012}$ & 
$-0.015_{-0.007 - 0.005}^{+0.004 + 0.005}$ & 
$0.001_{-0.000- 0.005}^{+0.001 + 0.005}$  \\ [1mm]
[\,4.3\,,\,8.68\,]  & &
$0.013_{-0.005 - 0.028}^{+0.009 + 0.028}$ & 
$-0.005_{-0.005 - 0.012}^{+0.002 + 0.013}$ & 
$0.000_{-0.000- 0.010}^{+0.000+ 0.009}$  \\ [1mm]
[\,10.09\,,\,12.89\,]  & &
$-0.001_{-0.001 - 0.017}^{+0.002 + 0.017}$ & 
$0.004_{-0.002 - 0.007}^{+0.001 + 0.007}$ & 
$-0.001_{-0.000- 0.007}^{+0.001 + 0.007}$  \\ [1mm]
[\,14.18\,,\,16\,]  & &
$0.000_{-0.000- 0.000}^{+0.000+ 0.000}$ & 
$0.004_{-0.002 - 0.001}^{+0.002 + 0.001}$ & 
$-0.002_{-0.001 - 0.001}^{+0.001 + 0.001}$  \\ [1mm]
[\,16\,,\,19\,]  & &
$0.000_{-0.000- 0.000}^{+0.000+ 0.000}$ & 
$0.002_{-0.001 - 0.000}^{+0.001 + 0.001}$ & 
$-0.002_{-0.001 - 0.001}^{+0.001 + 0.001}$  \\ [1mm]
[\,1\,,\,6\,]  & &
$0.040_{-0.012 - 0.013}^{+0.016 + 0.013}$ & 
$-0.014_{-0.007 - 0.005}^{+0.004 + 0.006}$ & 
$0.001_{-0.000- 0.005}^{+0.001 + 0.006}$  \\ [1mm]
%
%
\bottomrule[1.1pt]
\end{tabular} 
\end{center} 
\end{table}  


\begin{table}
\ra{1.35}
\rb{6mm}
\refstepcounter{table}
\label{tabSM6}
\footnotesize
{\ \ \textsf{\small Table \arabic{table}. Standard Model Predictions for CP-violating observables.}}
\begin{center}
\rowcolors{1}{}{lgris}
\begin{tabular}{@{}lcrrr@{}}
\toprule[1.1pt]
Bin (GeV$^2$) & & \cen{$10^2 \times \av{A_{2s}}$} & \cen{$10^2 \times \av{A_{2c}}$} & \cen{$10^2 \times \av{A_3}$}  \\ [1mm]
\hline
%
[\,1\,,\,2\,]  & &
$-0.095_{-0.085 - 0.018}^{+0.053 + 0.015}$ & 
$-0.387_{-0.142 - 0.048}^{+0.163 + 0.056}$ & 
$-0.002_{-0.001 - 0.026}^{+0.001 + 0.027}$ \\ [1mm]
[\,0.1\,,\,2\,]  & &
$-0.108_{-0.078 - 0.013}^{+0.045 + 0.013}$ & 
$-0.208_{-0.141 - 0.035}^{+0.121 + 0.038}$ & 
$0.000_{-0.000- 0.034}^{+0.000+ 0.034}$ \\ [1mm]
[\,2\,,\,4.3\,]  & &
$-0.016_{-0.029 - 0.008}^{+0.017 + 0.008}$ & 
$-0.479_{-0.115 - 0.034}^{+0.150 + 0.042}$ & 
$-0.007_{-0.006 - 0.017}^{+0.004 + 0.016}$ \\ [1mm]
[\,4.3\,,\,8.68\,]  & &
$0.066_{-0.032 - 0.007}^{+0.039 + 0.006}$ & 
$-0.402_{-0.121 - 0.023}^{+0.147 + 0.024}$ & 
$-0.016_{-0.009 - 0.013}^{+0.008 + 0.014}$ \\ [1mm]
[\,10.09\,,\,12.89\,]  & &
$0.049_{-0.031 - 0.003}^{+0.040 + 0.003}$ & 
$-0.169_{-0.033 - 0.008}^{+0.050 + 0.008}$ & 
$-0.014_{-0.010 - 0.006}^{+0.008 + 0.007}$ \\ [1mm]
[\,14.18\,,\,16\,]  & &
$0.002_{-0.001 - 0.000}^{+0.001 + 0.000}$ & 
$-0.004_{-0.002 - 0.000}^{+0.003 + 0.000}$ & 
$-0.001_{-0.001 - 0.000}^{+0.003 + 0.000}$ \\ [1mm]
[\,16\,,\,19\,]  & &
$0.002_{-0.001 - 0.000}^{+0.001 + 0.000}$ & 
$-0.004_{-0.002 - 0.000}^{+0.002 + 0.000}$ & 
$-0.002_{-0.001 - 0.000}^{+0.002 + 0.000}$ \\ [1mm]
[\,1\,,\,6\,]  & &
$-0.008_{-0.029 - 0.008}^{+0.017 + 0.007}$ & 
$-0.446_{-0.123 - 0.035}^{+0.155 + 0.040}$ & 
$-0.008_{-0.006 - 0.016}^{+0.004 + 0.015}$ \\ [1mm]
%
%
\midrule[1.1pt]
%
%
\rowcolor{white} & & \cen{$10^2 \times \av{A_4}$} & \cen{$10^2 \times \av{A_5}$} &  \cen{$10^2 \times \av{A_{6s}}$} \\ [1mm]
\hline
%
[\,1\,,\,2\,]  & &
$0.033_{-0.010 - 0.035}^{+0.033 + 0.031}$ & 
$-0.413_{-0.080 - 0.063}^{+0.068 + 0.069}$   & 
$-0.285_{-0.202 - 0.055}^{+0.145 + 0.054}$  \\ [1mm]
[\,0.1\,,\,2\,]  & &
$-0.008_{-0.011 - 0.030}^{+0.026 + 0.029}$ & 
$-0.240_{-0.062 - 0.052}^{+0.051 + 0.060}$   & 
$-0.140_{-0.060 - 0.068}^{+0.048 + 0.065}$  \\ [1mm]
[\,2\,,\,4.3\,]  & &
$0.131_{-0.029 - 0.023}^{+0.026 + 0.022}$ & 
$-0.531_{-0.115 - 0.051}^{+0.132 + 0.056}$   & 
$-0.468_{-0.402 - 0.057}^{+0.261 + 0.052}$  \\ [1mm]
[\,4.3\,,\,8.68\,]  & &
$0.187_{-0.028 - 0.011}^{+0.015 + 0.009}$ & 
$-0.428_{-0.052 - 0.022}^{+0.071 + 0.023}$   & 
$-0.467_{-0.284 - 0.033}^{+0.226 + 0.033}$  \\ [1mm]
[\,10.09\,,\,12.89\,]  & &
$0.099_{-0.040 - 0.006}^{+0.033 + 0.005}$ & 
$-0.169_{-0.071 - 0.010}^{+0.075 + 0.010}$   & 
$-0.213_{-0.186 - 0.011}^{+0.143 + 0.012}$  \\ [1mm]
[\,14.18\,,\,16\,]  & &
$0.003_{-0.002 - 0.000}^{+0.002 + 0.000}$ & 
$0.000_{-0.000- 0.000}^{+0.000+ 0.000}$   & 
$0.000_{-0.000- 0.000}^{+0.000+ 0.000}$  \\ [1mm]
[\,16\,,\,19\,]  & &
$0.003_{-0.002 - 0.000}^{+0.001 + 0.000}$ & 
$0.000_{-0.000- 0.000}^{+0.000+ 0.000}$   & 
$0.000_{-0.000- 0.000}^{+0.000+ 0.000}$  \\ [1mm]
[\,1\,,\,6\,]  & &
$0.132_{-0.025 - 0.022}^{+0.020 + 0.020}$ & 
$-0.505_{-0.082 - 0.045}^{+0.108 + 0.050}$   & 
$-0.461_{-0.348 - 0.050}^{+0.244 + 0.047}$  \\ [1mm]
%
%
\midrule[1.1pt]
%
%
\rowcolor{white} & & \cen{$10^2 \times \av{A_7}$} & \cen{$10^2 \times \av{A_8}$} & \cen{$10^2 \times \av{A_9}$} \\ [1mm]
\hline
%
[\,1\,,\,2\,]  & &
$0.466_{-0.199 - 0.059}^{+0.088 + 0.062}$ & 
$-0.341_{-0.087 - 0.033}^{+0.155 + 0.032}$ & 
$0.016_{-0.009 - 0.026}^{+0.011 + 0.026}$  \\ [1mm]
[\,0.1\,,\,2\,]  & &
$0.360_{-0.155 - 0.057}^{+0.078 + 0.058}$ & 
$-0.280_{-0.079 - 0.022}^{+0.127 + 0.023}$ & 
$0.015_{-0.006 - 0.036}^{+0.005 + 0.033}$  \\ [1mm]
[\,2\,,\,4.3\,]  & &
$0.335_{-0.159 - 0.049}^{+0.080 + 0.054}$ & 
$-0.223_{-0.062 - 0.025}^{+0.118 + 0.025}$ & 
$0.011_{-0.007 - 0.015}^{+0.009 + 0.017}$  \\ [1mm]
[\,4.3\,,\,8.68\,]  & &
$0.122_{-0.054 - 0.034}^{+0.012 + 0.034}$ & 
$-0.064_{-0.015 - 0.015}^{+0.042 + 0.016}$ & 
$0.003_{-0.003 - 0.012}^{+0.002 + 0.013}$  \\ [1mm]
[\,10.09\,,\,12.89\,]  & &
$0.025_{-0.015 - 0.012}^{+0.006 + 0.013}$ & 
$-0.008_{-0.004 - 0.006}^{+0.01 + 0.006}$ & 
$-0.001_{-0.001 - 0.007}^{+0.000+ 0.007}$  \\ [1mm]
[\,14.18\,,\,16\,]  & &
$0.000_{-0.000- 0.000}^{+0.000+ 0.000}$ & 
$0.001_{-0.000- 0.000}^{+0.000+ 0.000}$ & 
$-0.001_{-0.000- 0.000}^{+0.000+ 0.000}$  \\ [1mm]
[\,16\,,\,19\,]  & &
$0.000_{-0.000- 0.000}^{+0.000+ 0.000}$ & 
$0.000_{-0.000- 0.000}^{+0.000+ 0.000}$ & 
$0.000_{-0.000- 0.000}^{+0.000+ 0.000}$  \\ [1mm]
[\,1\,,\,6\,]  & &
$0.306_{-0.139 - 0.046}^{+0.057 + 0.050}$ & 
$-0.206_{-0.050 - 0.024}^{+0.104 + 0.023}$ & 
$0.010_{-0.007 - 0.014}^{+0.008 + 0.016}$  \\ [1mm]
%
%
\bottomrule[1.1pt]
\end{tabular} 
\end{center} 
\end{table}  


\begin{table}
\ra{1.35}
\rb{6mm}
\refstepcounter{table}
\label{tabSM7}
\footnotesize
{\ \ \textsf{\small Table \arabic{table}. Standard Model Predictions for CP-averaged observables.}}
\begin{center}
\rowcolors{1}{}{lgris}
\begin{tabular}{@{}lcrrr@{}}
\toprule[1.1pt]
Bin (GeV$^2$) & & \cen{$\av{A_T^{(3)}}$} & \cen{$\av{A_T^{(4)}}$} & \cen{$\av{A_T^{(5)}}$}  \\ [1mm]
\hline
%
[\,1\,,\,2\,]  & &
$0.190_{-0.046 - 0.014}^{+0.046 + 0.014}$ & 
$2.056_{-0.340 - 0.156}^{+0.556 + 0.179}$ & 
$0.301_{-0.030 - 0.010}^{+0.029 + 0.010}$ \\ [1mm]
[\,0.1\,,\,2\,]  & &
$0.351_{-0.028 - 0.019}^{+0.023 + 0.021}$ & 
$1.516_{-0.134 - 0.094}^{+0.162 + 0.105}$ & 
$0.470_{-0.003 - 0.008}^{+0.003 + 0.006}$ \\ [1mm]
[\,2\,,\,4.3\,]  & &
$0.592_{-0.053 - 0.027}^{+0.070 + 0.029}$ & 
$0.592_{-0.112 - 0.039}^{+0.119 + 0.041}$ & 
$0.441_{-0.034 - 0.009}^{+0.036 + 0.007}$ \\ [1mm]
[\,4.3\,,\,8.68\,]  & &
$1.067_{-0.015 - 0.063}^{+0.014 + 0.060}$ & 
$0.869_{-0.034 - 0.049}^{+0.020 + 0.056}$ & 
$0.284_{-0.061 - 0.010}^{+0.060 + 0.009}$ \\ [1mm]
[\,10.09\,,\,12.89\,]  & &
$1.195_{-0.314 - 0.040}^{+0.532 + 0.036}$ & 
$0.825_{-0.284 - 0.026}^{+0.222 + 0.029}$ & 
$0.104_{-0.009 - 0.005}^{+0.089 + 0.007}$ \\ [1mm]
[\,14.18\,,\,16\,]  & &
$1.442_{-0.823 - 0.024}^{+1.186 + 0.025}$ & 
$0.671_{-0.354 - 0.013}^{+0.629 + 0.012}$ & 
$0.132_{-0.040 - 0.005}^{+0.016 + 0.006}$ \\ [1mm]
[\,16\,,\,19\,]  & &
$2.004_{-1.153 - 0.029}^{+1.653 + 0.031}$ & 
$0.476_{-0.252 - 0.008}^{+0.443 + 0.007}$ & 
$0.138_{-0.055 - 0.003}^{+0.046 + 0.003}$ \\ [1mm]
[\,1\,,\,6\,]  & &
$0.578_{-0.050 - 0.026}^{+0.065 + 0.029}$ & 
$0.631_{-0.104 - 0.039}^{+0.106 + 0.042}$ & 
$0.492_{-0.012 - 0.004}^{+0.007 + 0.002}$ \\ [1mm]
%
%
\bottomrule[1.1pt]
\end{tabular} 
\end{center} 
\end{table}  


\newpage


\end{document}